\documentclass[11pt]{article}
\parskip \bigskipamount
\usepackage{latexsym,epsf,bm}
\usepackage{amsmath}
\usepackage{amsfonts}
\usepackage{amssymb}
\usepackage{caption}
\usepackage{subcaption}
\usepackage{graphicx}
\usepackage{tikz}

\usetikzlibrary{arrows.meta, angles, quotes}

\newcommand{\uvec}[1]{\ensuremath{\hat{\boldsymbol{#1}}}}

\usepackage{physics}
\usepackage{bm}
\usepackage{array}
\usepackage{ragged2e} %For left justification

\title{Quantum Noise Limited Phased Arrays for Single-Electron Cyclotron Radiation Emission Spectroscopy}

\begin{document}

\author{Stafford Withington$^1$, Christopher Thomas$^2$ and Songyuan Zhao$^2$ \\ \\
$^1$ Clarendon Laboratory, Department Physics, University Oxford, UK \\  $^2$ Cavendish Laboratory, Department Physics, University Cambridge, UK }

\maketitle

%\date{\today}

{\bf Abstract:} Neutrino oscillation experiments show that neutrinos have mass; however, the absolute mass scale is exceedingly difficult to measure and is currently unknown. A promising approach is to measure the energies of the electrons released during the radioactive decay of tritium. The energies of interest are within a few eV of the 18.6 keV end point, and so are mildly relativistic. By capturing the electrons in a static magnetic field and measuring the frequency of the cyclotron radiation emitted the initial energy can be determined, but end-point events are infrequent, the observing times short, and the signal to noise ratios low. To achieve a resolution of $<$ 10 meV,  single-electron emission spectra need to be recorded over large fields of view with highly sensitive receivers. The principles of Cylotron Radiation Emission Spectroscopy (CRES) have already been demonstrated by Project 8, and now there is considerable interest in increasing the FoV to $>$ 0.1 m$^3$. We consider a range of issues relating to the design and optimisation of inward-looking quantum-noise-limited microwave receivers for single-electron CRES, and present a single framework for understanding signal, noise and system-level behaviour. Whilst there is a great deal of literature relating to the design of outward-looking phased arrays for applications such as radar and telecommunications, there is very little coverage of the new issues that come into play when designing ultra-sensitive inward-looking phased arrays for volumetric spectroscopy and imaging.

\section{Introduction}

One direct approach to determining the absolute mass scale of the neutrino is to measure the energy distribution of the electrons released during the radioactive decay of tritium \cite{Asner_2015, Formaggio, Katrin}. The energies of interest are within a few eV of the 18.6 keV end point, and can in principle be measured by placing atomic tritium in a static magnetic field and observing the cyclotron radiation of the decay electrons. Single electron Cyclotron Radiation Emission Spectroscopy (CRES)  has been demonstrated, and there is now considerable interest in developing the technique further; particularly with respect to  increasing the instrumented volume to $>$ 0.1 m$^3$ whilst maintaining a high spectral resolution across the field of view (FoV) \cite{Project8a}. An 18.6 keV electron in a 1 T field orbits with a radius of 0.5 mm and radiates principally at a wavelength of about 11 mm (27 GHz). It appears as a source having a physical size much smaller than a wavelength. At these energies $\beta = v/c \approx$ 0.25, and so the radiation pattern is that of a rotating synchrotron beam having a front-to-back intensity ratio of just under 10. When viewed by a single antenna, the sweeping beam creates spectral harmonics, but if the radiation is observed in a narrow band around the fundamental tone, the source appears as an orthogonal pair of out-of-phase electric dipoles, which is the lowest-order multipole term in the emitted radiation. 

A central requirement of a CRES experiment is that the energies of the decay electrons must be measured to one part in 10$^5$, which can in principle be achieved by surrounding the active volume with a system of antennas, down-converting the signal from each antenna into the MHz frequency range, and using digital sampling and software defined Fourier transform spectroscopy to give information about the dynamical behaviour of each decay event.  In practice, the technique is challenging. To achieve the necessary spectral resolution, each electron must be observed for longer than 200 $\mu$s, and yet an 18.6 keV electron in linear motion travels a distance of 16 km during this time. Unless a perfectly circular orbit can be formed, each electron spirals a long distance in a short period of time, making continuous monitoring difficult. One solution is to add a longitudinal trap, and to observe the electron's  helical motion as it sloshes forwards and backwards in the trap.  However, additional sidebands are then created, which together with radiative decay and occasional inelastic scattering, makes data analysis difficult unless an appreciable signal to noise ratio (SNR) can be achieved.  

\begin{figure}[!ht]
 \centering
\includegraphics[width=100mm]{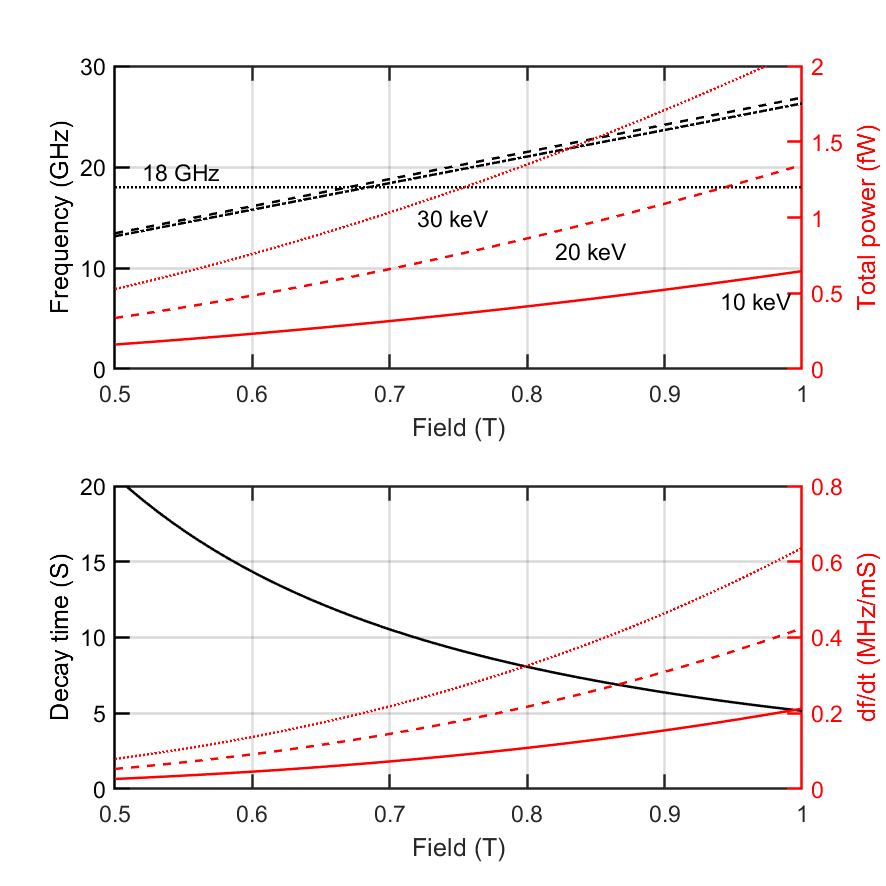}
\caption{\label{fig:basic_params_1} Top: Cyclotron frequency (black, left scale) and total power (red, right scale) as a function of magnetic field strength for electron energies of 
 10 keV (solid), 20 keV (dashed)  and 30 keV (dotted). The dotted horizontal black line corresponds to the frequency to which cryogenic microwave components are readily available (18 GHz). Bottom: Energy decay time (black, left scale) and the rate of change of frequency due to radiative decay (red, right scale). Again, these are shown for electron energies of  10 keV (solid), 20 keV (dashed)  and 30 keV (dotted).
 }
\end{figure}

CRES sets out to determine the energies of individual electrons through frequency measurements. The precision with which frequency can be measured depends on the SNR and the observing time available. A particular challenge is that the power radiated by an 18.6 keV electron in a 1 T field is of the order of 1 fW, and so the antenna system must be efficient, and the first stage of amplification highly sensitive. It is not possible to achieve the needed SNR by integrating for a long period of time, because only single-electron events are available. Ideally, the noise temperatures of the first stage amplifiers should approach the quantum limit (0.7 K at 28 GHz), and a low system noise temperature ($<$ 1.5 K) maintained despite the blackbody radiation emitted by warm surfaces, such as the trap and tritium-handling parts of the apparatus.  Figure~\ref{fig:basic_params_1}, top plot, shows the resonance frequency (black) and total power (red) as a function of magnetic field strength for electron energies of 10, 20 and 30 keV. The horizontal black dotted line is the frequency to which cryogenic microwave components having SMA connectors are readily available, which corresponds to a wavelength of 17 mm. The FoV, from the antennas' perspective, scales with wavelength, and so a large wavelength is beneficial, but this is traded off against power, which scales with frequency and therefore magnetic field.  The bottom plot shows the energy relaxation time (black), and the rate of change  of frequency caused by radiative decay (red). These parameters have a direct impact on the design of the readout system. 

\begin{figure}[!ht]
 \centering
\includegraphics[width=100mm]{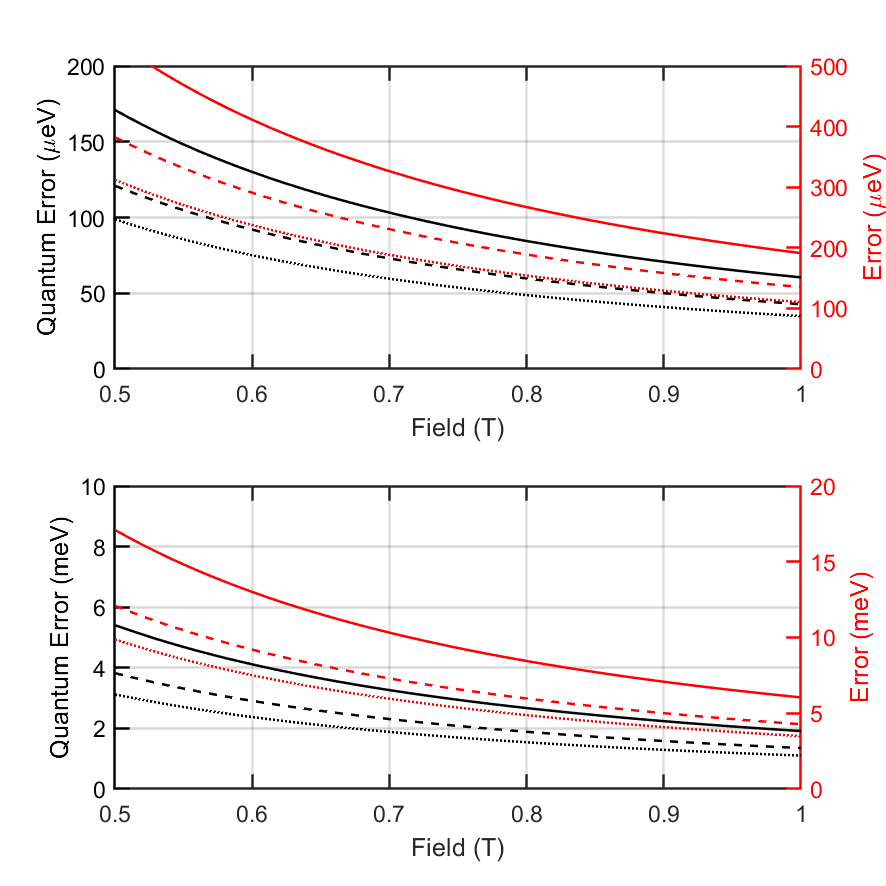}
\caption{\label{fig:basic_params_2} Limiting energy resolution of a system that uses quantum-noise limited amplifiers (black, left scale), and amplifiers having a noise temperature 10 times the quantum limit (red, right scale). The top plot is for a system having a total power collection efficiency of 1, and the bottom plot is for a total power collection efficiency of 0.001. All quantities are calculated for electron energies of 10 keV (solid), 20 keV (dashed)  and 30 keV (dotted). The electron-background scattering time is taken to be 1 ms.  }
\end{figure}

Figure~\ref{fig:basic_params_2} shows the Cram\'{e}r-Rao bound (CRB), which will be discussed later, on a measurement of electron energy  when a quantum-noise-limited amplifier is used (black), and when an amplifier having a noise temperature 10 times the quantum limit is used (red) . Both quantities are calculated for electron energies of 10, 20 and 30 keV. The system is identical to that modelled in Fig.~\ref{fig:basic_params_1} in all other respects. The top plot is for a total power collection efficiency of 1, and the bottom plot is for a total power collection efficiency of 0.001. It is assumed that the electron is observed in an uninterrupted way for 1 mS, which is taken to be the time to the first background scattering event.  The CRB is the lower bound of the maximum likelihood estimator of the initial energy of a chirped sinusoidal tone buried in noise. It is achieved asymptotically in the case of large SNR.  We shall return to this point later. Crucially, however, many other factors limit the actual resolution achieved---for example sampling and timing errors, particularly with respect to not knowing the exact time at which the pulse turns on. The energy-measurement error shown here corresponds to a noisy CRES detector being used in a perfect way. Overall, these plots indicate that excellent performance is in principle possible giving better than meV accuracy for a quantum-noise limited high-efficiency system. The design of a CRES system is, however, necessarily a delicate balance between competing requirements, and so careful studies are needed to ensure that the overall performance approaches this optimum.

A key difficulty is that the primary decay events are infrequent, and so the FoV of the instrument should ideally be of order 0.1 m$^3$ or more; also, the reception pattern of each antenna, or indeed of any synthesised beam achieved through phasing, must couple efficiently to the single-electron radiators regardless of where they appear in the instrumented volume.  Rather than using an array of antennas, another possibility is to use a high-Q cavity to enhance the coherent signal generated by each single-electron source. This approach has benefits, but care is needed because backaction can perturb behaviour, and the whole purpose of the experiment is to measure the energies of individual decay electrons at the time of their release. An electron can only radiate into field modes that are available, and so an electron in a high-Q cavity will not radiate in the same way as an electron in free space.

In this paper, we discuss a number of principles that can be used to guide the design of antenna-based CRES instruments. In \S \ref{sec:synch} we consider how a mildy relativistic electron appears when it is in a cylotron orbit and observed at its fundamental frequency. In \S \ref{sec:can_arr} we describe a specific configuration and use it to draw attention to several issues facing all antenna-based experiments. In  \S \ref{sec:full_elec_model} we describe a rigorous 2D electromagnetic model based on a scattering parameter methodology. The results are described in terms of cylindrical instrumented volume per unit length, which is valuable for design purposes.  It is convenient to use a 2D model because fundamental behavioural trends can be highlighted without the need to present the effects of polarisation in complicated 3D plots. Extending the model to 3D, and including polarisation, is straightforward for detailed numerical calculations, but the basic physical characteristics remain unchanged. The scattering parameter approach allows us, in \S \ref{sec:cres_receiver}, to model a complete CRES instrument, which takes into account correlated thermal noise at the ports of the antennas resulting from  ohmic loss and straylight coupling into the instrumented volume. The signal and noise models are then combined in \S \ref{sec:prob_data} and \S \ref{sec:fisher} to consider the bounds on the measurement of cyclotron frequency. In \S \ref{sec:fov} we use aspects of the model to highlight an information-theoretic limit, which places a constraint on the number of antennas needed to observe efficiently a given FoV. In \S \ref{sec:det_pos} we consider how the position of an electron can be determined from the digitised receiver outputs, and then in \S \ref{sec:synth} we discuss the same problem from the perspective of synthesised beams. 

Overall, we believe that this paper gives an insight into the range of issues encountered when trying to design and understand the behaviour of inward-looking phased arrays. By providing a single framework for understanding signal, noise and system-level behaviour, we hope to emphasise that antenna-system design cannot be separated from the matter of data analysis. We also believe that the work leads to valuable conceptual insights, and provides a number of powerful tools for comparing different antenna types and configurations.

\section{Relativistic cyclotron radiation pattern}
\label{sec:synch}

At $18.6$ keV the motion of an orbiting electron is mildly relativistic with $\beta = v/c = 0.25$. Radiation from such electrons inevitably contains synchrotron characteristics, such as relativistic beaming. In order to properly consider the radiation from energetic electrons, the relativistic Liénard-Wiechert potentials can be used:
\begin{align}
\Phi(\mathbf{r},t)&=\frac{1}{4\pi\epsilon_0}q\left[\frac{1}{R(1-\bm{\beta}\cdot\mathbf{n})}\right]_{\mathrm{ret}} \\
\mathbf{A}(\mathbf{r},t)&=\frac{\mu_0}{4\pi}qc\left[\frac{\bm{\beta}}{R(1-\bm{\beta}\cdot\mathbf{n})}\right]_{\mathrm{ret}}\,,
\end{align}
where $\bm{\beta}$ is the velocity vector divided by the speed of light $c$, $[ \,\, ]_\mathrm{ret}$ denotes evaluating quantities in retarded time $t'=t-R_{\mathrm{ret}}/c$, and $R$ is the distance from the electron to the observer along $\mathbf{n}$. The radiated fields can be readily obtained by substituting the Liénard-Wiechert potentials into the field strength tensor to give
\begin{align}
\mathbf{E}(\mathbf{r},t)&=\frac{q}{4\pi\epsilon_0}\left[\frac{\mathbf{n}\times\{(\mathbf{n}-\bm{\beta})\times\dot{\bm{\beta}}\}}{cR(1-\bm{\beta}\cdot\mathbf{n})^{3}}\right]_{\mathrm{ret}}\, \label{eq:radiating_E}\\
\mathbf{B}(\mathbf{r},t)&=\frac{\mathbf{n}_{\mathrm{ret}}\times\mathbf{E}}{c}\,.
\end{align}

\begin{figure}[!ht]
\noindent \centering
\includegraphics[width=8.0cm]{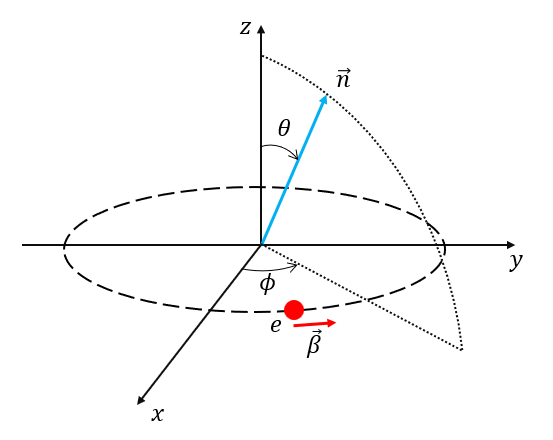}
\caption{\label{fig:cyclotron_set_up} Coordinate system for the analysis of an electron in circular motion.}
\end{figure}

These formula can be applied to the case where the electron is in circular motion, as shown in Fig.~\ref{fig:cyclotron_set_up}. Without loss of generality, the electron here performs circular motion in the $x-y$ plane around the $z-$axis. We have made the assumption that radiation does not appreciably change the energy of the electron within the time frame of the analysis. In the above, 
$\mathbf{n}=[\sin\theta \cos\phi,\sin\theta\sin\phi,\cos\theta]$ is the unit vector to the observer,  $\mathbf{r}_\mathrm{e}=r_\mathrm{0}[\cos(\omega_0t),\sin(\omega_0t),0]$ is the electron's position,  ${\bm \beta}=\beta_\mathrm{0}[-\sin(\omega_0t),\cos(\omega_0t),0]$ is the velocity,  and $\dot{\bm \beta}=-\beta_\mathrm{0}\omega_0[\cos(\omega_0t),\sin(\omega_0t),0]$ the acceleration. $\omega_0=eB/m_\mathrm{e,\gamma}$ is the orbital frequency,  $r_\mathrm{0}=m_\mathrm{e,\gamma}\beta_0c/(eB)$ is the orbital radius, $m_\mathrm{e,\gamma}=\gamma m_\mathrm{e}$ is the relativistic mass of the electron, and $m_\mathrm{e}$ is the rest mass.

Substituting the above identities into (\ref{eq:radiating_E}) gives the following field:
\begin{align}
\mathbf{E}(\mathbf{r},t)=& \frac{q}{4\pi\epsilon_0}\frac{\beta_0\omega_0}{cR}   \frac{1}{[1+\beta_0\sin\theta\sin(\omega_0t-\phi)]^3}\notag\\
& \Bigl[ \cos\theta\cos(\omega_0t-\phi)\,\mathbf{e}_\theta + \sin(\omega_0t-\phi)\,\mathbf{e}_\phi+\beta_0\sin\theta\,\mathbf{e}_\phi \Bigr]  \biggr]_{\mathrm{ret}} \label{eq:Radiation_explicit}\,.
\end{align}

\begin{figure}[!ht]
\noindent \centering
\includegraphics[width=8.0cm]{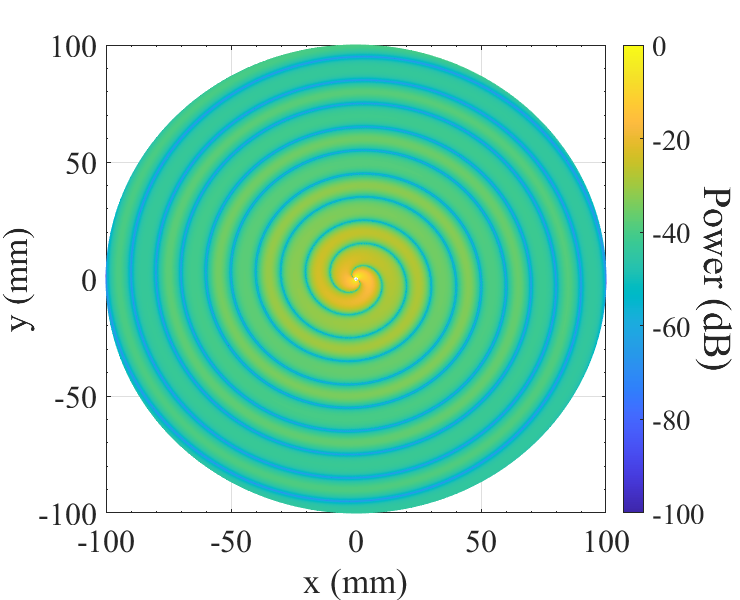}
\caption{\label{fig:cyclotron_power} Normalised power radiated by an electron orbiting in the $x-y$ plane. In this simulation,  $\omega_0=15\,\mathrm{GHz}$ and $\beta_0=0.25$.}
\end{figure}

The above result can be verified against \cite{Katoh_2017}. Figure~\ref{fig:cyclotron_power} shows the normalised power in the $x-y$ plane radiated by an electron in a circular orbit where $\omega_0=15\,\mathrm{GHz}$ and $\beta_0=0.25$. The total power radiated can be calculated using the Liénard extension of the Larmor formula
\begin{align}
  P&=\frac{Z_0 q^2}{6\pi}\gamma^4\left[\dot{\beta}^2+\gamma^2\left(\beta\cdot\dot{\beta}\right)^2\right]\, \label{eq:power_Larmor} \\
  &=\frac{Z_0}{6\pi}\left(\frac{qr_0\omega_0^2}{c}\right)^2\gamma^4\,,
\end{align}
where $Z_0$ is the impedance of free space. 

% Figures path % C:\Work\Postdoc\Investigations\Cyclotron source study

The full field equations could be used for detailed optimisation of CRES antenna systems, but to gain conceptual insight, it is useful to focus on specific terms and their effects. Even for  $\beta_0 = 0.25$, synchrotron radiation exhibits beaming, where, in the laboratory frame, the forward lobe is enhanced and the backward lobe is suppressed. This effect can  be understood by considering the multiplicative beaming term in (\ref{eq:Radiation_explicit}):
\begin{align}
\label{eqn:beaming_term}
  \frac{1}{[1+\beta_0\sin\theta\sin(\omega_0t-\phi)]^3} \,.
\end{align}

\begin{figure}[!ht]
\noindent \centering
\includegraphics[width=8cm]{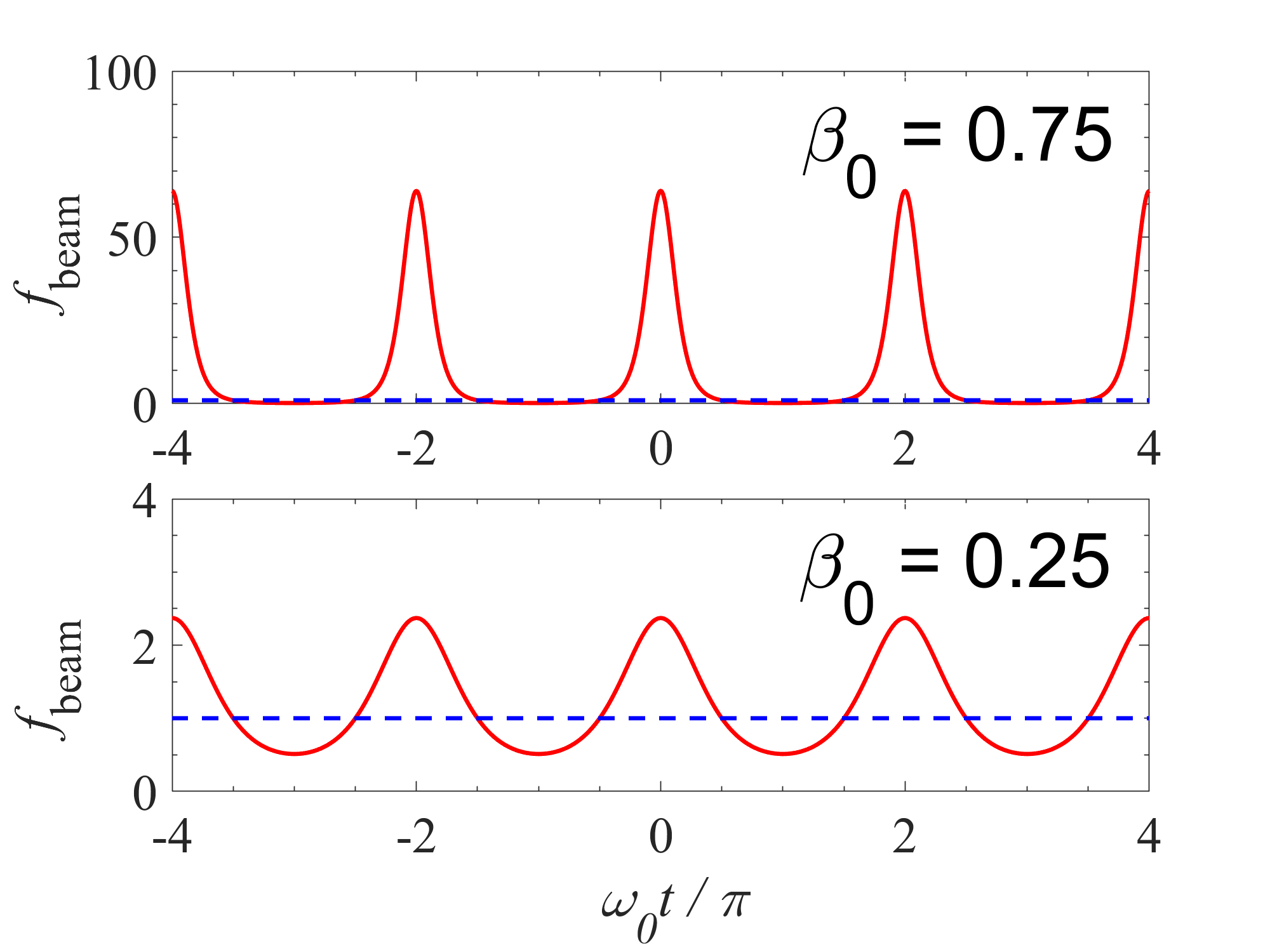}
\caption{\label{fig:cyclotron_beaming} Magnitude of the beaming factor as observed at $\theta = \pi/2$ and $\phi=\pi/2$. The time axis refers to retarded time. Top: Beaming factor well into the relativistic regime with $\beta_0 = 0.75$. Bottom: Beaming factor in the intermediate regime with $\beta_0 = 0.25$. The dotted blue line shows the non-relativistic limit, with $\beta_0 = 0$, of unity.}
\end{figure}
Figure~\ref{fig:cyclotron_beaming} shows the magnitude of the beaming factor as observed at $\theta = \pi/2$ and $\phi=\pi/2$, i.e. in the direction of the forward lobe at $t=0$. When the velocity of the electron is highly relativistic, the beaming factor enhances the radiation strongly over a narrow angular window, as shown in the upper plot for $\beta_0 = 0.75$. The bottom plot shows the beaming factor of an electron with an intermediate velocity, $\beta_0 = 0.25$, as in the case of a neutrino mass experiment. The periodicity of the beaming factor is the same as that of the radiation, with $\omega_0t=2\pi$, since an electron will again face the same observation point after making a full orbit.

\begin{figure}[!ht]
\noindent \centering
\includegraphics[width=8.0cm]{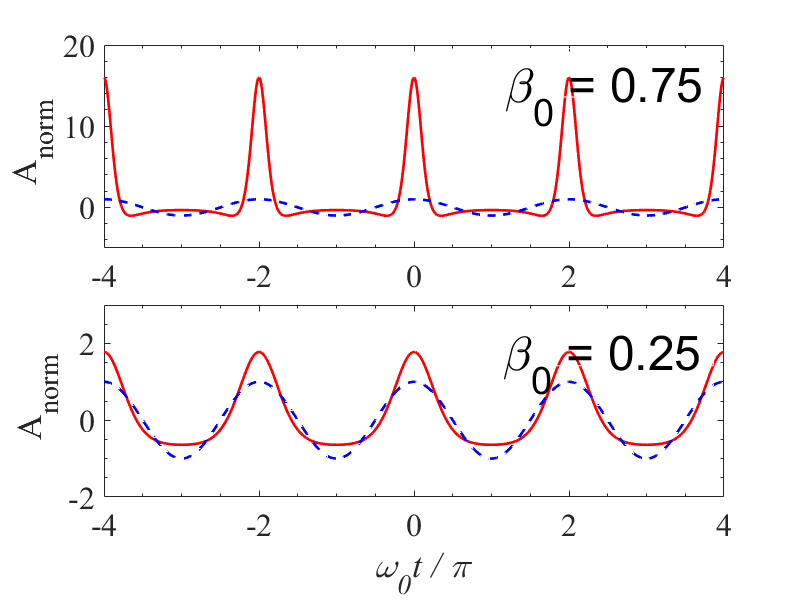}
\caption{\label{fig:cyclotron_beaming_field} Magnitude of the normalised electric field as observed at $\theta = \pi/2$ and $\phi=\pi/2$. The time axis refers to retarded time. Top: Electric field well into the relativistic regime with $\beta_0 = 0.75$. Bottom: Electric field in the intermediate regime with $\beta_0 = 0.25$. The dotted blue line shows the non-relativistic limit, with $\beta_0 = 0$, of unit magnitude.}
\end{figure}

Figure~\ref{fig:cyclotron_beaming_field} shows the magnitude of normalised electric field observed at $\theta = \pi/2$ and $\phi=\pi/2$. Normalisation is applied with respect to constant terms in (\ref{eq:Radiation_explicit}) such that the field in the non-relativistic limit ($\beta_0 = 0$) has a magnitude of unity. Comparing Fig.~\ref{fig:cyclotron_beaming_field} with Fig.~\ref{fig:cyclotron_beaming}, we observe that the magnitude of the field is enhanced when the beaming factor is greater than unity, and the magnitude is reduced when the beaming factor is less than unity. The waveform of the electric field is no longer purely sinusoidal. This distortion results in frequency components other than the non-relativistic cyclotron frequency.

\begin{figure}[!ht]
\noindent \centering
\includegraphics[width=8.0cm]{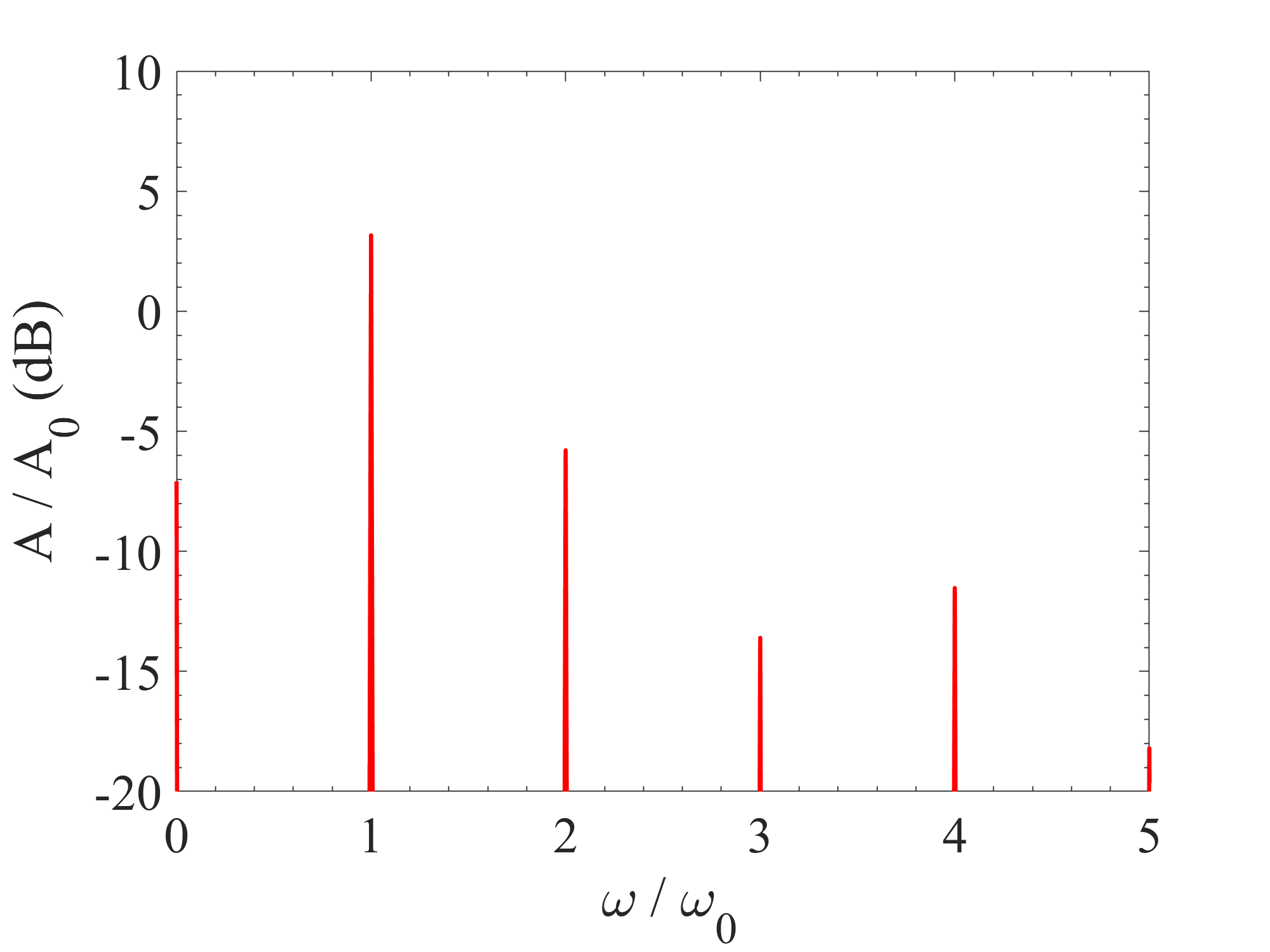}
\caption{\label{fig:cyclotron_frequencies} Normalised spectrum of the electric field radiated by a cyclotron electron with $\beta_0 = 0.25$, as observed at $\theta = \pi/2$ and $\phi=\pi/2$.}
\end{figure}

The relativistic beaming term is a periodic function of time and imparts additional frequency components to the radiation spectrum, compared with the non-relativistic case. Figure~\ref{fig:cyclotron_frequencies} shows the frequency spectrum of the radiated electric field when $\beta_0 = 0.25$, normalised against the spectrum when $\beta_0=0$. It can be seen that the radiated electric field contains components that are integer multiples of the fundamental non-relativistic cyclotron frequency $\omega_0$. In a CRES experiment only the fundamental frequency is observed, and the harmonics are a source of loss. Here we take the view that for even relatively small magnetic fields, $>$ 0.2 T, it would not be realistic to observe a number of harmonics simultaneously, even though it may be  beneficial theoretically: we shall comment on this point later. In the next section, we shall show that when $\beta_0$ is small, the fraction of power in the fundamental frequency can be approximated by
\begin{align}
  \frac{\langle P_\mathrm{fund}\rangle}{\langle P_\mathrm{total}\rangle}=\frac{1}{\gamma^4}\,, \label{eq:Fund_energy}
\end{align}
which is $0.88$ when $\beta_0=0.25$, which is already starting to eat into the efficiency budget.

\subsection{Radiation pattern}

It is important to know the radiation pattern at the fundamental frequency since it contains most of the power when $\beta_0=0.25$. The form of this radiation can be obtained in two ways: (i) we can extract the fundamental frequency terms in (\ref{eq:Radiation_explicit}) for small $\beta_0$, or (ii) we can build up the overall radiation pattern using electromagnetic multipoles.

To first order in $\beta_0$, the term that contains the fundamental frequency in (\ref{eq:Radiation_explicit}) is given by
\begin{align}
  \mathbf{E}_\mathrm{fund}=& \frac{q}{4\pi\epsilon_0}\frac{\beta_0\omega_0}{cR}\left[\cos\theta\cos(\omega_0t-\phi)\,\mathbf{e}_\theta\right.\notag\\
  &+\left.\sin(\omega_0t-\phi)\,\mathbf{e}_\phi\right]\,. \label{eq:approximate_Fundamental}
\end{align}
This is a good approximation since the next lowest-order radiating term is third order in $\beta_0$.

Alternatively, the radiation pattern can be approximated using the lowest order of multipole radiation (that is non-zero), i.e. electric dipoles in the case of CRES experiments. The motion of the electron can be approximated by two sinusoidal dipoles along the $x$ and $y$ axes, $\pi/2$ out of phase with each other (and also a monopole, which does not radiate). The dipole moment of this system is given by
\begin{align}
  \mathbf{p}=q\mathbf{r}_\mathrm{e}=p_\mathrm{0}[\cos(\omega_0t),\sin(\omega_0t),0]\,,
\end{align}
where $p_\mathrm{0}= qr_0$ is the magnitude of the electric dipole moment. The corresponding electric field can be calculated using
\begin{align}
  \mathbf{E}(\mathbf{r},t)=\frac{1}{4\pi\epsilon_0 Rc^2}  \Bigl[ \mathbf{n}\times(\mathbf{n}\times\ddot{\mathbf{p}}) \Bigr]_{\mathrm{ret}}\,,
\end{align}
yielding the same radiating field as (\ref{eq:approximate_Fundamental}). Beyond the rotating dipole approximation, the full radiation pattern contains higher order terms such as quadrupole, magnetic multipoles, etc. The powers in these higher order multipole terms are smaller, compared with the power from the rotating dipole radiation, by at least a factor of $\beta_0^2$ \cite{Likharev_2018}.

One might be tempted to approximate the cyclotron field using a current loop. The magnetic dipole of a current loop is given by
\begin{align}
  \mathbf{m}=I\mathbf{A}=\frac{q}{2}\frac{\beta_0^2 c^2}{\omega_0}\hat{\mathbf{z}}\,,
\end{align}
and the radiating field can be obtained by
\begin{align}
  \mathbf{E}(\mathbf{r},t)=\frac{1}{4\pi\epsilon_0 Rc^3}  \Bigl[ (\mathbf{n}\times\ddot{\mathbf{m}}) \Bigr]_{\mathrm{ret}}\,.
\end{align}
However, this term is non-radiating as the magnetic dipole does not vary with time. Artificially imposing $\mathbf{m}=\mathbf{m}_0 \cos(\omega_0t)$ results in a simple dipole radiation field that is unrelated to $\mathbf{E}_\mathrm{fund}$ in terms of radiation pattern, and is a factor of $\beta_0/2$ smaller than $\mathbf{E}_\mathrm{fund}$ in a peak-to-peak comparison.

\begin{figure}[!ht]
\noindent \centering
\includegraphics[width=8.0cm]{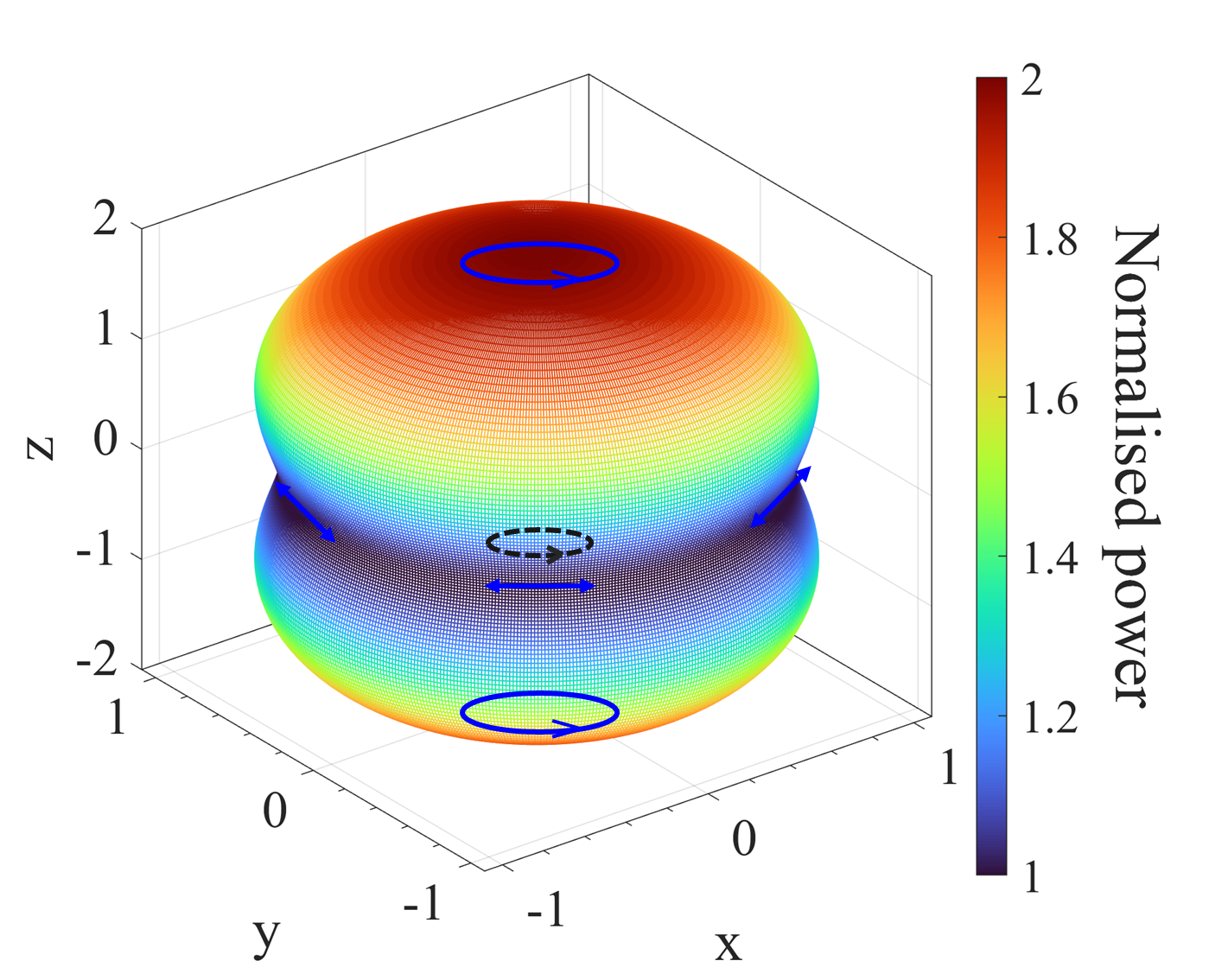}
\caption{\label{fig:rotating_dipole_radiation} Normalised power pattern of the radiation from a rotating dipole. The dotted black line and arrow indicates the cyclotron orbit. The blue arrows show the polarisation of the radiated electric field.}
\end{figure}

It is useful to visualise the radiation pattern associated with the rotating dipole approximation. It can be shown that the field (\ref{eq:approximate_Fundamental}) leads to a time-averaged energy flux of
\begin{align}
  \langle S_\mathrm{fund} \rangle = \frac{Z}{2}\frac{p_0\omega_0^2}{4\pi Rc}(1+\cos^2\theta)\,, \label{eq:Energy_flux}
\end{align}
and a total power of
\begin{align}
  \langle P_\mathrm{fund} \rangle = \frac{Z}{6\pi}\left(\frac{p_0\omega_0^2}{c}\right)^2\,, \label{eq:Fund_energy_calculation}
\end{align}
where the averaged power is conveniently twice that of an ordinary sinusoidal electric dipole. Comparing the total relativistic synchrotron power calculated using (\ref{eq:power_Larmor}) with the rotating dipole approximation in the fundamental frequency (\ref{eq:Fund_energy_calculation}), we find that their ratio is given by (\ref{eq:Fund_energy}). 

The radiation pattern is shown in Fig.~\ref{fig:rotating_dipole_radiation}, where we have normalised against the constants in front of the bracket in (\ref{eq:Energy_flux}). Unlike a simple dipole, the radiation pattern does not reduce to zero in any direction. This can be understood intuitively by noting that the null direction of a simple dipole is along its displacement. In the cyclotron case, there is no common direction that is along the displacement of both dipoles, which are orthogonal to each other. Further, we note that the strongest direction of radiation is along the $z-$axis, and weakest around the orbit in the $x-y$ plane. This can also be understood intuitively because a simple dipole radiates most strongly in directions orthogonal to its displacement. The $z-$axis is orthogonal to both dipoles and thus the radiation pattern peaks along the $z-$axis. In Fig.~\ref{fig:rotating_dipole_radiation}, the polarisation of the electric field is indicated by blue arrows. Over the $x-y$ plane, the radiation is linearly polarised in the $x-y$ plane and orthogonal to the direction of radiation. Along the $z-$axis, the radiation is circularly polarised.

The radiation pattern is of crucial importance when designing CRES antenna systems. For example, it implies that the best antenna system might comprise two helical antennas in-line with the static magnetic field: or as will be seen, arrays of helical antennas. However, for practical reasons, it is important to consider a ring of antennas in the $x-y$ plane. Because most electrons will move backwards and forwards longitudinally, in a trap, a cylindrical array of antennas seems to be the most natural geometry, even though power is radiated preferentially along the axis of the static field. These and related issues will be considered in later sections.  

\subsection{Geometrical limits on efficiency}
\label{sec:geometry}

Imagine an array of antennas on a cylindrical surface, where the ends are left open to allow a beam of tritium or indeed a beam of electrons  to pass through, and place an electron in a cyclotron orbit at the centre. It is valuable to consider the upper limit on the power that can be collected under this simple geometrical constraint. A limit can be found by integrating the total power in the fundamental frequency travelling radially across a cylindrical surface 
having half-length $L$:
\begin{align}
  \int_0^{2\pi}\int_{-L}^{L}\langle S \rangle R \sin^2\theta \, dz d\phi\,,
\end{align}
where $\langle S \rangle$ is the magnitude of the Poynting vector $\mathbf{E}\times\mathbf{H}$, $R=\sqrt{r^2+z^2}$ is the distance from the on-axis source to the point of observation, $r$ is the cylinder's radius, $z$ is the longitudinal position, and $\phi$ is the azimuthal angle. This corresponds to a configuration where there is perfect coupling between the antenna array and the radiated field, regardless of the direction of travel and polarisation of the field.

We can also take into account, for the purpose of illustration, some assumed antenna pattern, where the phase front of the antenna's reception pattern is not spherical. In other words, the antenna has some intrinsic directionality, which depends on the amplitude and phase of the reception field pattern over the antenna's reference surface. For a sinusoidal form, typical of an idealised open-ended waveguide polarised along the circumference of the cylinder, the integral becomes
\begin{align}
  \int_0^{2\pi}\int_{-L}^{L}\langle S \rangle R \sin^3\theta \, dz d\phi\,.
\end{align} 
Other antennas have different functional forms, but this expression is suficient to show that even more power can be lost easily.

\begin{figure}[!ht]
\noindent \centering
\includegraphics[width=8.0cm]{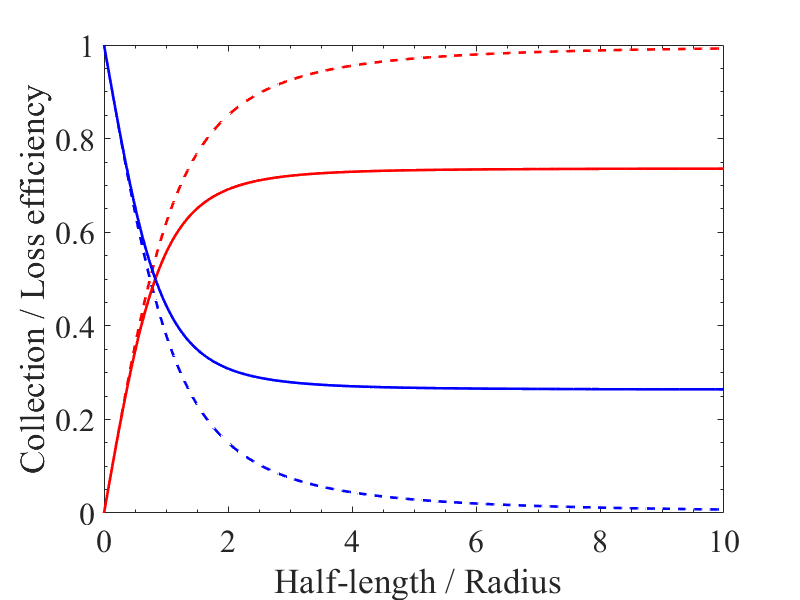}
\caption{\label{fig:energy_collection} Upper limit of energy collection efficiency (red)  and loss (blue) as a function of antenna array half-length. The dashed lines are for perfect coupling between the antenna and the radiated field, and the solid lines illustrate the effects of the directivity.}
\end{figure}

Figure~\ref{fig:energy_collection} shows the collection efficiency (red) and implied loss (blue) as a function of normalised cylinder half-length. The powers have been normalised to the total power radiated at the fundamental frequency, given by (\ref{eq:Fund_energy_calculation}). The  half-length of the cylinder $L$ is normalised by its radius $r$. The dashed lines correspond to all of the power crossing the cylindrical surface, and the solid lines show the effects of a typical antenna (open-ended waveguide) where the coupling efficiency depends on the direction of the incoming wave. The maximum collection efficiency approaches unity as the length-to-radius ratio is increased, which is intuitive because when the cylinder is long, the end-caps cover only a small solid angle when viewed from the source, and only a small amount of power is lost.
Overall,  Fig. \ref{fig:energy_collection} reflects the functional form of Fig. \ref{fig:rotating_dipole_radiation}.
 
Comparing the dashed lines (maximum field coupling) with the solid lines (partial field coupling), it can be seen the collection efficiencies are similar at small lengths.  Both efficiencies are around a half ($0.62$ for full field coupling and $0.56$ for partial field coupling) when the normalised half-length is unity. The efficiencies deviate more significantly for directive antennas on long cylinders, where in this case the power collection efficiency asymptotically approaches $0.74$. This is again intuitive because the phase front of the radiation is spherical. With short cylinders, the antennas are operating on-axis, whereas in the case of long cylinders available power is lost, and indeed scattered off of the antennas, due to the phase front of the field being poorly matched to that of the antennas. It can be seen that for a simple ring of antennas, say $L/r = 0.1$ with an on-axis source, the efficiency is down to 5~\%, which implies very low SNRs, even before any other imperfections are taken into consideration. To get high coupling efficiencies an array of wide-beam antennas is needed, where the array has a normalised half-length of greater than 5. This observation already implies that a large number of antennas is needed to achieve even a small field of view (FoV). There is little difference, in this case, if directive or non-directive  antennas are used in an array whose total length is equal to the diameter of the cylinder: 50\% efficiency. For long arrays, however, the directive nature of the antennas must be taken into consideration. An array having a total length to diameter ratio of greater than 2 can achieve collection efficiencies approaching 90~\%. Crucially, these values correspond to an electron at the centre, and so the whole matter of FoV comes into play, which will be discussed later. 

It should be noted that whereas the dashed red line shows the collection efficiency of a cylidrical surface, the dashed blue line is also the maximum collection efficiency of antennas arranged on two circular parallel plates, and indicates that there is cross over in efficiency when the radius is equal to the half-length of the cylinder.  A spherical arrangement of horns would also approach the maximum collection efficiency shown, but then polarisation becomes important. Goemetrical inefficiency is hard to avoid, and is another factor that degrades the efficiency of the system overall.

\begin{table*}[!ht]
\centering
\renewcommand{\arraystretch}{1.5}
\begin{tabular}{ | >{\raggedleft}m{2cm}||p{3.5cm}|p{3.5cm}|p{7cm}| }
 \hline
 \multicolumn{4}{|c|}{\textbf{Typical CRES Parameters}} \\
 \hline
\textbf{Parameter}\,\, & $f_0=15\,\mathrm{GHz}$ & $f_0=27\,\mathrm{GHz}$  & \textbf{Description}\\
\hline
 $E_\mathrm{K}\,\,$   & \multicolumn{2}{l|}{$18.6\,\mathrm{keV}$  }   &Kinetic energy of electron \\
  \hline
 $\gamma\,\,$   & \multicolumn{2}{l|}{$1.0364$}    &Relativistic factor \\
  \hline
 $\beta_0\,\,$   & \multicolumn{2}{l|}{$0.263$}    &Electron velocity ratio \\
  \hline
 $B\,\,$   & $0.555\,\mathrm{T}$ &$1.00\,\mathrm{T}$    &Magnetic flux density \\
      \hline
 $r_0\,\,$   & $0.84\,\mathrm{mm}$ &$0.46\,\mathrm{mm}$   &Orbit radius \\
  \hline
 $f_0\,\,$   & $15\,\mathrm{GHz}$ & $27\,\mathrm{GHz}$    &Orbit frequency \\
      \hline
 $\lambda_0\,\,$   & $20\,\mathrm{mm}$ & $11\,\mathrm{mm}$   &Radiation wavelength \\
          \hline
 $P_\mathrm{total}\,\,$   &$0.37\,\mathrm{fW}$ & $1.2\,\mathrm{fW}$   &Total radiated power \\
  \hline
 $P_\mathrm{fund}\,\,$   & $0.32\,\mathrm{fW}$ & $1.0\,\mathrm{fW}$    &Radiated power in the fundamental rotating dipole \\
 \hline
 $\epsilon_\mathrm{fund}\,\,$   & \multicolumn{2}{l|}{$0.87$}    & Proportion of power in the fundamental frequency\\
  \hline
 $\epsilon_\mathrm{geom}\,\,$   & \multicolumn{2}{l|}{$0.62$}     & Power in the fundamental frequency passing through a cylindrical surface having length equal to diameter\\
 \hline
 $P_\mathrm{n, ref}\,\,$   & \multicolumn{2}{l|}{$8.3\times10^{-3}\,\mathrm{fW}$}    & Noise power corresponding to a noise temperature of $20\,\mathrm{K}$ in spectral bin width of $30\,\mathrm{kHz}$\\
 \hline
 $T_\mathrm{fund}\,\,$   & $7.7\times10^2\,\mathrm{K}$  & $2.4\times10^3\,\mathrm{K}$    &Radiation power at the fundamental frequency in terms of equivalent noise temperature, assuming a spectral bin width of $30\,\mathrm{kHz}$ \\
 \hline
 $T_\mathrm{HEMT, f_0}\,\,$   & $5\,\mathrm{K}$ &  $7\,\mathrm{K}$  &Noise temperature of typical cryogenic HEMT amplifier. \\
 \hline
  $SNR \,\, $ &   154 &  343 &  Implied signal to noise ratio for $B \tau = 1$.\\
 \hline
 $T_\mathrm{SQL, f_0}\,\,$   & $0.72\,\mathrm{K}$ & $1.3\,\mathrm{K}$   & Standard Quantum Limit of linear amplifier and background in terms of noise temperature \\
 \hline
\end{tabular}
\caption{Key numerical values typical of a CRES experiment. Two frequencies are shown corresponding to static magnetic field strengths of 0.55 T and 1 T. }
\label{table:Parameters_QTNM}
\end{table*}

The overall challenge is illustrated by Table~\ref{table:Parameters_QTNM}, which shows various numerical values typical of a CRES system. In an experiment, the total radiated power experiences several loss factors before reaching the first stage of amplification: (i) only the power in the fundamental frequency is detected, resulting in an efficiency factor of  $\epsilon_\mathrm{fund}$; (ii) radiation in the fundamental frequency may not cross the antenna reference surface at all $\epsilon_\mathrm{geom}$; (iii) the incident radiation only partially couples electromagnetically to the detection mode of each antenna, depending on factors such as wavefront shape and polarisation $\epsilon_\mathrm{ant}$; and (iv) the signal experiences insertion and cabling loss between the antenna and the first stage of amplification $\epsilon_\mathrm{loss}$. These loss factors are important because the signal is subsequently corrupted by noise. 

The primary challenge is that it is difficult to get high coupling efficiencies, the SNRs are potentially low, and  there is no opportunity to integrate because of the need to observe single-electron events. In CRES experiments, the SNR is limited mainly by the noise temperature of the first stage amplifier. By operating with a low-temperature radiometric background, and with cryogenic amplifiers, ideally quantum-noise-limited, appreciable SNR's can be achieved, as illustrated by the last 5 rows of Table~\ref{table:Parameters_QTNM}. The advantages of operating with cryogenic amplifiers are however easily thrown away as the inefficiencies mount, and the move to large FoV's, whilst retaining high coupling efficiency,  seems problematic.

\section{Canonical antenna arrangement}
\label{sec:can_arr}

We now start to consider in more detail the problems associated with designing and optimising inward looking arrays of antennas. There is a considerable amount of literature on desiging outward looking arrays for radar, direction finding and telecommunications, but the problem of designing inward looking arrays and the new issues that come into play have received little attention. An added complication is that each receiver system must be ultra-low noise, and yet the noise received by one antenna includes noise radiated by the others, which can be appreciable.

In what follows, we shall adopt a canonical arrangement based on a cylindrical array of antennas. To aid physical interpretation, we shall consider a 2D system that is, strictly speaking, translationary invariant in the axial $z$ direction. This model serves to highlight the challenges associated  with achieving efficient coupling without getting into the complications of full 3D calculations. In a 2D system, radiators must be represented by line currents. An axial line current radiates a cylindrical wave, with the $H$-vector everywhere transverse to $z$, and the $E$-vector everywhere parallel to $z$. In a CRES experiment an electron's radiation pattern, however, has the $E$-vector everywhere transverse to $z$, and the $H$-vector everywhere parallel to $z$. Nevertheless, a line-current source captures the essential features of a full 3D system, and lends itself to highly efficient and intuitive analysis methods. The extension to 3D is straightforward and discussed at the end of the paper. In order to ensure that the power radiated by a line current,  $P = L I_0^{2} Z_0 \omega_0 / 8 c$, is the same as that radiated by an orbiting electron at the fundamental frequency, $ \langle P_\mathrm{fund} \rangle = \left( Z/6\pi \right) \left( p_0\omega_0^2/c \right)^2 $, where $ p_0 = r_e q$, the following identity is useful:
\begin{align}
\label{eqn:equivalent}
I_0 & = \sqrt{ \frac{2}{3 \pi L} \frac{p_0^2\omega_0^3}{c} },
\end{align}
where $L$ is the half-length of the cylindrical system.

When designing an antenna system, the first idea might be to surround the single-electron source with a collection of electric dipole antennas. Assume that these are placed on a circle that defines an instrumented area (in 2D). Already it can be appreciated that the coupling efficiency will be low because, in the far field, the radiation pattern of the electron has a diverging phase front, and the radiation pattern of each antenna has its own diverging phase front. The radii of curvature have different signs, and so the two will never couple efficiently. Simulations of a system of dipole antennas show that for even small instrumented volumes, the overall power coupling efficiency of all of the antennas taken together, unphased, is tenths of a percent. A further problem is that, due to inherent scattering, the beam pattern of each antenna in free space is not is not the same as the beam pattern of each antenna in the presence of the others. Each antenna port is likely to have a complicated reception pattern,  which can display spatial structure on the scale size of $\lambda/2$, and even smaller. This spatial structure can be surprisingly complicated, depending on the Q factors of the unintended cavity modes, and this will degrade severally the uniformity of the frequency resolution of the instrument, and lead to multiple blind spots.  In addition, the input impedance of each antenna is modified by the presence of the others, and in general will have a reactive part that changes rapidly with frequency. The ports of the array can be described by an impedance matrix, which in an ideal world would be diagonal, real-valued and matched to the input impedance of the first-stage amplifiers. Designing of a collection of inward-looking antennas that couple to any point source efficiently and that have maneagable real input impedances over the required band is very difficult, even when iterative electromagnetic modelling software is used. There is another problem because a large fraction of the noise power seen by each port of the interacting antenna system will originate from warm absorptive surfaces of other parts of the apparatus.  It follows that the antenna array should be designed to couple as weakly as possible to its environment.

\begin{figure}
\noindent \centering
\includegraphics[trim = 3cm 16cm 7cm 2cm, width=60mm]{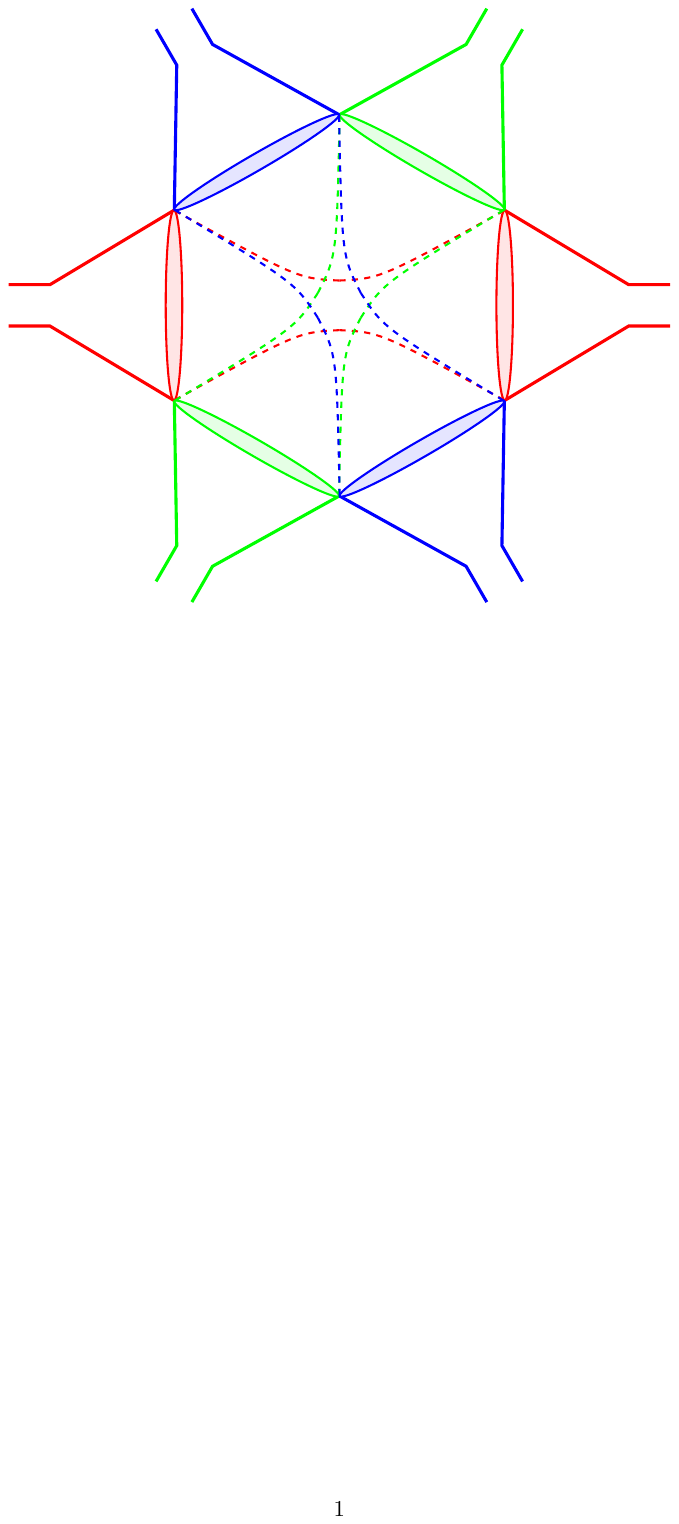}
% left,bottom,right,top
\caption{Rectangular horns packed around a circle. The horns are arranged in pairs, indicated through colour. Each horn has a phase transforming lens to (i) match the phase front of the reception pattern to the phase front of the radiating electron at the centre; (ii) match the phase front of each horn to the phase front of its opposite partner. For illustrative purposes, lenses are shown, but other schemes could be used.}
\label{figure1}
\end{figure}

Figure \ref{figure1} shows a canonical scheme where the apertures of a set of rectangular horns are arranged around the circumference of a circle. Each horn has a phase transforming lens, symbollically, that focuses the reception pattern of the horn onto the centre of the circle, where for the moment we will assume the electron is located. The purpose of this configuration is to ensure that the phase front of each horn is reasonably well matched to the phase front of the radiating electron such that high coupling efficiencies can be obtained. Although lens-horn antennas are used to illustrate the basic principles, numerous other schemes are of interest, such as profiling the walls of each horn to create a negative radius of curvature and therefore converging beam; using dipole-fed curved reflectors having cylindrical, parabolic or elliptical cross sections, and perhaps even curved patch antennas. The  crucial point is that the radius of curvature of the phase front of each antenna should be matched as best as possible to the phase front of the radiation pattern of the source. 

The horns are arranged in confocal pairs so that each horn couples to another horn with high efficiency. With straightforward design, high coupling efficiencies can be achieved, which results in a high degree of rejection of noise from the thermal environment. For example, a coupling loss of -25 dB from one antenna port to its opposite partner would add only 1~K to the system noise temperature for a 300~K environment. Noise radiated into the system from the inputs of the receivers will be discussed later. A quantum-noise limited receiver would not be degraded significantly in a 100 K environment. Speculatively, this level of isolation may even allow squeezing amplifiers to be used to beat the quantum limit. Crucially, however, this advantage relies on near-perfect antenna coupling, in the sense that every antenna couples efficiently to a combination of all of the others, so that no power is lost to the environment. An additional benefit is that because the inter-antenna coupling is high, the antenna pattern of each port will be the same as the antenna pattern that would be measured in free space. Also, the input impedance of each antenna will be the same as that measured in free space, and the impedance matrix of the whole set of ports will be diagonal, reducing the need for iterative numerical design procedures. A final consideration is that if the beams of the antennas and electron are matched, the electron effectively radiates into a free-space environment, minimising the effects of backaction. However, as will be seen, an electron at an off-axis position couples to each of the antennas differently, and power that is reflected will couple to resonanaces having a variety of Q factors depending on leakage. 

There is a bigger problem, however, because an electron appears as a point-like source, and so high efficiency coupling requires antennas with strongly converging beams, but then if the electron is not at the design centre of the system, the coupling efficiency will be degraded, and one is left wondering about the FoV of a beam-matched system. Many other considerations then come to mind: What is the optimum number of antennas that should be placed around the periphery of the instrumented volume to achieve best performance? Should the radius of curvature of the aperture fields of the horns be the same as the radius of the circle on which they are placed?  In \S \ref{sec:full_elec_model} we describe 2D simulations that capture the elements of these basic issues, but before proceeding, it is instructive to consider whether it is possible to achieve, even in principle, high-efficiency antenna-antenna coupling. 

In Gaussian-mode optics, the field from an antenna is propagated through free space using a set of Gaussian-Hermite or Gaussian-Laguerre polynomials, which are solutions of the paraxial wave equation \cite{Goldsmith_1998}. This method is highly effective, and can be applied surprisingly well to even wide-angle beams. It enables complicated multi-element optical systems to be modelled easily. First, the aperture field, or effective aperture field, of the source antenna is decomposed into a set of Gaussian modes. For a rectangular horn, Gaussian-Hermite modes are favoured, where the scale-size and phase-front radius of curvature of the lowest order mode are chosen to maximise the power in that mode. With multiple modes, the form of the beam can be traced as it diffracts, but even the lowest-order Gaussian mode is valuable on its own.  For a Gaussian-mode beam, 
\begin{eqnarray}
\label{eqn:gaussian}
w(z) & = w_0 \left[ 1 + \left( z / z_c \right)^2 \right]^{1/2} \\ \nonumber
R(z)  & = z \left[ 1 +  \left( z_c / z \right)^2 \right] \\ \nonumber
\psi (z) & = \tan^{-1} \left( z / z_{c} \right),
\end{eqnarray}
where $z_c = \pi w_0^{2} / \lambda$ is the Rayleigh distance. $w(z)$ characterises scale size,  $R(z)$ the phase-front radius of curvature, and the modal phase slippage $\psi (z)$ the spatial form of the field. A differential phase slippage between two surfaces of $\Delta \psi=$~0 corresponds to an image of the first surface appearing at the second, $\Delta \psi = \pi$ to an inverted image, and $\Delta \psi = \pi/2$ to the spatial Fourier transform. 

If rectangular horn apertures are placed on a circle having radius $R_2$, the width of the aperture must be $a = 2 R_2 \sin (\Delta \alpha/2 )$, where $\Delta \alpha$ is the full angular width.  The TE$_{10}$ mode of a rectangular horn couples 99\% of its power into a Gaussian beam having a beam width of $w_h = 0.35a$ at the aperture, and a radius of curvature of $R_h$ at the aperture. $R_h$ is the radius of curvature of the aperture field lines terminating normally on the conducting walls. The design choice $R_h = R_2$ does not necessarily mean that the waist of the beam lies at the centre of  $ {\cal C}_{2}$, and so we need to find the value of $R_h$ that places the beam waist at the centre of $ {\cal C}_{2}$ . 

Defining $z_h = \pi \omega_h^2 /\lambda$, it can be shown that
\begin{equation}
\label{eqn:horn_rad}
R_h  = z_h^2 \left[ \frac{1}{2 R_2'} \pm \left( \frac{1}{(2 R_2')^2} - \frac{1}{z_h^2}\right)^{1/2} \right]
\end{equation}
where $R_2' = R_2 \cos  (\Delta \alpha/2 )$. The beam waist at the centre is then given by
 \begin{equation}
\label{eqn:horn_waist}
w_0^2 = \frac{\lambda}{\pi} \frac{R_h^2 / z_h}{1 + \left( R_h / z_h \right)^2}.
\end{equation}

\begin{figure}
\noindent \centering
\includegraphics[trim = 1cm 1cm 8cm 16.5cm, width=65mm]{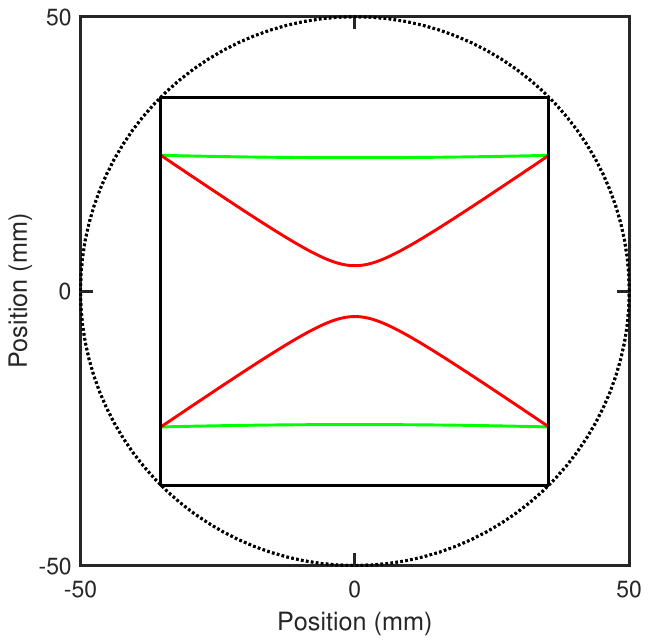}
% left,bottom,right,top
\caption{4 horn apertures (black lines) arranged around a circle having a radius of  50mm. The wavelength is 10 mm. The near-field Gaussian-mode solution is shown in green, and the far-field solution in red. The near-field waist is 24.3mm and the far-field waist 4.6mm corresponding to a strongly diffracting beam.
}
\label{figure7}
\end{figure}

\begin{figure}
\noindent \centering
\includegraphics[trim = 1cm 1cm 8cm 16.5cm, width=65mm]{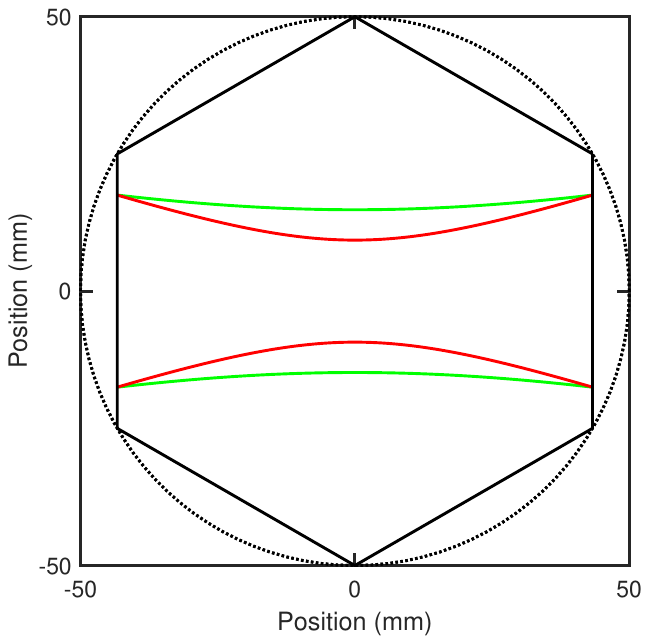}
% left,bottom,right,top
\caption{6 horn apertures, black lines, arranged around a circle having a radius of 50mm. The wavelength is 10 mm. The near-field Gaussian-mode solution is shown in green, and the far-field solution in red. The near-field waist is 14.8 mm and the far-field waist 9.2 mm.}
\label{figure8}
\end{figure}

(\ref{eqn:horn_rad}) shows that there are two solutions: The positive sign corresponds to the horn being in the near field of the beam's focus $R_2 < z_c$, and the negative sign to the horn being in the far field  $R_2 > z_c$. In the first case, the beam propagates in near-collimated form across  the observed region; the spatial form of the beam appearing on the far side of the circle is much the same as the spatial form of the field across the aperture of the horn. In this near-field case, the horns will couple to each other well, the field of view will be large, but the coupling efficiency to the source reduced. In the second case, the beam converges on a focus and then opens out again, and in principle the coupling can be large because an inverted image is formed on the far side. In each case, the total phase slippage across the circle is $\Delta \psi = 2 \tan^{-1} \left( R_2 / z_{c} \right)$, and the closer this is to 0 (non-inverted image) or $\pi$ (inverted image) the greater the coupling. In any case, because 99 \% of the power is in the lowest-order mode anyway, very high horn-horn coupling can be achieved. It should be noted that for both solutions $R_h > R_2$, showing that profiled horns can in principle deliver both solutions. 

\begin{figure}
\noindent \centering
\includegraphics[trim = 1cm 1cm 8cm 16.5cm, width=65mm]{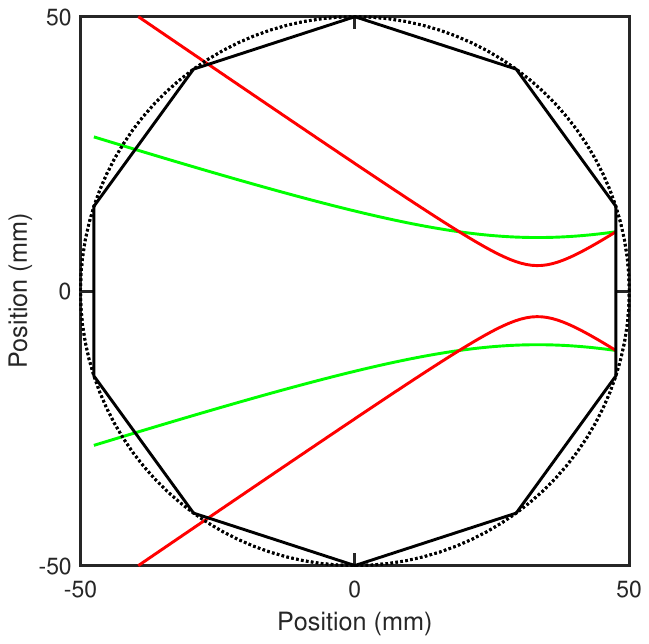}
% left,bottom,right,top
\caption{10 horn apertures arranged around a circle having a radius of  50mm. The wavelength is 10 mm. The near-field Gaussian-mode solution is shown in green, and the far-field solution in red.  The waist was set to be at a distance of 0.3 $R_2$ from the right hand horn aperture. A large number of horn apertures can be accommodated by moving the focus off axis, but at the expense of single-horn to single-horn coupling. }
\label{figure9}
\end{figure}

Figures \ref{figure7} and  \ref{figure8} show the near-field and far-field Gaussian-mode solutions for 4 and 6 horns on a circle having a radius of 50 mm and operating wavelength of 10 mm. The near-field solution appears to give a greater field of view, but at the cost of reduced coupling to an electron. The far-field solution with 4 horns appears to be a good arrangement, because the coupling to the CRES signal can be maximised, and the horn-horn coupling can be maximised to reduce the coupling to thermal background noise. There is a problem however because (\ref{eqn:horn_rad}) only has a solution if the argument of the square root is positive. If one attempts to increase the number of horns, keeping $R_2$ constant, there is some limit for which solutions exist because the requirement $w_h > \sqrt{2 R_2' \lambda / \pi}$  is violated. For the $R_2 = $ 50 mm example, shown here, the maximum number of antennas is 6. It will be shown in \S \ref{sec:fov} that this falls well short of the number of degrees of freedom that must be collected to ensure a high source coupling efficiency regardless of where the electron is located in $R_2$. If single-horn to single-horn coupling is essential, $R_2$ must be much larger than the region that can be observed efficiently.

Figure \ref{figure9} shows the effect of moving the focus of each horn off axis. As the focus is moved towards the aperture, or behind it as for an ordinary horn with a diverging beam, the number of horns that can be accommodated increases, leading eventually to a circle of ordinary antennas having diverging beams. In this case, the coupling efficiency to a point-like source  is low for the reasons described previously. The coupling to the environment may still be low as long as the beam from each antenna is collected efficiently by all of the others, which is more likely to be the case if only a small number of receiving antennas is involved. It seems that for a small FoV, a good solution can be found, but for a large FoV one is forced back into having a large number of antennas with diverging beams. 

\section{Full electromagnetic model}
\label{sec:full_elec_model}

To characterise the behaviour of a CRES experiment, it is beneficial to create a scattering-parameter model of the whole of the antenna system. In due course, it will be seen why a scattering parameter model is suited to this kind of problem; however, to our knowledge, the method of analysis described has not been reported previously.

The scattering matrix relating the travelling waves at the ports of the antennas could be found numerically through a Huygens-Fresnel diffraction integral, where the field at the aperture of one antenna is propagated to the others and field-coupling calculations performed. Given the cyclic nature of the system, this calculation would only need to be done for one source antenna. However, one gets into questions about what obliquity factors should be used, the role of evanescent fields and the effects of multiple internal scattering. In this section, we describe an electromagnetic model  that allows the scattering parameters at the ports of the antennas to be calculated even when evanescent fields and multiple internal scattering are present, and crucially allows the noise correlation matrix to be determined, including the contribution from the thermal background in which the CRES array is embedded. Our model is described in 2D, but its extension to 3D is straightforward, and will be discussed later.

\usetikzlibrary {arrows.meta,bending}
\begin{figure}
\centering
\begin{tikzpicture}[scale=0.6]

%\draw[help lines] (0,0) grid (10,10);

\draw [line width=1.2, red] plot [smooth cycle] coordinates {(7,5) (6,6) (5,7) (3.5,6.5) (3,5) (3.5,3.5) (5,3) (6.5,3.5) };
\draw [line width=1.2,black,dashed] (4.5,5) circle (3.9);
\draw [line width=1.2,black] (4.5,5) circle (1);
\draw [line width=1.1,black,arrows = {-Stealth[scale=0.9,bend]}] (7.5,7.5) -- (8.5,8.5);

\node at (5,5.8) [anchor=south]{${\cal C}_1$};
\node at (5,7) [anchor=south]{${\cal C}_2$};
\node at (5,9) [anchor=south]{${\cal C}_3$};
\node at (8.5,8.5) [anchor=south]{$\infty$};

\end{tikzpicture}
\caption{\label{fig:stratton_chu} $ {\cal C}_{1}$ (black) bounds the field of view; ${\cal C}_{2}$ (red) is an arbitrary surface on which antennas are placed; and  ${\cal C}_{3}$ (black dashed) is a surface at infinity, which allows the embedding to be calculated. 
}
\end{figure}
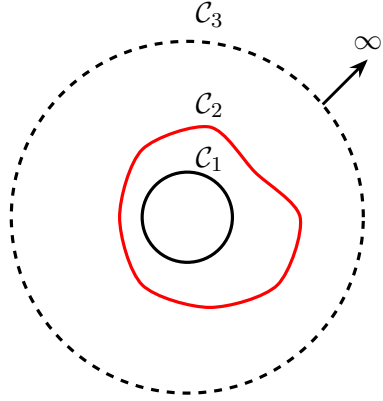

Imagine a fictitious contour  $ {\cal C}_{2}$ around which an array of antennas is placed: Fig.~\ref{fig:stratton_chu}. These antennas need to monitor a FoV, which is bounded by some other contour  $ {\cal C}_{1}$. Fundamentally, we are interested in power that flows across  $ {\cal C}_{2}$, either from the sources within  $ {\cal C}_{1}$, or from sources, such as noise, which are incident on  $ {\cal C}_{2}$ from the region outside of  $ {\cal C}_{2}$. Therefore we are interested in the field components on and tangential to $ {\cal C}_{2}$.  It can be shown, using Maxwell's equations, that for a closed 2D boundary on a plane (taking a planar cut through a cylindrical system having an arbitrary cross section)
\begin{equation}
\label{eqn:stratton_chu_gen}
\begin{aligned}
\alpha E_{z} (\mathbf{r}_{1}) & =  \frac{ k_{0} }{4}    \int_{{\cal C}_2}  H_0^{(1)} (k_0 |\mathbf{r}_1 - \mathbf{r}_2|) Z_{o} H_{t} (\mathbf{r}_{2}) dL  \\
&  -  \frac{i k_0}{4} \int_{{\cal C}_2}   \frac{\uvec{n} (\mathbf{r}_{2})  \cdot (\mathbf{r}_1 - \mathbf{r}_2)}{|\mathbf{r}_1 - \mathbf{r}_2|}
H_1^{(1)} (k_0 |\mathbf{r}_1 - \mathbf{r}_2|)  E_{z} (\mathbf{r}_{2})  dL \\
& - \frac{k_{0} Z_{o}}{4} \int_{{\cal C}_1}  H_0^{(1)} (k_0 |\mathbf{r}_1 - \mathbf{r}_2|)  J_{z} (\mathbf{r}_{2}) \, dS,
\end{aligned}
\end{equation}
where $k_0 = \omega / c$, $c$ is the speed of light in vacuum, $H^{(1)}_n (z)$ is the Hankel function of the first kind of order $n$, and
\begin{equation}\label{eqn:def_alpha}
	\alpha = \begin{cases}
		1 & r_1 < R_{2} \\
		\frac{1}{2} & r_1 = R_{2}.
	\end{cases}
\end{equation}
 $\mathbf{r}_{2}$ denotes source points, $\mathbf{r}_{1}$ denotes observation points, and  $\uvec{n} (\mathbf{r}_{2})$ is the outward surface normal.

Equation (\ref{eqn:stratton_chu_gen}) is a Stratton-Chu-like formulation of the boundary-valued problem, which expresses the electric field in $ {\cal C}_{2}$, $E_{z} (\mathbf{r}_{1})$, in terms of the tangential field components,  $H_{t} (\mathbf{r}_{2})$ and $ E_{z} (\mathbf{r}_{2})$, on  $ {\cal C}_{2}$, and any current sources $ J_{z} (\mathbf{r}_{2})$ in  $ {\cal C}_{1}$. When $\mathbf{r}_{1}$ is on the boundary, and explicitly for a circular cross section,
\begin{equation}
\label{eqn:stratton_chu_boundary}
\begin{aligned}
	E_z (\phi_1) &= 	\frac{k_0 R_{2} }{2} \int_0^{2 \pi} H_0^{(1)} \bigl( 2 k_0 R_{2} \bigl| \sin \bigl( \tfrac{\phi_1 - \phi_2}{2} \bigr) \bigr| \bigr) \{ Z_0 H_\phi (\phi_2) \} \, d\phi_2 \\
	&+ \frac{i k_0 R_{2} }{2} \int_0^{2 \pi} 	\bigl| \sin \bigl( \tfrac{\phi_1 - \phi_2}{2} \bigr) \bigr| H_1^{(1)} \bigl( 2 k_0 R_{2} \bigl| \sin \bigl( \tfrac{\phi_1 - \phi_2}{2} \bigr) \bigr| \bigr) 	E_z (\phi_2) \, d\phi_2 \\
	&- \frac{k_0 Z_0}{2} \int_{r_1=0}^{R_{1}} \int_{\phi_2=0}^{2\pi} H_0^{(1)} \Bigl( k_0 \sqrt{(R_{2} - r_2)^2 + 4 R_{2} r_2 \sin^2 \bigl( \tfrac{\phi_1 - \phi_2}{2} \bigr)} \Bigr) J_z(r_2, \phi_2) \, r_2 dr_2 d\phi_2.
\end{aligned}
\end{equation}
$(r_2, \phi_2)$, and $(r_1,\phi_1)$ are the positions of source and observation points in cylindrical coordinates. $ \sqrt{(r_1 - r_2)^2 + 4 r_1 r_2 \sin^2 \bigl( \tfrac{\phi_1 - \phi_2}{2} \bigr)}$ is the distance between the observation and current-source points in cylindrical coordinates. Equation (\ref{eqn:stratton_chu_gen}) applies to a general contour, but for now concentrate on the circular contour described by  (\ref{eqn:stratton_chu_boundary}).

 (\ref{eqn:stratton_chu_boundary}) is best solved numerically due to the appearance of $E_z (\phi)$ on each side of the equation. For numerical purposes assume that fields on $ {\cal C}_{2}$, generically called $F(\phi)$, are represented by a set of top-hat functions $p (\phi - \phi_n)$:
\begin{equation}\label{eqn:rooftop_expansion}
	F(\phi) \approx \sum_{n=0}^{N-1} f_n p (\phi - \phi_n)
\end{equation}
where
\begin{equation}
\begin{aligned}
	& \phi_n = n \Delta \phi \\
	& \Delta \phi = \frac{2 \pi}{N} \\
	& p(\phi) = \begin{cases}
		1 & |\phi| < \frac{\Delta \phi}{2} \\
		0 & \text{otherwise}.
	\end{cases}
\end{aligned}
\end{equation}
It follows that
\begin{equation}\label{eqn:fn}
	f_n = F(\phi_n)
\end{equation}
giving a  vector of expansion coefficients $\mathbf{f}$. In addition, for a set of line currents
\begin{equation}
	J_z (\mathbf{r}) = \sum_{n=0}^{K-1} i_n \delta (\mathbf{r} - \mathbf{r}_n),
\end{equation}
where the $i_n$ have units of current. As described previously, these currents can be scaled to give the same power as that radiated by an orbiting electron, (\ref{eqn:equivalent}).

When expressed in this sampling basis (\ref{eqn:stratton_chu_boundary}) gives
\begin{equation}
\label{eqn:discretised_stratton_chu}
	\mathbf{e} = Z_0 \mathcal{A} \cdot \mathbf{h}
		+ \mathcal{B} \cdot \mathbf{e}
		+  k_0 Z_0 \mathcal{C} \cdot \mathbf{i},
\end{equation}
where
\begin{eqnarray}\label{eqn:def_a_matrix}
A_{mn} & = \frac{k_0 R_2 }{2} \int_{\phi_n - \Delta \phi /2}^{\phi_n + \Delta \phi /2} H_0^{(1)} \bigl( 2 k_0 R_2 \bigl| \sin \bigl( \tfrac{\phi_m - \phi_2}{2} \bigr) \bigr| \bigr) 
p (\phi_2 - \phi_n) \, d\phi_2 \\ \nonumber
 &  \approx \begin{cases}
		\frac{k_0 R_2 \Delta \phi}{2}
		H_0^{(1)} \bigl( 2 k_0 R_2 \bigl| \sin \bigl( \tfrac{\phi_m - \phi_n}{2} \bigr) \bigr| \bigr)
		 & m \neq n \\
	\tfrac{i k_0 R_2}{\pi} \Bigl\{
		-\tfrac{i \pi}{2} +  \ln \bigl( \tfrac{k_0 R_2}{4} \bigr) + \gamma
		+ \ln \Delta \phi - 1 \Bigr\} \Delta \phi
		& m = n,
	\end{cases}
\end{eqnarray}
\begin{eqnarray}\label{eqn:def_b_matrix}
B_{mn} & = \frac{i k_0 R_2}{2} \int_{\phi_n - \Delta \phi / 2}^{\phi_n + \Delta \phi / 2}\bigl| \sin \bigl( \tfrac{\phi_m - \phi_2}{2} \bigr) \bigr| H_1^{(1)} 
\bigl( 2 k_0 R_2 \bigl| \sin \bigl( \tfrac{\phi_m - \phi_2}{2} \bigr) \bigr| \bigr) p (\phi_2 - \phi_n) \, d\phi_2 \\ \nonumber
  & \approx \begin{cases}
	   \frac{i k_0 R_2 \Delta \phi}{2} 
		\bigl| \sin \bigl( \tfrac{\phi_m - \phi_n}{2} \bigr) \bigr|
		H_1^{(1)} \bigl( 2 k_0 R_2 \bigl| \sin \bigl( \tfrac{\phi_m - \phi_n}{2} \bigr) \bigr| \bigr)
	& m \neq n \\
	\frac{1}{2 \pi} \Delta \phi + \frac{i}{96} (k_0 R_2)^2 {\Delta \phi}^3
		& m = n,
	\end{cases}
\end{eqnarray}
and
\begin{equation}
C_{mn} = - \frac{1}{2}	H_0^{(1)} \Bigl( k_0 \sqrt{(R_2 - r_n)^2 + 4 R_2 r_n \sin^2 \bigl( \tfrac{\phi_m - \theta_n}{2} \bigr)} \Bigr).
\end{equation}
The diagonal elements of $\mathcal{A}$ and  $\mathcal{B}$ involve a complication because the Hankel functions are singular when $\phi_{1} = \phi_2$.  The singularities are however integrable, and different approaches can be taken to evaluating the integrals, leading to different levels of accuracy. It is important to consider the boundary carefully in the region of the singularity with different expressions relating to piecewise linear contours and circular contours of the kind described here. The expressions for the diagonal elements in (\ref{eqn:def_a_matrix}) and  (\ref{eqn:def_b_matrix}) were evaluated over the circle using a small argument approximation for the Hankel function.

Rather than solving (\ref{eqn:discretised_stratton_chu}) directly it is helpful to cast it into the form of directed energy flow. We continue to use the terminology `energy flow' even though the formulation includes evanescent processes, where energy sloshes forwards and backwards across the boundary. Rather than using $E_{z}(\phi)$ and $H_{\phi}(\phi)$ directly, it is more interesting to work with the derived fields
\begin{eqnarray} \label{eqn:trav_waves}
a(\phi) & = \frac{1}{2 \sqrt{Z_{0}}} \left[ E_{z} (\phi) + Z_{0}  H_{\phi} (\phi) \right]  \\ \nonumber
b(\phi) & =  \frac{1}{2 \sqrt{Z_{0}}} \left[ E_{z} (\phi) - Z_{0}  H_{\phi} (\phi) \right].
\end{eqnarray}
 It can be shown that $ a(\phi)$ can only do work on the region inside of $ {\cal C}_{2}$, and $ b(\phi)$ can only do work on the region outside of $ {\cal C}_{2}$. In fact, using Poynting's Theorem, it is straightforward to show that the total power flowing across a cylindrical extension of $ {\cal C}_{2}$ of length $L$ is 
\begin{eqnarray}
\label{eqn:power_surface}
\frac{R_{2} L}{2} \int_{0}^{2 \pi} |a(\phi)|^{2} - | b(\phi)|^{2} d \phi,
\end{eqnarray}
which is an energy-flow picture of the fields on  $ {\cal C}_{2}$.

Using the discretised and vectorised forms of (\ref{eqn:trav_waves}), 
\begin{eqnarray}\label{eqn:e_from_a}
	\mathbf{e} & = \{ \mathbf{a} + \mathbf{b} \} \sqrt{Z_0} \\ \nonumber
	Z_0 \mathbf{h} & = \{ \mathbf{a} - \mathbf{b} \} \sqrt{Z_0},
\end{eqnarray}
(\ref{eqn:discretised_stratton_chu}) becomes
\begin{eqnarray}
\label{eqn:field_network_equation}
	\mathbf{b} & = \mathcal{S}_f \cdot \mathbf{a} + k_0 \sqrt{Z_0} \mathcal{K}_f \cdot \mathbf{i} \\ \nonumber
& = \mathcal{S}_f \cdot \mathbf{a}	+  \mathbf{c},
\end{eqnarray}
where $\mathbf{c}$ is a set of source waves deriving from the current, and
\begin{equation}
\label{eqn:field_scattering_matrix}
	\mathcal{S}_f = -(\mathcal{I} + \mathcal{A} - \mathcal{B}) ^{-1}
		\cdot (\mathcal{I} - \mathcal{A} - \mathcal{B})
\end{equation}
and
\begin{equation}\label{eqn:field_propagator_matrix}
	\mathcal{K}_f = (\mathcal{I} + \mathcal{A} - \mathcal{B}) ^{-1} \cdot \mathcal{C}.
\end{equation}

(\ref{eqn:field_network_equation}) is a scattering network representation of directed energy flow across  $ {\cal C}_{2}$ with respect to the surface normal. There is, however, a twist because both reactive and dissipative energy can be transferred across $ {\cal C}_{2}$. These correspond to evanescent and radiative processes even though the localised nature of the scattering description hides this fact. It is a mistake to assume that the free-space region beyond $ {\cal C}_{2}$  that gives rise to $\mathbf{a}$ is fully described by a matched source having impedance $Z_{0}$. Represent the region looking outwards from
$ {\cal C}_{2}$ in terms of a new scattering martrix $\mathcal{S}_e$, such that
\begin{equation}
\label{eqn:ext_field_equation}
\mathbf{b}'  = \mathcal{S}_e \cdot \mathbf{a}'	+  \mathbf{c}'.
\end{equation}
When the two are regions are connected $\mathbf{a}' = \mathbf{b}$ and $\mathbf{a} = \mathbf{b}'$, but how do we determine  $\mathcal{S}_e$?

Imagine a new region bounded by the inner surface  $ {\cal C}_{2}$ and a new outer surface  $ {\cal C}_{3}$. The Stratton-Chu-like method can be applied to this new region, but if $ {\cal C}_{3}$ is taken out to infinity, the field components on $ {\cal C}_{3}$ tend to zero as a consequence of the radiation condition:  strictly, this is the radiation condition for a very weakly absorbing medium. The remaining integrals over  $ {\cal C}_{2}$ can be derived from those already considered by the interchanges
\begin{equation}
\begin{gathered}
	\mathbf{a} \rightarrow \mathbf{b}' \\
	\mathbf{b} \rightarrow \mathbf{a}' \\
	\uvec{n} \rightarrow -\uvec{n}.
\end{gathered}
\end{equation}
Applying these swaps gives (\ref{eqn:ext_field_equation}), where on recognising that $\uvec{n} \rightarrow -\uvec{n}$ results in $A \rightarrow - A$ and $B \rightarrow - B$,
\begin{equation}
\label{eqn:s_infinity}
	\mathcal{S}_e = -(\mathcal{I} + \mathcal{A} + \mathcal{B})^{-1} \cdot (\mathcal{I} - \mathcal{A} + \mathcal{B}).
\end{equation}

Equation (\ref{eqn:ext_field_equation}) represents the region outside of  $ {\cal C}_{2}$ in terms of a set of scattering parameters on  $ {\cal C}_{2}$: it includes both reactive and radiative processes. It is interesting to note that whilst $\mathcal{S}_f$ must be unitary for a lossless CRES region, $\mathcal{S}_e$ is generally not lossless because of the loss of power to radiated fields. The unitarity of $\mathcal{S}_f$ is a good way of checking the integrity of modelling software, particularly the precision with which the singularities have been removed. As will be demonstrated,  singular value decomposition (SVD) of $\mathcal{S}_e$ can be used to distinguish between non-local evanescent and radiative fields, which in turn implies that the Hermitian and anti-Hermitian parts of $\mathcal{S}_e$ must be related to evanescent and radiative processes respectively.

\usetikzlibrary {arrows.meta,bending}
\begin{figure}
\centering
\begin{tikzpicture}

%\draw[help lines] (0,0) grid (8,7);

\draw [line width=1.4,fill=black] (5,1) circle (0.05);
\draw [line width=1.4,fill=black] (5,2) circle (0.05);
\draw [line width=1.4,fill=black] (5,3) circle (0.05);
\draw [line width=1.4,fill=black] (5,4) circle (0.05);
\draw [line width=1.4,fill=black] (5,5) circle (0.05);
\draw [line width=1.4,fill=black] (5,6) circle (0.05);

\draw [line width=1.4,fill=black] (3,1) circle (0.05);
\draw [line width=1.4,fill=black] (3,2) circle (0.05);
\draw [line width=1.4,fill=black] (3,3) circle (0.05);
\draw [line width=1.4,fill=black] (3,4) circle (0.05);
\draw [line width=1.4,fill=black] (3,5) circle (0.05);
\draw [line width=1.4,fill=black] (3,6) circle (0.05);

\draw [line width=1.2,green,arrows = {-Stealth[scale=0.9,bend]}] (3,5) arc (+60:-60:-0.56);
\draw [line width=1.2,green,arrows = {-Stealth[scale=0.9,bend]}] (3,5) arc (-60:+60:-0.56);
\draw [line width=1.2,green,arrows = {-Stealth[scale=0.9,bend]}] (3,5) arc (-60:+60:-1.7);
\draw [line width=1.2,green,arrows = {-Stealth[scale=0.9,bend]}] (3,3) arc (+60:-60:-0.58);
\draw [line width=1.2,green,arrows = {-Stealth[scale=0.9,bend]}] (3,3) arc (-60:+60:-0.58);
\draw [line width=1.2,green,arrows = {-Stealth[scale=0.9,bend]}] (3,3) arc (+60:-60:-1.7);
\draw [line width=1.2,green,arrows = {-Stealth[scale=0.9,bend]}] (3,1) arc (+60:-60:-0.58);
\draw [line width=1.2,green,arrows = {-Stealth[scale=0.9,bend]}] (3,1) arc (+60:-60:-1.7);
\draw [line width=1.2,green,arrows = {-Stealth[scale=0.9,bend]}] (3,1) arc (+60:-60:-2.9);

\draw [line width=1.2,green,arrows = {-Stealth[scale=0.9,bend]}] (5,6) arc (+60:-60:0.56);
\draw [line width=1.2,green,arrows = {-Stealth[scale=0.9,bend]}] (5,4) arc (-60:+60:0.56);
\draw [line width=1.2,green,arrows = {-Stealth[scale=0.9,bend]}] (5,2) arc (-60:+60:1.7);
\draw [line width=1.2,green,arrows = {-Stealth[scale=0.9,bend]}] (5,4) arc (+60:-60:0.56);
\draw [line width=1.2,green,arrows = {-Stealth[scale=0.9,bend]}] (5,4) arc (+60:-60:1.7);
\draw [line width=1.2,green,arrows = {-Stealth[scale=0.9,bend]}] (5,6) arc (+60:-60:1.7);
\draw [line width=1.2,green,arrows = {-Stealth[scale=0.9,bend]}] (5,2) arc (+60:-60:0.56);
\draw [line width=1.2,green,arrows = {-Stealth[scale=0.9,bend]}] (5,2) arc (-60:+60:0.56);
\draw [line width=1.2,green,arrows = {-Stealth[scale=0.9,bend]}] (5,6) arc (60:-60:2.9);

\draw [line width=1.4,fill=black] (1,2) circle (0.05);
\draw [line width=1.2,red,arrows = {-Stealth[scale=0.9,bend]}] (1,2) -- (3,2);
\draw [line width=1.4,fill=black] (1,4) circle (0.05);
\draw [line width=1.2,red,arrows = {-Stealth[scale=0.9,bend]}] (1,4) -- (3,4);
\draw [line width=1.4,fill=black] (1,6) circle (0.05);
\draw [line width=1.2,red,arrows = {-Stealth[scale=0.9,bend]}] (1,6) -- (3,6);

\draw [line width=1.4,fill=black] (7,1) circle (0.05);
\draw [line width=1.2,red,arrows = {-Stealth[scale=0.9,bend]}] (7,1) -- (5,1);
\draw [line width=1.4,fill=black] (7,3) circle (0.05);
\draw [line width=1.2,red,arrows = {-Stealth[scale=0.9,bend]}] (7,3) -- (5,3);
\draw [line width=1.4,fill=black] (7,5) circle (0.05);
\draw [line width=1.2,red,arrows = {-Stealth[scale=0.9,bend]}] (7,5) -- (5,5);

\draw [line width=1.2,green,dashed] (3.6,1) -- (4.4,1);
\draw [line width=1.2,green,dashed] (3.6,2) -- (4.4,2);
\draw [line width=1.2,green,dashed] (3.6,3) -- (4.4,3);
\draw [line width=1.2,green,dashed] (3.6,4) -- (4.4,4);
\draw [line width=1.2,green,dashed] (3.6,5) -- (4.4,5);
\draw [line width=1.2,green,dashed] (3.6,6) -- (4.4,6);

\node at (3,1) [anchor=west]{$a_1$};
\node at (3,2) [anchor=west]{$b_1$};
\node at (3,3) [anchor=west]{$a_2$};
\node at (3,4) [anchor=west]{$b_2$};
\node at (3,5) [anchor=west]{$a_N$};
\node at (3,6) [anchor=west]{$b_N$};

\node at (5,1) [anchor=east]{$b'_1$};
\node at (5,2) [anchor=east]{$a'_1$};
\node at (5,3) [anchor=east]{$b'_2$};
\node at (5,4) [anchor=east]{$a'_2$};
\node at (5,5) [anchor=east]{$b'_N$};
\node at (5,6) [anchor=east]{$a'_N$};

\node at (1,2) [anchor=east]{$n_1$};
\node at (1,4) [anchor=east]{$n_2$};
\node at (1,6) [anchor=east]{$n_N$};

\node at (7,1) [anchor=west]{$n'_1$};
\node at (7,3) [anchor=west]{$n'_2$};
\node at (7,5) [anchor=west]{$n'_N$};

\draw [line width=1.2,blue,arrows = {-Stealth[scale=0.9,bend]}] (1,4.8) -- (3,6);
\draw [line width=1.2,blue,arrows = {-Stealth[scale=0.9,bend]}] (1,4.8) -- (3,4);
\draw [line width=1.2,blue,arrows = {-Stealth[scale=0.9,bend]}] (1,4.8) -- (3,2);

\node at (1,4.8) [anchor=east]{$i$};

\node at (4,0.1) [anchor=south]{On ${\cal C}_2$};
\node at (1,0.1) [anchor=south]{Inside ${\cal C}_2$};
\node at (7,0.1) [anchor=south]{Outside ${\cal C}_2$};

\node at (3,6.2) [anchor=south]{${\cal S}_f$};
\node at (5,6.2) [anchor=south]{${\cal S}_e$};
\node at (1.6,5.2) [anchor=south]{${\cal K}_f$};

\end{tikzpicture}
\caption{\label{fig:mode_coupling} Signal flow graph of processes on the reference surface ${\cal C}_2$. ${\cal S}_f$ is the scattering matrix looking towards the CRES region.  $a$ and $b$ are the directed fields looking inwards. ${\cal S}e$ is the scattering matrix looking outwards towards free space.   $a'$ and $b'$ are the directed fields looking outwards.  A current source $i$ is distributed through ${\cal K}f$ into outgoing fields (blue). $n$ are noise sources (red) originating from the internal background environment, coupling energy out of the CRES region. $n'$ are noise sources originating from the external background environment, coupling energy into the CRES region. The noise souces are generally correlated.
}
\end{figure}
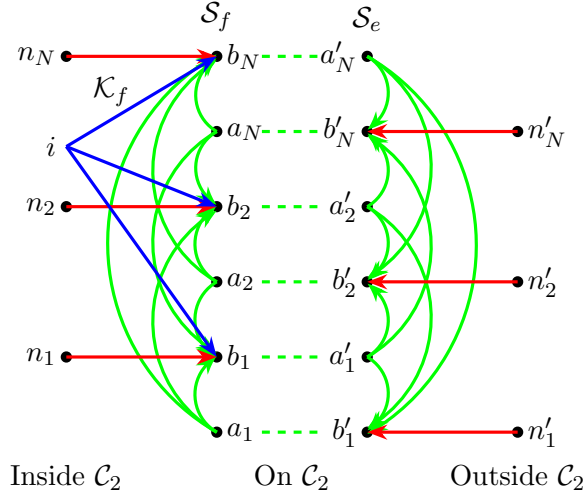

The next step is to combine (\ref{eqn:field_network_equation}) and (\ref{eqn:ext_field_equation}), which is best understood through the signal-flow-graph shown in Fig.~\ref{fig:mode_coupling}. The source region (\ref{eqn:field_network_equation}) is connected to the load region (\ref{eqn:ext_field_equation}) using $\mathbf{a}' = \mathbf{b}$ and $\mathbf{b}' = \mathbf{a}$ to give
\begin{eqnarray}
\label{eqn:embed_free}
\mathbf{b} = \left( \mathcal{I}  - \mathcal{S}_f  \mathcal{S}_e \right)^{-1} \left( \mathbf{c} + \mathcal{S}_f  \mathbf{c}' \right) \\ \nonumber 
\mathbf{a} =  \left( \mathcal{I}  - \mathcal{S}_e  \mathcal{S}_f \right)^{-1} \left(  \mathcal{S}_e \mathbf{c} +  \mathbf{c}' \right) .
\end{eqnarray}
These expressions allow the vectors $\mathbf{a}$ and $\mathbf{b}$ to be found for any internal $\mathbf{c}$ and external $\mathbf{c}'$ sources. Because  $\mathbf{a}$ and  $\mathbf{b}$ are known, the fields within within $ {\cal C}_{2}$ can be found. The terms  $\left( \mathcal{I}  - \mathcal{S}_f  \mathcal{S}_e \right)^{-1}$ and $\left( \mathcal{I}  - \mathcal{S}_e  \mathcal{S}_f \right)^{-1}$ are representative of resonant loops, and occur because if evanescent fields are excited, near this virtual surface,  energy sloshes backwards and forwards across  $ {\cal C}_{2}$. Simulations show that  (\ref{eqn:embed_free})  does indeed give the correct results for free-space embedding, even when evanescent fields are present.

\section{Horn antenna system}
\label{sec:horn_antennas}

The next step is to place antennas on $ {\cal C}_{2}$. Ideally, the exact scattering parameter representation of the antenna system would be known, but various approximations are possible, which capture the essential features. We shall assume that the antennas are horns, even though other antennas can be modelled in the same way.  Assume that incoming and outgoing travelling waves on the transmission line of antenna $n$ are described by new amplitudes  $a''_{n}$ and $b''_{n}$ respectively. The incoming wave on the $n$'th transmission line distributes itself over  $ {\cal C}_{2}$ according to the possibly complex-valued antenna transmission aperture field, $U_n (\phi)$, and so
\begin{eqnarray}
\label
{eqn:a_due_to_antenna}
a (\phi) & = \frac{1}{\sqrt{R_2 L}} \sum_{n} \alpha_n U_n (\phi) a''_n,
\end{eqnarray}
which is simply statement of linear dependence. $\alpha_n$ is a complex-valued coefficient, which accounts for any overall loss or phase factor. 

The power flowing into $ {\cal C}_{2}$ as a consequence of all of the  $a''_n$'s is, according to (\ref{eqn:power_surface}),
\begin{eqnarray}
\frac{R_{2} L}{2} \int_{0}^{2 \pi} | a(\phi)|^{2} d \phi  = \frac{1}{2} \sum_{mn} a^{'' \ast}_m a''_n  \alpha_m^\ast \alpha_n  \int_{0}^{2 \pi}  U_m^\ast (\phi)  U_n (\phi) d \phi  =  \frac{1}{2} \sum_{n}  | \alpha_n |^2 | a''_n |^2, 
\end{eqnarray}
where we have used 
\begin{equation}
\label{eqn:u_normalisation}
	\int_0^{2 \pi} U^*_m (\phi) U_n^{\vphantom *} (\phi) \, d\phi = \delta_{mn}.
\end{equation}
which follows because the ports of the antennas are spatially distinct, and so their antenna patterns are orthogonal. The normalisation in (\ref{eqn:a_due_to_antenna}) ensures that $| a''_n |^2 / 2$ is the power flowing into the port of antenna $n$. 

After sampling, (\ref{eqn:a_due_to_antenna}) can be written
\begin{equation}
\label{eqn:a_from_a_prime}
\mathbf{a}  = \frac{1}{\sqrt{R_2 L}} \mathcal{U}  \cdot \mathcal{\alpha} \cdot \mathbf{a}'',
\end{equation}
where each column of $\mathcal{U}$ contains the sampled transmission pattern of a single antenna, {\it defined over the whole of} ${\cal C}_2$:
\begin{equation}
\label{eqn:def_u_matrix}
\mathcal{U}_{mn} = U_n (\phi_m),
\end{equation}
and $\mathcal{\alpha}$ is a diagonal matrix of modal transmission factors.  $\mathcal{U}$  is injective because there are generally fewer antennas than degrees of freedom in $\mathbf{a}$, and so it has a left inverse, but no right inverse. 
 
The reception pattern of an antenna needs more thought. The spatio-temporal form of the various electromagnetic fields are  being described through $F(\mathbf{r}) e^{-i \omega t}$, where $F(\mathbf{r})$ is some complex-valued function of position. $F(\mathbf{r}) e^{+i \omega t}$ has the same spatial form, but with all wave-like processes travelling backwards in a time-reversed manner. Ultimately, however, only $\mbox{Re}[ F(\mathbf{r}) e^{-i \omega t}]$ has physical significance, and the same time-reversed behaviour results from $\mbox{Re}[ F^{\ast}(\mathbf{r}) e^{-i \omega t}]$, which has the advantage of preserving the temporal factor. Thus time-reversed behaviour can be achieved by using $F^{\ast}(\mathbf{r})$ rather than $F(\mathbf{r})$.  It follows that the reception pattern of an antenna is given by  $V_n (\phi) = U_n^\ast (\phi)$. 

To calculate the received signal, we must form the overlap integral between the incoming field and the reception pattern of each antenna:
\begin{equation}
\label{eqn:b_due_to_antenna}
b''_ n  =  \alpha_n^{\ast}  \sqrt{R_2 L} \int U_n (\phi)^{\ast}  b(\phi)  \, d \phi.
\end{equation}
Equation (\ref{eqn:b_due_to_antenna}) can also be derived through reciprocity. The total power collected by the antennas is
\begin{equation}
\label{eqn:pow_due_to_antenna}
\frac{1}{2} \sum_{n}  | b''_ n |^2  = \frac{R_{2} L}{2} \sum_{n} | \alpha_n |^2   \int_{0}^{2 \pi}  \int_{0}^{2 \pi}  U_n (\phi)^{\ast}  U_n (\phi')  b(\phi)   b(\phi')^\ast \,  d \phi  \, d \phi'. 
\end{equation}
Even if we ignore the loss factors, such that  $\alpha_n$ is purely imaginary, the power received may be less than that incident on the apertures. The problem is that the mapping that represents the antenna system is not mathematically complete: $\sum_{n}  U_n (\phi)^{\ast} U_n (\phi')   \neq \delta (\phi - \phi') $. The reception pattern comprises a single mode, which is generally different to the field actually incident on the aperture; most of the incident power is reflected. It should be noted that the situation is different to power detectors having free-space resistive absorbers, because these can abosorb power in multiple spatial modes simultaneously. If the incoming field matches the reception pattern of each antenna, over its respective domain, all of the power is collected, as described by (\ref{eqn:pow_due_to_antenna}). Finally, the net power delivered to the load of the antenna is
\begin{equation}\label{eqn:antenna_input_power}
	P_n =  \frac{1}{2} |b''_n|^2 - \frac{1}{2} |a''_n|^2,
\end{equation}
where usually for matched terminations, $|a''_n|^2 = 0, \, \forall n$. Imperfect loads can be modelled using reflection coefficients in the usual way, but now standing waves may be set up within ${\cal C}_2$.

After sampling  (\ref{eqn:b_due_to_antenna}) can be written
\begin{equation}
\label{eqn:b_from_a_prime}
\mathbf{b}''  = \sqrt{R_2 L} \mathcal{\alpha}^{\dagger} \cdot \mathcal{U}^\dagger \cdot \mathbf{b} = \sqrt{R_2 L} \mathcal{\alpha}^{\dagger} \cdot \mathcal{V} \cdot \mathbf{b}.
\end{equation}
The reception operator $\mathcal{V}$ does not have a left inverse, because the spatial form of the field at the aperture cannot be recovered from the complex amplitude of the wave at the port.

Substituting (\ref{eqn:a_from_a_prime})  into (\ref{eqn:field_network_equation}) and  using (\ref{eqn:b_from_a_prime}) gives
\begin{equation}
\label{eqn:antenna_scattering_network}
\mathbf{b}'' = \mathcal{S} \cdot \mathbf{a}''		+  \tfrac{1}{2} \sqrt{k_0 L Z_0} \, \mathcal{K} \cdot \mathbf{i}
\end{equation}
where
\begin{eqnarray}
\label{eqn:s_matrix_antennas}
	\mathcal{S} & =  \mathcal{\alpha}^{\dagger} \cdot  \mathcal{U}^\dagger \cdot \mathcal{S}_f \cdot \mathcal{U}  \cdot \mathcal{\alpha} \\ \nonumber
	\mathcal{K} & = \sqrt{4 k_0 R_2} \,  \mathcal{\alpha}^{\dagger} \cdot \mathcal{U}^\dagger \cdot \mathcal{K}_f,
\end{eqnarray}
$\mathcal{K}$ has been normalised to the total power radiated by a line current: $k_0 L Z_0 I_0^2 / 8$. 

$S$ is the scattering matrix of the CRES system referenced to the ports of the antennas.  Notice that because $\mathcal{S}_f$ is unitary, $\mathcal{S}$ is unitary if the antenna system couples to itself in a lossless way: $ \mathcal{U}^\dagger \cdot \mathcal{U}  = \mathcal{I}$, where $\mathcal{I}$ is the $N \times N$ identity matrix, and $N$ is the number of horns. If the antenna system, $\mathcal{U}$, is not unitary,  $\mathcal{S}$ is not unitary and noise power can appear the outputs of the antennas from stray radiation coupling in from the environment in which  ${\cal C}_2$ is embedded. Later we will use  $S$ to determine the noise correlation matrix of thermal noise. Equation (\ref{eqn:antenna_scattering_network}) is a central element of our model. It assumes that the antennas terminate the outgoing fields on $ {\cal C}_{2}$ in such a way that $\mathcal{S}_e = 0$. It applies when none of the signal incident on the antennas is reflected, which is an idealisation because evanescent fields cannot be terminated in this way.

For illustrative purposes we shall use the TE$_{10}$ mode of a profiled rectangular horn, and assume that the radius of curvature of the phase front is $R_2$. Then
\begin{equation}
\label{eqn:horn_cosine}
	U_n (\phi) = \begin{cases}
		\sqrt{\tfrac{2}{\Delta \alpha}}
			\cos \Bigl( \tfrac{\pi \{ \phi - \alpha_n \}}{\Delta \alpha} \Bigr)
			& 2 |\phi - \alpha_n | < \Delta \alpha \\
		0 & \text{otherwise},
	\end{cases}
\end{equation}
where $\alpha_n$ sets the centre of the aperture and $\Delta \alpha$ the angular width.  For an electron source, the horn would need to be rotated such that the polarisation is perpendicular to $z$ and then the field distribution would be uniform. $U_n (\phi)$ is real because the flare angle of the horn places a phase front over $R_2$. $U_n (\phi)$ could be complex to take into account any mismatch between the radius of curvature of the horn and $R_2$. Alternatively, and we have done this in our simulations, the surface ${\cal C}_2$ can be chosen to follow the phase front of the field at the apertures of the horns.

\begin{figure}
%C:\Users\Stafford Withington\Documents\Projects starting in Oxford March 2023\QTNM\PSWF for CRES\CRES_PSWF_ANT_COUPLING_V1
     \centering
     \begin{subfigure}[b]{1.0\textwidth}
         \centering
         \includegraphics[trim = 1cm 1cm 8cm 15cm, width=70mm]{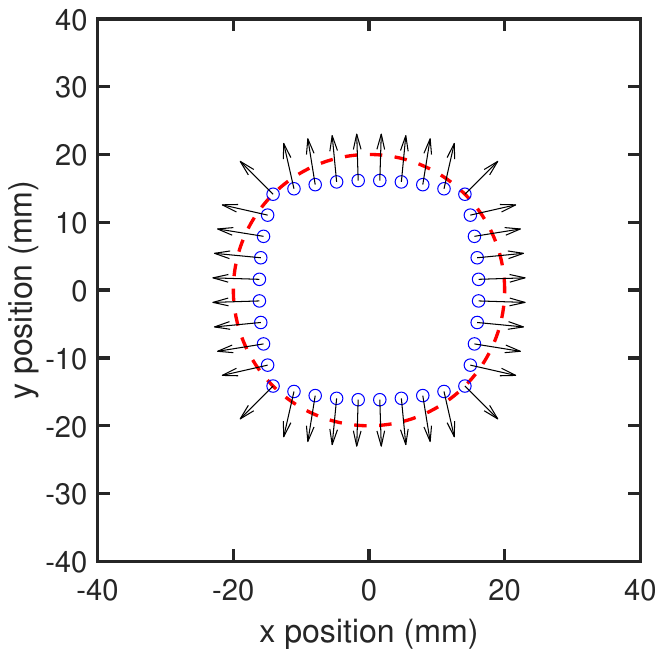}
% left,bottom,right,top         
     \end{subfigure}
     \\   
     \begin{subfigure}[b]{1.0\textwidth}
     \centering
     \includegraphics[trim = 1cm 1cm 8cm 15cm, width=70mm]{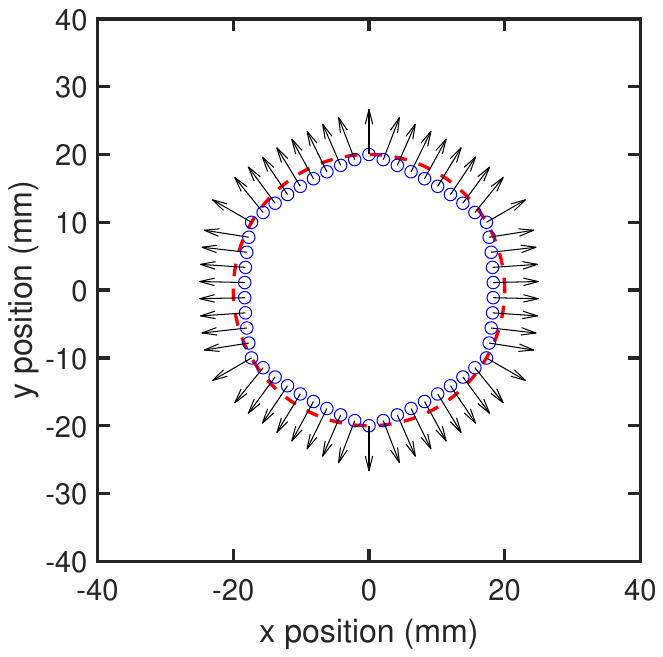}
% left,bottom,right,top         
     \end{subfigure}
     \\   
     \begin{subfigure}[b]{1.0\textwidth}
     \centering
     \includegraphics[trim = 1cm 1cm 8cm 15cm, width=70mm]{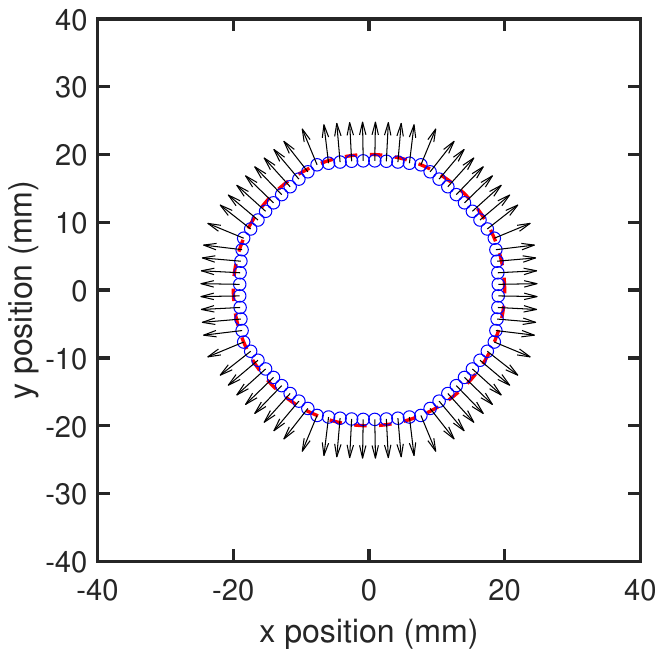}
% left,bottom,right,top         
     \end{subfigure}
\caption{Illustration of the sampling strategy for 4 (top), 6 (middle) and 8 (bottom) horns having aperture radii of curvature, 50 mm, different to the radius of curvature of the reference circle, 20 mm (red dashed line). Sample points (blue circles) and unit normals (black arrows) are shown, but in the simulations many more sample points, typically 1600, were used. }
\label{sampling_fig}
\end{figure}

As a numerical example, consider the case where horns having cosine aperture field distributions, consistent with the polarisation of the 2D case, are arranged around a circle. Figure \ref{sampling_fig} gives an indication of the sampling strategy used. The plots correspond to  4 (top), 6 (middle) and 8 (bottom) horns having aperture radii of curvature, 50 mm, different to the radius of curvature of the reference circle, 20 mm (red dashed line). Sample points (blue circles) and unit normals (black arrows) are shown, but in the simulations many more sample points, typically 1600, were used. In the simulations, ${\cal C}_2$ was chosen to follow the phase fronts of the horns rather than the reference circle. This requires care when integrating the singularities, but has the advantage that the horn reception patterns are real over the surface of integration even though the flare angle can be chosen to adjust the focus point.

Table \ref{coupling4} shows the results of a simulation with $\lambda =$ 11 mm, $R2$ = 40 mm and the radius of curvature of the horns $Rh = $ 40 mm.  This arrangement is essentially the same as Fig.~\ref{figure7}. The top block of the Table shows $| {\cal S}_{mn} |^2$ at the ports of the antennas. In this simple case, the basic scattering matrix ${\cal S}_f$ was used, corresponding to the evanescent waves being terminated by $Z_0$ and no reflections from the horn apertures in those modes orthogonal to the reception pattern. More complicated simulations are possible, as described in the text. In the Table, it is clear that the reflection coefficients at the horn apertures are low (- 34 dB), the coupling to the opposite antenna is high, and the coupling to neighboring antennas low. This occurs because apertures of the horns, in wavelengths, are large, and the beams quite narrow. The bottom block of the Table shows $[{\cal I} -  {\cal S} {\cal S}^{\dagger} ]$. As will be described in the next section, the diagonal elements give the classical noise per unit bandwidth emanating from the antenna ports, assuming that the system is enclosed in a thermal environment having temperature $T_p$ = 1 K. These values can be simply scaled by the actual environmental temperature to give the noise temperature of the ports. So here, if $T_p =$ 50 K, the added noise contribution to each receiver would only be 2.5 K, emphasising the benefit of having high-efficiency coupling. The off-diagonal elements give the real-valued correlations between the noise waves leaving different antenna ports, expressed as a temperature for a reference temperature of 1 K. The noise waves are highly uncorrelated as a consequence of the system being well matched at the ports. Internal reflections will degrade this number and increase the correlated noise.

Tables \ref{coupling6} and \ref{coupling8} show the same information but with the number of horns increased to 6 and 8 respectively. As the apertures get smaller, the beams become wider, and each horn couples  to a number of horns on the other side of the reference circle. As the coupling to the opposite horn reduces, the noise increases, but not badly because a large fraction of the beam still couples into the other receivers, rather than being terminated on the environment. Notice that the correlations increase also.  Other simulations, not shown here, indicate that if the radius of curvature of the phase front is adjusted in accordance with \S~\ref{sec:can_arr}, the coupling efficiencies can be improved futher.  For example in the 6-horn case, the coupling efficiency can be increased to over 92 \% by increasing the radii of curvature of the horns to 120 mm, yielding a high resilience to thermal background noise.

\begin{table}[h]
\begin{centering}
\begin{tabular}{ |c | c c c c  |} \hline
    &   1		    &      2      &         3    &       4     \\ \hline 
1  &  0.0004  &  0.0003  &  0.9495  &  0.0003   \\
2  &  0.0003  &  0.0004  &  0.0003  &  0.9495   \\
3  &  0.9495  &  0.0003  & 0.0004   & 0.0003    \\
4  &  0.0003  &  0.9495  &  0.0003  &  0.0004   \\ \hline
    &       1       &        2      &      3        &     4          \\ \hline
1  & 0.0495   &   0.0233 & -0.0043  &  0.0233   \\
2  &  0.0233  &   0.0495 &  0.0233  &  -0.0043  \\
3  & -0.0043  &   0.0233 &  0.0495  &  0.0233   \\
4  &  0.0233  & -0.0043  &  0.0233  &  0.0495   \\ \hline
\end{tabular}
\caption{Simulation where 4 cosine horns, operating at a wavelength of $\lambda =$ 11 mm, were arranged around a circle having R2 = 40 mm, and also Rh = 40 mm. The top block shows the squared moduli of the elements of ${\cal S}$. The bottom block shows ${\cal I} - {\cal S}{\cal S}^{\dagger}$, which are the correlations in the noise waves leaving the ports when the system is embedded in an environment having a physical temperature of 1K.}
\label{coupling4}
\end{centering}
\end{table}

\begin{table}[h]
\begin{centering}
\begin{tabular}{ |c | c c c c c c |} \hline
    &   1		    &      2      &         3    &       4   & 5    &   6  \\ \hline 
1  &  0.0003  &  0.0000  &  0.0033  &  0.8337  &  0.0033  &  0.0000 \\
2  &  0.0000  &  0.0003  &  0.0000  &  0.0033  &  0.8337  &  0.0033 \\
3  & 0.0033   &  0.0000  &  0.0003  &  0.0000  &  0.0033  &  0.8337 \\
4  & 0.8337  &  0.0033   &  0.0000  &  0.0003  &  0.0000  &  0.0033 \\
5  & 0.0033  &  0.8337   & 0.0033   & 0.0000   & 0.0003   &  0.0000 \\
6  & 0.0000  &  0.0033   & 0.8337   & 0.0033   & 0.0000   &  0.0003 \\ \hline
  &    1            &       2        &      3        &     4         &       5       &    6      \\ \hline
1  &  0.1594  &  0.0760  & -0.0080   & -0.0106  & -0.0080  &  0.0760  \\
2  &  0.0760  &  0.1594  &  0.0760   & -0.0080   & -0.0106  & -0.0080 \\  
3   & -0.0080  &  0.0760  &  0.1594  &  0.0760  & -0.0080  &  -0.0106 \\
4  &  -0.0106  &  -0.0080 &   0.0760 &   0.1594  &  0.0760  & -0.0080 \\
5 &  -0.0080   & -0.0106  &  -0.0080 &   0.0760  &  0.1594  &  0.0760 \\
6 &   0.0760  &  -0.0080  & -0.0106  & -0.0080   & 0.0760  &  0.1594 \\ \hline
\end{tabular}
\caption{Simulation where 6 cosine horns, operating at a wavelength of $\lambda =$ 11 mm, were arranged around a circle having R2 = 40 mm, and also Rh = 40 mm. The top block shows the squared moduli of the elements of ${\cal S}$. The bottom block shows ${\cal I} - {\cal S}{\cal S}^{\dagger}$, which are the correlations in the noise waves leaving the ports when the system is embedded in an environment having a physical temperature  of 1K.}
\label{coupling6}
\end{centering}
\end{table}

\begin{table}[h]
\begin{centering}
\begin{tabular}{ |c | c c c c c c c c |} \hline
    &   1		    &      2      &         3    &       4   & 5    &   6  &    7    &   8 \\ \hline 
1    & 0.0002  &  0.0000 &   0.0000 &   0.0231 &   0.6216 &   0.0231 &   0.0000 &    0.0000 \\ 
2   &    0.0000  &  0.0002  &  0.0000 &   0.0000 &   0.0231  &  0.6216  &  0.0231 &   0.0000 \\
3   &    0.0000  &  0.0000   & 0.0002  &  0.0000 &   0.0000  &  0.0231  &  0.6216  &  0.0231 \\
4   &   0.0231   & 0.0000  &  0.0000   & 0.0002  &  0.0000  &  0.0000  &  0.0231  &  0.6216 \\
5   &    0.6216   &  0.0231 &   0.0000 &   0.0000 &   0.0002 &   0.0000 &   0.0000 &   0.0231 \\
6   &    0.0231   & 0.6216   & 0.0231  &  0.0000 &   0.0000  &   0.0002  &  0.0000 &   0.0000 \\
7   &    0.0000  &  0.0231  &  0.6216  &  0.0231  &  0.0000  &  0.0000   &  0.0002  &  0.0000 \\
8   &    0.0000  &  0.0000  &  0.0231   & 0.6216  &  0.0231  &  0.0000   &  0.0000  &  0.0002 \\ \hline
    &      1          &      2         &     3         &      4        &      5        &       6        &        7     &  8\\ \hline
1  &   0.3319  &   0.1485  &  -0.0154  &  -0.0041 &  -0.0153  &  -0.0041 &  -0.0154 &   0.1485 \\
2 &  0.1485  &  0.3319  &  0.1485 &  -0.0154 &  -0.0041 &  -0.0153 &  -0.0041 &  -0.0154 \\ 
3 &  -0.0154 &   0.1485 &   0.3319  &  0.1485 &  -0.0154 &  -0.0041 &  -0.0153 &  -0.0041 \\
4  & -0.0041 &  -0.0154 &   0.1485 &   0.3319 &   0.1485 &  -0.0154 &  -0.0041 &  -0.0153 \\
5  & -0.0153 &  -0.0041 &  -0.0154 &   0.1485 &   0.3319 &   0.1485 &  -0.0154 &  -0.0041 \\
6  & -0.0041 &   -0.0153 &  -0.0041  & -0.0154  &  0.1485  &  0.3319  &  0.1485  &  -0.0154 \\
7  &  -0.0154  &  -0.0041 &  -0.0153 &  -0.0041 &  -0.0154 &   0.1485 &   0.3319 &   0.1485 \\
8   &  0.1485 &  -0.0154  &  -0.0041 &  -0.0153 &  -0.0041 &  -0.0154 &   0.1485 &   0.3319  \\ \hline 
\end{tabular}
\caption{Simulation where 8 cosine horns, operating at a wavelength of $\lambda =$ 11 mm, were arranged around a circle having R2 = 40 mm, and also Rh = 40 mm. The top block shows the squared moduli of the elements of ${\cal S}$. The bottom block shows ${\cal I} - {\cal S}{\cal S}^{\dagger}$, which are the correlations in the noise waves leaving the ports when the system is embedded in an environment having a physical temperature  of 1K.}
\label{coupling8}
\end{centering}
\end{table}

\newpage

The above analysis does not take into account power reflected back into $ {\cal C}_{2}$ off of the antennas, which for illustration we shall take to be the apertures of horns. To include these reflections, it is necessary to form an embedding matrix  $\mathcal{H}$, similar to ${\cal S}_e \in \mathbb{C}^{N \times N}$, but which takes into account the presence of the horns.  Now, however, ${\cal H}  \in \mathbb{C}^{(N+M) \times (N+M)}$,  because the ports of the $M$ horns add additional degrees of freedom to the $N$ spatial sample points on ${\cal C}_2$. ${\cal H}$ has the block form
\begin{equation}
\label{eqn:hmatrix}
\begin{bmatrix}
{\bf b}'' \\
{\bf b}'
\end{bmatrix}
=
\begin{bmatrix}
{\cal H}_{11} & {\cal H}_{12} \\
{\cal H}_{21} & {\cal H}_{22} 
\end{bmatrix}
\cdot
\begin{bmatrix}
{\bf a}'' \\
{\bf a}'
\end{bmatrix},
\end{equation}
where ${\bf a}''$ and ${\bf b}''$  are the waves incident on and travelling away from the ports of the horns respectively.  ${\bf a}'$ and ${\bf b}'$ are the same as those of Fig \ref{fig:mode_coupling}, but are now constrained by the boundary condition imposed on ${\cal C}_2$ by the apertures of the horns. The submatrix ${\cal H}_{11} \in \mathbb{C}^{M \times M}$ describes the intrinsic scattering properties of the horns at their terminals, independent of coupling through the radiation patterns. The diagonal elements are the horn input reflection coefficients, and the off-diagonal elements account for any intrinsic, not through  ${\cal C}_2$, cross talk between the horns in an array. For ideal, matched horns ${\cal H}_{11} = {\cal O}$. Each column of ${\cal H}_{21} \in \mathbb{C}^{N \times M}$ is a horn transmission aperture field, and is the same as $\mathbf{U}$, described previously (\ref{eqn:def_u_matrix}). Likewise,  each row of ${\cal H}_{12} \in \mathbb{C}^{M \times N}$ is a horn reception aperture field, where  ${\cal H}_{12} = {\cal H}_{21}^{\dagger}$.  Finally,  ${\cal H}_{22} \in \mathbb{C}^{N \times N}$ is a block matrix, where each diagonal block comprises the spatial scattering parameters looking into the aperture of each horn.  ${\cal H}_{22}$ is block diagonal if it is assumed that a wave incident on the aperture of one horn does not appear travelling away from the aperture of another. 

There is now an important consideration. We shall assume, reasonably, that the reception mode of each horn is orthogonal to all of the modes that can scatter power at the aperture. In other words, there is some incoming field that is transmitted perfectly to the output port, and this mode is orthogonal to all of the modes that are reflected back at the aperture. ${\cal H}_{22}$ therefore has an $M$-dimensional null space at best. This restriction can be removed if the reception pattern is itself partially scattered even when the ouput port is terminated in $Z_0$.

If the antenna has no dissipative losses ${\cal H}$ is unitary: $ {\cal H}^{\dagger}  {\cal H} =  {\cal I}_N$.  If ${\cal H}_{11} = {\cal O}$, the unitary condition multiplies out to give 3 independent expressions
\begin{eqnarray}
\label{eqn:unitary}
{\cal H}_{21}^{\dagger} {\cal H}_{21}  &  = {\cal I}_{M \times M} \\ \nonumber
{\cal H}_{21}^{\dagger} {\cal H}_{22} &  = {\cal O}_{M \times N} \\ \nonumber
{\cal H}_{22}^{\dagger} {\cal H}_{22}  &  = {\cal I}_{N \times N}.
\end{eqnarray}
The first identity describes propagating incoming waves at the ports to the apertures and back again in a time-reversed manner, ensuring that no power is lost: lossless transmission properties. The second identity describes scattering a field off of the aperture, and then time-reversed back propagation through the horn to the input port. If the modes available for scattering are orthogonal to the transmission modes, as discussed above, there are no fields for which this process can occur. The last identity describes reflecting fields at the aperture off of the aperture and then time-reversing the process. All possible reflections at the aperture must be lossless also. Therefore unitarity overall can be derived from knowledge of the individual scattering processes ${\cal H}_{21}$ and ${\cal H}_{22}$. Equations (\ref{eqn:unitary}) provide useful tests for simulation software.

The next step is to consider how to combine the horn scattering matrix  ${\cal H}$ with the CRES system scattering matrix  ${\cal S}_f$. Equation (\ref{eqn:hmatrix}) can be expanded, assuming ${\cal H}_{11} = {\cal O}$, to give 
\begin{eqnarray}
{\bf b}'' & = &  {\cal H}_{12} {\bf a}' \\ \nonumber
{\bf b}'  & = & {\cal H}_{22} {\bf a}' + {\cal H}_{21} {\bf a}'' ,
\end{eqnarray}
but also ${\bf a}' = {\cal S}_f {\bf b}' + {\bf c}$, and these can be combined to give
\begin{eqnarray}
\label{eqn:embed_horn_exp}
{\bf b}'' & =  {\cal H}_{12} \left[ {\cal I} - {\cal S}_f {\cal H}_{22} \right]^{-1} {\cal S}_f  {\cal H}_{21} {\bf a}'' +  {\cal H}_{12} \left[ {\cal I} - {\cal S}_f {\cal H}_{22} \right]^{-1} {\bf c},
\end{eqnarray}
which replaces (\ref{eqn:embed_free}). 

The first term in (\ref{eqn:embed_horn_exp}) gives the scattering matrix of the whole system looking into the ports of the antennas, and the second term describes the signals generated by a CRES electron. The  inverse $ \left[ {\cal I} - {\cal S}_f {\cal H}_{22} \right]^{-1}$ describes the formation of multiple loops, characterising the appearance of resonant modes trapped in the CRES region by horn-aperture scattering: there are potentially many of them: $N-M$ at most. These processes include energy sloshing backwards and forwards in the localised near field, and also processes involving energy bouncing backwards and forwards across ${\cal C}_2$. If this term is expanded, the different orders correspond to different numbers of round-trip paths. Expanding to second order:
\begin{eqnarray}
\label{eqn:embed_horn}
{\bf b}'' & \approx &  \left\{ {\cal H}_{12}  {\cal S}_f  {\cal H}_{21}  {\bf a}''    +  {\cal H}_{12} {\bf c} \right\}
 +  \left\{  {\cal H}_{12}  {\cal S}_f {\cal H}_{22} {\cal S}_f  {\cal H}_{21} {\bf a}''   +    {\cal H}_{12}  {\cal S}_f {\cal H}_{22} {\bf c} \right\}.
\end{eqnarray}
The first term, ${\cal H}_{12}  {\cal S}_f  {\cal H}_{21}$, describes the direct path comprising transmission, propagation, and reception; the second term,  ${\cal H}_{12}  {\cal S}_f {\cal H}_{22} {\cal S}_f  {\cal H}_{21}$, describes transmisson, propagation in the CRES region, scattering off of the horns, propagation in the CRES region and then reception. These paths, which describe single scattering events at the horns, add to the direct paths to give Fano-type resonances. The higher orders describe multiple loops around the same paths. Notice that if ${\cal S}_f {\cal H}_{22} = {\cal O}_N$, standing waves would be eliminated. For example,  ${\cal H}_{22} = {\cal O}_N$ would correspond to each horn dissipating perfectly all of the modes orthogonal to the reception pattern. Also, according to the term, ${\cal H}_{12}  {\cal S}_f {\cal H}_{22} {\cal S}_f  {\cal H}_{21}$ resonant modes may be isolated from the ports of the antennas because there may be resonant modes that fall with the null space of ${\cal H}_{21}$; these may have very high Q factors. 

\begin{figure}
\centering
\includegraphics[trim = 1cm 1cm 9cm 15cm, clip,width=70mm ]{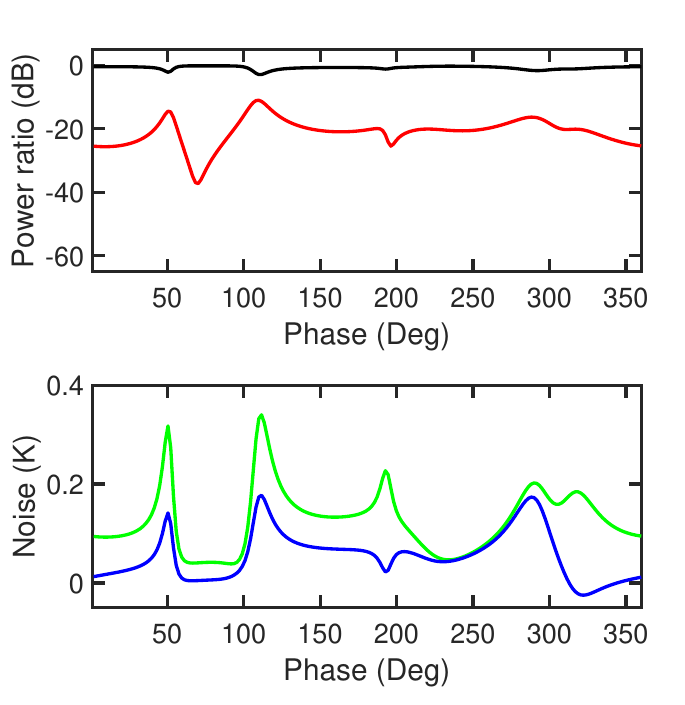}
% left,bottom,right,top
\caption{Simulation of a 6 element array of cosine horns with aperture reflections. Wavelength  $\lambda =$ 11 mm, R2 = 40 mm and Rh= 40 mm. Top plot: Power transmission factor between opposite horns (black), and input reflection coefficient of each horn (red). Bottom plot: noise temperature of the wave leaving each port assuming that the system is embedded in a 1 K environment (green) , and the correlated noise temperature between two neighbouring ports (blue). Both are plotted as a function of the phase of the aperture reflection coefficient.
}
\label{fig:ref_1}
\end{figure}
\begin{figure}
\centering
\includegraphics[trim = 1cm 1cm 9cm 15cm, clip,width=70mm ]{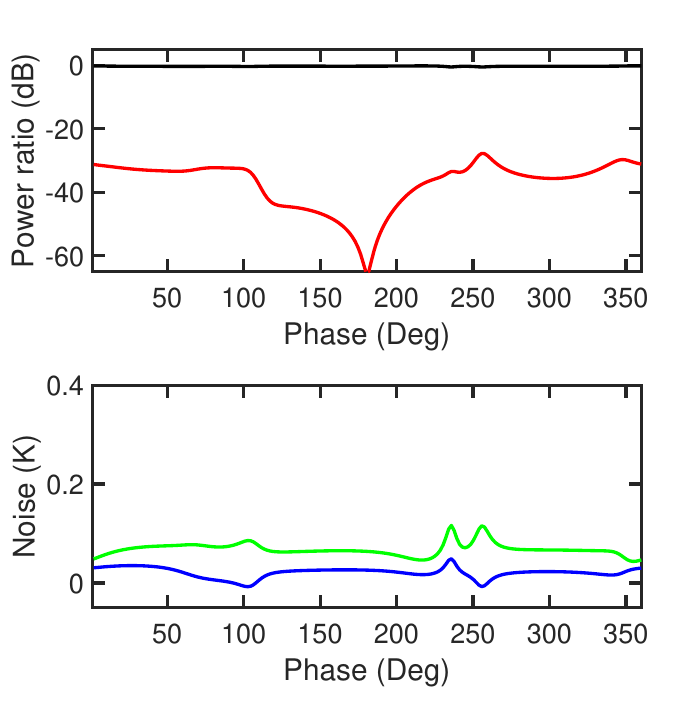}
% left,bottom,right,top
\caption{Simulation of a 6 element array of cosine horns with aperture reflections. Wavelength  $\lambda =$ 11 mm, R2 = 40 mm and Rh= 100 mm. The system is identical to Fig. \ref{fig:ref_1}, but the phase front of the horn is profiled to provide a better match to the opposite partner. Top plot: Power transmission factor between opposite horns (black), and input reflection coefficient of each horn (red). Bottom plot: noise temperature of the wave leaving each port assuming that the system is embedded in a 1 K environment (green) , and the correlated noise temperature between two neighbouring ports (blue). Both are plotted as a function of the phase of the aperture reflection coefficient.}
\label{fig:ref_2}
\end{figure}

The appearance of resonant trapped modes is a significant problem for all inward-looking antennas observing large FoVs. The first term in  (\ref{eqn:embed_horn_exp}) leads to imperfect input impedances, and as will be described, correlations in radiated thermal noise. The appearance of trapped modes in the second term, will lead to a non-uniform response over the FoV, and a non-uniform energy resolution.  Also, it provides a mechanism for backaction on the radiating electron; in fact this term can be used to calculate the field appearing at the position of the electron. These issues are of considerable significance, and it would be desirable to use the notion of minumum-scattering antennas when designing a CRES system. In any case, it may be possible to control the effects trapped modes by using considerations of the above kind.

How should ${\cal H}_{22}$ be populated in numerical simulations? The answer depends on the exact design. Horn mode-matching design software will give ${\cal H}_{22}$ directly as a part of the design process. For waveguide horns, a simpler approach is to construct ${\cal H}_{22}$ from all of the modes that are cut off, and modes that are not compatible with waveguide propagation at all (for example, those fields that are not compatible with the boundary conditions imposed by metallic rectangular waveguide). For illustrative simulations, the phases of the reflection coefficients can be chosen to span the range of possibiities. For horns, some modes may penetrate deep inside the horn before being reflected, whereas others are reflected straight away.  In the most general case, ${\cal H}_{22}$ can be written as an SVD, but here we shall use the eigenvector expansion
\begin{equation}
{\cal H}_{22} = \sum_{j=1}^{N-M} {\bf h}_j \Gamma_j {\bf h}_j^{\dagger},
\end{equation}
where ${\bf h}_j$ is a column vector that contains the spatial form of reflected mode $j$, and $\Gamma_j$ the associated complex-valued reflection coefficient.

Another possibility is that some part of ${\cal C}_2$ comprises a surface having surface impedance $Z_s$: for example a metallic sheet or resistive absorber. In this case, the terminating scattering matrix takes the form of a diagonal reflection matrix having elements $(Z_s - Z_0) / (Z_s + Z_0)$. In fact the formulation is then essentially a scattering matrix representation of the Electric Field Integral Equation (EFIE).

Figures \ref{fig:ref_1} and  \ref{fig:ref_2} show typical simulations. A ring of 6 waveguide horn antennas, operating at $\lambda =$ 11 mm and having cosine reception fields, were arranged around a circle having a radius of $R_2 = $ 40 mm. The three lowest-order Fourier modes, above that of the cosine reception pattern,  were reflected at the aperture with reflection coefficients of unity, and reflection phases of $\phi$. In each figure, the top plot shows (black) the power transmission factor from each horn to that of its opposite partner, and (red) the input return loss of each port: both as a function of the phase of the aperture reflection coefficient. Fano resonances are evident. In each case, the bottom plot shows (green) the noise temperature of the wave leaving each port when the system is embedded in an environment having a physical temperature of 1 K, and (black) the correlated contribution between each port and one of its neighbours. In Figure \ref{fig:ref_1} the radius of curvature of the phase front of every horn was 40 mm, equal to $R_2$, and in Figure \ref{fig:ref_2} every phase front was increased to 100 mm, in order to improve the coupling between each horn and its opposite partner. The increased coupling is evident from the improvement in the input return loss, and this is accompanied by a fall in the coupling to the environment, and a decrease in the correlations, as described previously. In a real system, many more modes would be reflected, and these would have different phases, which may help considerably due to the dephasing of the different round-trip paths. 

The central conclusion of this section is that an inward-looking antenna array can be modelled in terms of a scattering matrix that describes power flow at the ports of the antennas, which is itself derived through a scattering parameter representation of energy flow within the FoV. Although the method has been illustrated using horns, it applies equally well to other kinds of antenna. The generic form (\ref{eqn:embed_horn_exp}) is found, with ${\cal H}$ depending on the antennas and assumptions used. 

\section{CRES receiver}
\label{sec:cres_receiver}

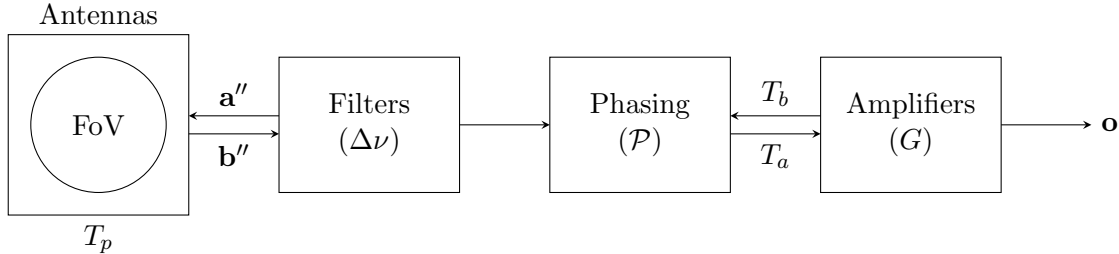
\begin{figure}
\centering
\begin{tikzpicture}[scale=1.2]
\draw[white] (-2, -2) rectangle (12, 2);
\node[anchor=south] at (0, 1) {Antennas};
\draw[shift={(-1, -1)}] (0, 0) rectangle (2, 2);
\node[anchor=north] at (0, -1) {$T_p$};

\draw [fill=none] (0,0) circle (0.75);
\node[anchor=center] at (0, 0) {FoV};

\draw[-stealth] (1, -0.1) to (2, -0.1);
\node[anchor=north] at (1.5, -0.1) {$\mathbf{b}''$};

\draw[stealth-] (1, 0.1) to (2, 0.1);
\node[anchor=south] at (1.5, 0.1) {$\mathbf{a}''$};

\node[anchor=center, align=center, text width=2cm] at (3, 0) {Filters \\ ($\Delta \nu$)};
\draw[shift={(-1, -0.75)}] (3, 0) rectangle (5, 1.5);
\draw[-stealth] (4, 0) to (5, 0);
\node[anchor=center, align=center, text width=2cm] at (6, 0) {Phasing \\ ($\mathcal{P}$)};
\draw[shift={(-1, -0.75)}] (6, 0) rectangle (8, 1.5);

\node[anchor=center, align=center, text width=2cm] at (9, 0) {Amplifiers \\ ($G$)};
\draw[shift={(-1, -0.75)}] (9, 0) rectangle (11, 1.5);

\draw[-stealth] (7, -0.1) to (8, -0.1);
\node[anchor=north] at (7.5, -0.1) {$T_a$};

\draw[stealth-] (7, 0.1) to (8, 0.1);
\node[anchor=south] at (7.5, 0.1) {$T_b$};

\draw[-stealth] (10, 0) to (11, 0);

\node[anchor=west] at (11, 0) {$\mathbf{o}$};

\end{tikzpicture}
\caption{\label{fig:receiver_system} Typical multi-channel CRES system.}
\end{figure}

The antenna model can be combined with a generic model of the signal chain to create a description of a complete CRES instrument: Fig. \ref{fig:receiver_system}. The signals from the antennas first pass through a band-defining filter network, which limits the bandwidth to $\Delta \nu$ around some central frequency $\nu_0$, and then passes through a phasing network, described by matrix $\mathcal{P}$. This network forms a weighted linear combination of the signals to create synthesised beams that are matched as best as possible to the source. Different networks correspond to different coherent reception patterns. If the phasing is carried out digitally, after down-conversion and sampling, multiple phasing networks can be realised simultaneously, ensuring that the system is sensitive to all electrons regardless of where they are situated. For the moment, the phasing network is placed ahead of the amplifiers, but this will be relaxed later. We will assume that the phasing network is lossless and reciprocal, so that $\mathcal{P}^\dagger = \mathcal{P}^{-1}$, which requires that there are as many degrees of freedom in the output of $\mathcal{P}$  as there are in the outputs of the antennas. The combined signals then pass through a set of ideal amplifiers, having power gain $G$ and input noise temperature $T_a$, before being individually down-converted and digitised at a sampling frequency of $f_s$.

In this section, to be consistent with the convention used in electronic system engineering, it will be convenient to work in terms of signals oscillating with positive time dependence, i.e. that vary as $e^{i \omega t}$ rather than $e^{-i \omega t}$. As a result, $\mathcal{S}$ and $\mathcal{K}$ are the conjugates of the results derived in the previous section. The output signals from the antennas are given by  (\ref{eqn:antenna_scattering_network}), or equivalently
(\ref{eqn:embed_horn_exp}) if embedding is explicitly included.

The time-dependent signal vector appearing at the output of the whole signal chain is given by
\begin{equation}
\label{eqn:signal_chain}
\mathbf{o} (t) = \tfrac{1}{2} \sqrt{k_0 L Z_0 G} \, \mathcal{P} \, \mathcal{K} \, \mathbf{i} \exp^{i 2 \pi f t} + \mathbf{n}_{\rm ext} (t) + \mathbf{n}_{\rm env} (t) + \mathbf{n}_{\rm amp} (t).
\end{equation}
Non-equal gains could be incorporated easily by the inclusion of a gain matrix.

$\mathbf{n}_{\rm ext} (t)$ is a stochastic random vector representing the noise that enters the CRES region through the ports of the antennas; such as noise emanating from the inputs of the amplifiers, or from the loads of any circulators used ahead of the amplifiers. $\mathbf{n}_{\rm env} (t)$ is a stochastic random vector representing the noise produced by  ohmic losses in the antennas and the coupling of stray electromagnetic fields to the thermal environment. $\mathbf{n}_{\rm amp} (t)$ describes the noise produced by the amplifiers.  Because the noise contributions are uncorrelated, the overall covariance matrix of the noise at the output of the instrument can be written
\begin{equation}
\label{eqn:cov_over}
\Sigma = \Sigma_{\rm ext} + \Sigma_{\rm env} + \Sigma_{\rm amp}.
\end{equation}

Assuming stationary statistics, the covariance matrix of the amplifier noise is
\begin{equation}
\label{eqn:cov_amp}
\Sigma_{\rm amp} = 2 G  k_b T_a \Delta \nu \, \mathcal{I},
\end{equation}
where all of the amplifiers are identical, and the factor of 2 arises because we are using a normalisation where   $\langle | \mathbf{n}_{\rm amp} |^{2} \rangle / 2$ is the power. For a sufficiently sensitive amplifier, the noise temperate approaches the quantum noise temperature $\hbar \omega / 2 k$.

The covariance matrix of the noise associated with the external sources is more subtle. In signal flow graph theory the noise produced by each amplifier needs to be represented by two noise sources: one of  which describes a noise wave that is effectively incident on the amplifier and gives rise to the noise temperature; the other is a noise wave that travels away from the input of the amplifier \cite{Withington_2023}. These two noise sources are generally correlated, but it can be shown that they are uncorrelated if the amplifier is matched internally for miniumum noise---if this were not the case, a matching network could be placed ahead of the amplifier to destructively interfere the two noise waves, achieving a lower noise temperature. Equally, they are uncorrelated if the amplifier includes an input circulator, because the noise travelling away from the input originates in the load of the circulator. 

For simplicity, we shall assume that a circulator is used, and so the noise waves travelling away from the inputs are uncorrelated and thermal in original as determined by the physical temperature of the loads. In a CRES instrument, noise leaving the input of one amplifier passes through the CRES region and appears at the input of the other amplifiers, thereby potentially increasing system noise temperature. Including a cooled circulator has the benefit that this additional noise can be kept to a minimum. The covariance of this noise at the output is as follows  
\begin{equation}
\label{eqn:cov_ext}
%\Sigma_{\rm ext} =  2 G k_b T_b \Delta \nu \, \mathcal{P} \mathcal{S} \mathcal{S}^{\dagger} \mathcal{P}^{\dagger}.
\Sigma_{\rm ext} =  2 G  \left( \frac{\hbar \omega}{2} \right) \coth\left( \frac{ \hbar \omega}{2 k_b T_b} \right)   \Delta \nu \, \mathcal{P} \mathcal{S} \mathcal{S}^{\dagger} \mathcal{P}^{\dagger}.
\end{equation}
where $T_b$ the the radiometric temperature of the noise travelling away from the input of each amplifier. We have also assumed that $\mathcal{P}$ is unitary. Ordinarily if $T_b$ is very low this term can be ignored, but when quantum noise-limited amplifiers are used, this contribution can be significant. 

The benefit of knowing the scattering matrix of the system is that Bosma's theorem \cite{Bosma} can be used to calculate the covariance matrix of the noise associated with the the environment,
\begin{equation}
\label{eqn:cov_env}
%\Sigma_{\rm env} =  2 G k_b T_p \Delta \nu \, \mathcal{P} \left[ \mathcal{I} - \mathcal{S} \mathcal{S}^{\dagger} \right] \mathcal{P}^{\dagger},
\Sigma_{\rm env} =  2 G   \left( \frac{\hbar \omega}{2} \right) \coth\left( \frac{ \hbar \omega}{2 k_b T_p} \right)     \Delta \nu \, \mathcal{P} \left[ \mathcal{I} - \mathcal{S} \mathcal{S}^{\dagger} \right] \mathcal{P}^{\dagger},
\end{equation}
where $T_p$ is the radiometric temperature of the environment. By `environment' we mean the noise that enters the antennas because of leakage of stray thermal radiation into ${\cal C}_2$. Notice that if  $\mathcal{S}$ is unitary, there is no environmental noise, as discussed previously; indeed this is the principle reason for ensuring that the antenna patterns couple efficienctly to each other. 

Here the vacuum fluctuations in the thermal background field have been included to ensure that the quantum limit is handled correctly at low temperatures, where CRES systems will operate. If the transmission patterns of the antennas contain ohmic loss, the temperature of the loss has to be equal to that of the environment for (\ref{eqn:cov_env}) to hold. If they are not equal, the noise from the antennas can be separated out, but remember that these losses both lead
to additional noise at the ports of the antennas and radiate noise into the CRES region. In any case, these are second-order considerations because the losses in the antennas are small.

Notice that if the loads of the circulators are at the same temperature as the the environment $T_b = T_p$,
\begin{equation}
\label{eqn:cov_uniform}
%\Sigma_{\rm ext} + \Sigma_{\rm env} =  2 G  k_b T_p}     \Delta \nu  \, \mathcal{I}, 
\Sigma_{\rm ext} + \Sigma_{\rm env} =  2 G    \left( \frac{\hbar \omega}{2} \right) \coth\left( \frac{ \hbar \omega}{2 k_b T_p} \right)     \Delta \nu  \, \mathcal{I}, 
\end{equation}
the noise at the outputs of the different channels are completely uncorrelated, which is a feature of all matched lossy microwave networks, and the analysis that follows simplifies. This decorrelation is valuable because it enhances the ability of software-defined beam systhesis methods to recover the position of the source. For now, we will not make this assumption.

\section{Probability of data set}
\label{sec:prob_data}

We now have a signal and noise model of the complete CRES system, and so can use Fisher information theory to place bounds on experimental outcomes. Suppose that the output signal is sampled at time points $t_n$ for $n=0, 1, \cdots N-1$, such that  $\mathbf{o}_n = \mathbf{o} (t_n)$. It can be shown easily that for all band-limited signals, where the noise power spectral density is white over the band $\Delta \nu$, the sample values are uncorrelated if the waveform is sampled at the Nyquist rate $f_s = \Delta \nu$: $\langle x(t_m) x(t_n) \rangle = C_0 \Delta \nu \delta_{mn}$. In other words, a measurement at time $t_n$ is statistically independent of any measurement at $t_m \neq t_n$. It follows that the probability of achieving a specific set of data when a series of $N$ samples is taken is given by a product of Gaussian distributions:
\begin{equation}
\label{eqn:probability_o}
	\text{Pr}(\mathbf{o}_0, \mathbf{o}_1, \dots \mathbf{o}_N | f, \mathbf{i})
	= \prod_{n=1}^{N} \frac{1}{\sqrt{||2 \pi \Sigma||}}
		\exp \Bigl( -\tfrac{1}{2} \{ \mathbf{o}_n - \boldsymbol{\mu}_n \}^\dagger
			\cdot \Sigma^{-1} \cdot \{ \mathbf{o}_n - \boldsymbol{\mu}_n \} \Bigr).
\end{equation}
This multi-dimensional distribution is dependent on parameters of the model, and here we indicate frequency $f$ and current $\mathbf{i}$, representing electrons in our 2D model,  explicitly. The mean values come from (\ref{eqn:signal_chain})
\begin{equation}
\label{eqn:mean_yn}
	\boldsymbol{\mu}_n =
		\tfrac{1}{2} \sqrt{k_0 L G Z_0} \, 
			\mathcal{P} \cdot \mathcal{K} \cdot \mathbf{i} \,
		e^{2 \pi i f n / \Delta \nu}
\end{equation}
and the covariance matrix is given by (\ref{eqn:cov_over}). (\ref{eqn:probability_o}) can be used to calculate Fisher information.

In doing so it will be necessary to calculate $\Sigma^{-1}$, and although this could be done numerically, it is beneficial to consider its functional form. $\Sigma_{\rm amp}$ is already diagonal, and can be handled easily. $\Sigma_{\rm env}$ and $\Sigma_{\rm ext}$ are more awkward. Let $\mathcal{S}$ have the singular value decomposition
\begin{equation}
\label{eqn:svd_of_s}
\mathcal{S} = \mathcal{L} \cdot \mathcal{E} \cdot \mathcal{R}^\dagger,
\end{equation}
where by definition $\mathcal{L}^\dagger  \mathcal{L} = \mathcal{L} \mathcal{L}^\dagger  = \mathcal{I}$ and $\mathcal{R}^\dagger  \mathcal{R} = \mathcal{R} \mathcal{R}^\dagger  = \mathcal{I}$. 

(\ref{eqn:cov_ext}) can then be written
\begin{equation}
\label{eqn:cov_ext_1}
\Sigma_{\rm ext} =  2 G  \left( \frac{\hbar \omega}{2} \right) \coth\left( \frac{ \hbar \omega}{2 k_b T_b} \right)  \Delta \nu\, \mathcal{P} \mathcal{L}  \mathcal{E}^2  \mathcal{L}^{\dagger} \mathcal{P}^{\dagger},
\end{equation}
and likewise for (\ref{eqn:cov_env})
\begin{equation}
\label{eqn:cov_env_1}
\Sigma_{\rm env} =  2 G  \left( \frac{\hbar \omega}{2} \right) \coth\left( \frac{ \hbar \omega}{2 k_b T_p} \right)     \Delta \nu \, \mathcal{P} \mathcal{L} \left[ \mathcal{I} - \mathcal{E}^2 \right]  \mathcal{L}^{\dagger}  \mathcal{P}^{\dagger},
\end{equation}
Combining the terms, and taking the classical limit  $\hbar \omega \ll 2 k T$ for brevity, 
\begin{equation}
\label{eqn:cov_inv}
\Sigma^{-1} = \frac{1}{2 G  k_b (T_a + T_p)  \Delta \nu}  \mathcal{P} \mathcal{L} \left[ \mathcal{I} + \frac{(T_b-T_p)}{(T_a+T_p)} \mathcal{E}^2 \right]^{-1}  \mathcal{L}^{\dagger}  \mathcal{P}^{\dagger},
\end{equation}
where the term in square brackets is diagonal and so trivial to invert. When $T_b = T_p$, such that no correlations are present, the inversion is particularly simple, 
\begin{equation}
\label{eqn:cov_inv_simp}
\Sigma^{-1} = \frac{1}{2 G  k_b (T_a + T_p)  \Delta \nu} \mathcal{I}
\end{equation}
and (\ref{eqn:probability_o}) reduces to a straightforward expression; or in the quantum limit
\begin{equation}
\label{eqn:cov_inv_simp2}
\Sigma^{-1} = \frac{1}{2 G  h \nu  \Delta \nu} \mathcal{I}.
\end{equation}

\section{Fisher information analysis} 
\label{sec:fisher}

For a CRES system, we wish to know the uncertainty in a measurement of frequency, actually initial energy, for a given source location, or in our 2D model, current distribution. The Cram\'{e}r-Rao Bound (CRB) implies that the uncertainty $\Delta f$ in any unbiased estimator of $f$ satisfies
\begin{equation}
\label{eqn:cramer_rao_bound}
\Delta \nu^2 \geq \frac{1}{F (f)}
\end{equation}
where $F$ is the Fisher information of the samples with respect to $f$. For what follows, it will be useful to rearrange (\ref{eqn:cramer_rao_bound}) as
\begin{equation}\label{eqn:cramer_rao_bound_v2}
	\frac{f^2}{\Delta f^2} \leq f^2 F (f)
\end{equation}
and use $(f / \Delta f)^2$ as our metric of sensitivity. The larger $(f / \Delta f)^2$ the better the sensitivity.

The Fisher information of multivariate Gaussians is
\begin{equation}\label{eqn:fi_v1}
	F(f) = \sum_{n=0}^{N-1}
		\frac{\partial \boldsymbol{\mu}^\dagger_n}{\partial f}
		\cdot \Sigma^{-1} \cdot
		\frac{\partial \boldsymbol{\mu}^{}_n}{\partial f}.
\end{equation}

Using (\ref{eqn:mean_yn}), 
\begin{equation}
\label{eqn:deriv}
	\frac{\partial \boldsymbol{\mu}^{}_n}{\partial f}
		 = \frac{\pi i n \sqrt{k_0 L G Z_0}}{\Delta \nu} \, \, \mathcal{P}
		 \cdot \mathcal{K} \cdot \mathbf{i}
		 \, e^{2 \pi i f n / \Delta \nu},
\end{equation}
and $F(f)$ evaluates to
\begin{equation}\label{eqn:fi_v2}
	F(f) = \frac{\kappa_a \pi^2 k_0 L Z_0 |\mathbf{i}|^2 \text{Tr}[\mathcal{H}]}
		{2 k_b \left(T_a + T_p \right) {\Delta \nu}^3}
		\sum_{n=0}^{N-1} n^2.
\end{equation}
where
\begin{equation}
	\kappa_a = \frac{
		\mathbf{i}^\dagger \cdot \mathcal{K}^\dagger \cdot \mathcal{L}
		\cdot \mathcal{H} \cdot \mathcal{L}^\dagger \cdot \mathcal{K} \cdot \mathbf{i}
	}{
		|\mathbf{i}|^2 \text{Tr}[\mathcal{H}]
	}
\end{equation}
and
\begin{equation}
\label{eqn:h_factor}
\mathcal{H} = \Bigl[  \mathcal{I} + \frac{(T_b-T_p)}{(T_a+T_p)} \mathcal{E}^2 \Bigr]^{-1}.
\end{equation}
For notational brevity, we have resorted to the classical limit for thermal noise power. In the limit of large $N$,
\begin{equation}
	\sum_{n=0}^{N-1} n^2 = \frac{(2 N - 1) (N - 1) N}{6} \rightarrow \frac{N^3}{3}.
\end{equation}

Defining $\tau = N / \Delta \nu$ as the total sampling time, we arrive at
\begin{equation}
\label{eqn:fi_v3}
	F(f) \sim \frac{ \kappa_a \pi^2 k_0 L \tau^3 Z_0 |\mathbf{i}|^2 \text{Tr}[\mathcal{H}]}
		{6 k_b \{T_a + T_p \}}.
\end{equation}

Substituting (\ref{eqn:fi_v3}) into (\ref{eqn:cramer_rao_bound_v2}) and rearranging, we find the best achievable reciprocal fractional frequency sensitivity is
\begin{equation}
\label{eqn: ccrb}
   \frac{f^2}{\Delta f^2} \sim
	\frac{4}{3} \pi^2 
	\times (f \tau)^2
	\times (\tau \Delta \nu)
	\times \kappa_a
	\times \frac{k_0 L Z_0 |\mathbf{i}|^2 \text{Tr}[\mathcal{H}]}
		{8 k_b \left( T_a + T_p \right) N_a \Delta \nu}
	\times N_a.
\end{equation}
where $N_a$ is the number of antennas. Clearly, an optimum experiment is one for which this ratio is maximised.

The right-hand side has been split into factors with the following significance: (i)  $(f \tau)$ is the number of cycles of oscillation captured. (ii) $\tau \Delta \nu$ is the radiometer factor; the number of statistically independent samples in the observation time. (iii) $\kappa_a$ characterises the effect of the antenna system for the given current distribution $\mathbf{i}$. Ordinarily only a single line current is present having unit position vector  $\uvec{i}_n$, and then
\begin{equation}
\label{eqn:antenna_factor_map}
	\kappa_a (\uvec{i}_n) = \frac{
		\uvec{i}_n^\dagger \cdot \mathcal{K}^\dagger
		 \cdot \mathcal{L} \cdot
		\mathcal{H} 
		\cdot \mathcal{L}^\dagger \cdot \mathcal{K}
		\cdot \uvec{i}_n^{\vphantom{\dagger}}
	}{
		\text{Tr}[\mathcal{H}]
	}
\end{equation}
By considering $\kappa_a (\uvec{i}_n)$ for different source positions it is possible to map out the spatial frequency sensitivity for line sources at different locations, as it is the only term that depends on source position. The antenna factor $\kappa_a (\uvec{i}_n)$ does not result from a straightforward combination of antenna beams, but instead includes a noise weighting through $\mathcal{H}$. If $T_b = T_p$ such that the thermal noise emerging from the ports is uncorrelated, 
\begin{eqnarray}
\label{eqn:antenna_factor_map_uncorr}
	\kappa_a (\uvec{i}_n) & = \frac{1}{N_a}
		\uvec{i}_n^\dagger \cdot \mathcal{K}^\dagger \cdot \mathcal{K}
		\cdot \uvec{i}_n^{\vphantom{\dagger}} \\ \nonumber
	 & = \frac{1}{N_a} \text{diag}_n \left[\mathcal{K}^\dagger \cdot \mathcal{K}\right].
\end{eqnarray}
where $\text{diag}_n$ indicates the $n$'th diagonal element. There is a penalty in this term for dividing a given collected power across a large number of antennas. (iv) The final factor is the effective power SNR per antenna, which is then multiplied by the number of antennas. If  $T_b = T_p$ this factor becomes 
\begin{equation}
\left( \frac{k_0 L Z_0 |\mathbf{i}|^2 }{8} \right) \left( \frac{1}{ k_b \left( T_a + T_p \right) \Delta \nu}\right) N_a,
\end{equation}
which is the total power radiated by the line current divided by the noise power of a single channel, multiplied by the number of antennas. The $N_a$ in (iii) and (iv) cancel to make the result independent of the number of antennas; which is beneficial because a large number of antennas is needed to get a large FoV, as will be seen.  This analysis demonstrates the importance of avoiding correlations in the noise if at all possible, because it is correlations in the signal, which are not in the noise, that lead to benefit. As expected, the frequency sensitivity increases with the intrinsic SNR, the total observation time and the antenna coupling factor.

As a final consideration, suppose that we choose to put the phasing network after the amplifiers. If the amplifiers have identical gains, and if the phasing network is lossless, it is straightforward to see that the covariances (\ref{eqn:cov_amp}), (\ref{eqn:cov_ext}) and (\ref{eqn:cov_env}) remain unchanged. Thus as long as the a complete set of vectors is formed from all of the channels available, it is possible to place the phasing network after the amplifiers, or equivalently to synthesise a complete set of beams in software, which is straightforwardly done. Basically, every position in the FoV can then be observed simultaneously because each phased output can correspond to a localised region within the FoV.

\section{Field of view}
\label{sec:fov}

\subsection{FoV from a field perspective}
\label{sec:fov_field}

An important question is how many electromagnetic degrees of freedom connect the FoV, $\mathcal{C}_1$, to the surface where the antennas are located, $\mathcal{C}_2$, as this determines the number of antennas needed to collect all of the information available: see for example \cite{Miller_2000}. In this section we will assume that only propagating modes are of interest, even though evanescent modes are include in the full model of \S \ref{sec:full_elec_model}. In this sense,  the number of degrees of freedom derived in this section are lower bounds.

\begin{figure}
%file:///C:/Users/Stafford Withington/Documents/Projects starting in Oxford March 2023/QTNM/CRES Imaging Paper Final Figure Code/CRES_Gaussian_beam_stratton_test_emb.m
\noindent \centering
\includegraphics[trim = 1cm 1cm 10cm 16cm, width=70mm]{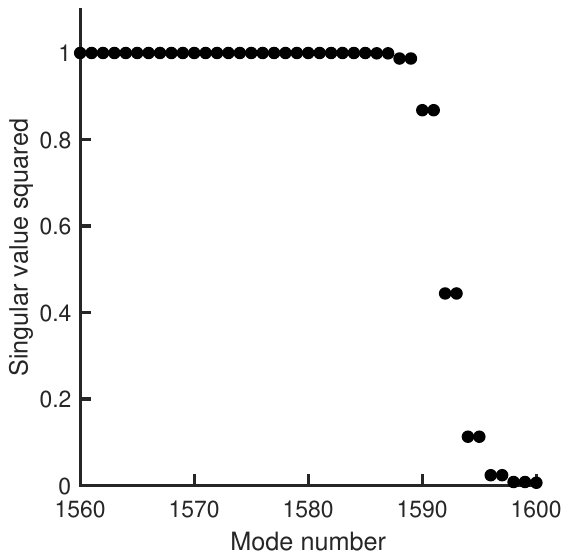}
% left,bottom,right,top
\caption{Last few singular values of the embedding scattering matrix  $\mathcal{S}_{e}$. The wavelength is 10mm, and the radius of the reference circle is 4 mm. 1600 field sample points were used. }
\label{semb_fig1}
\end{figure}

The question can be answered in a variety of ways. First we note that, with a suitable choice of $R$,  (\ref{eqn:s_infinity}) is the scattering matrix looking out from the reference surface $\mathcal{C}_1$  towards infinity were the radiation condition applies, and therefore the singular values of $\mathcal{S}_{e}$ must be unity for evansecent modes, and less than unity for modes that lose energy in the form of propagating waves. Figure \ref{semb_fig1} shows the singular values of the last few modes on a reference circle having $R_1 =$ 10 mm at a wavelength of $\lambda =$ 10 mm. 1600 sample points were used on $R_1$ to ensure dense sampling.  As can be seen, most of the modes on $R_1$ do not propagate, but around 13 radiative modes are available.

\begin{figure}
%file:///C:/Users/Stafford Withington/Documents/Projects starting in Oxford March 2023/QTNM/PSWF for CRES/CRES_Gaussian/CRES_Gaussian_beam_stratton_test.m
\noindent \centering
\includegraphics[trim = 2cm 1cm 7cm 16cm, width=70mm]{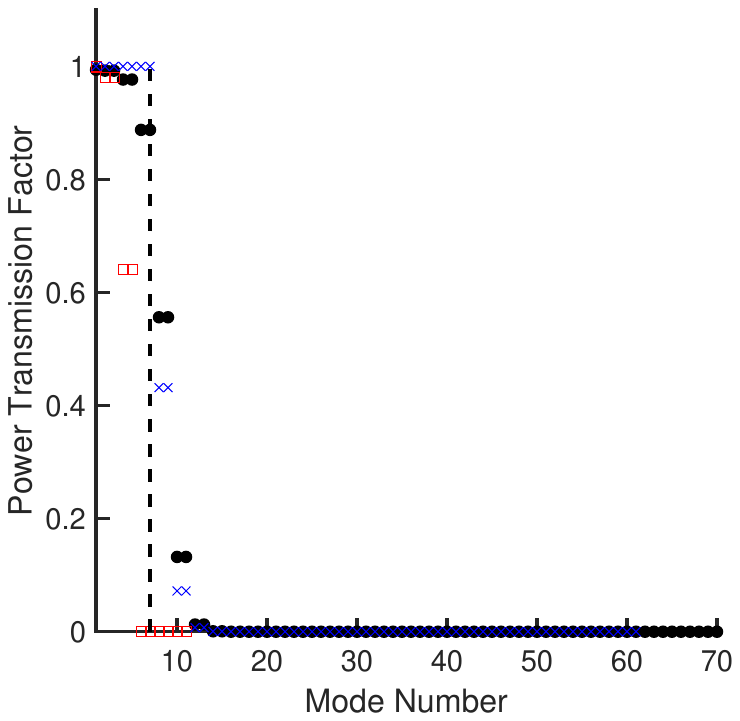}
% left,bottom,right,top
\caption{First few singular values of the embedding scattering matrix loss matrix $\mathcal{I} - \mathcal{S}_{e} \mathcal{S}_{e}^\dagger$ (black circles). The wavelength is 10mm, and the radius of the reference circle is 5 mm. The impedance method (red squares), and the propagation method (blue crosses) described in the text. These bound the actual solution. The vertical dashed line indicates the threshold $2 k_0 R_1$.}
\label{trans_fig1}
\end{figure}

\begin{figure}
%file:///C:/Users/Stafford Withington/Documents/Projects starting in Oxford March 2023/QTNM/PSWF for CRES/CRES_Gaussian/CRES_Gaussian_beam_stratton_test.m
\noindent \centering
\includegraphics[trim = 2cm 1cm 7cm 16cm, width=70mm]{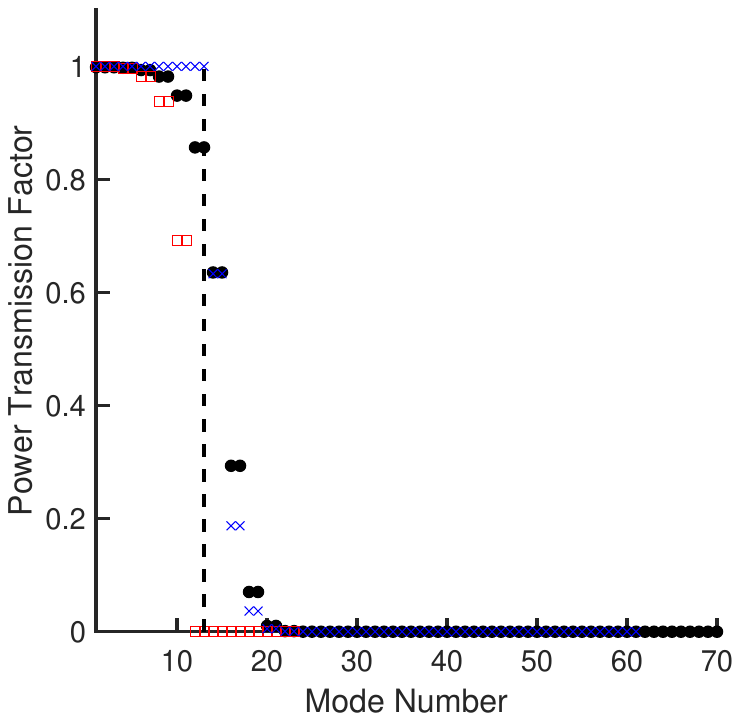}
% left,bottom,right,top
\caption{First few singular values of the embedding scattering matrix  $\mathcal{I} - \mathcal{S}_{e} \mathcal{S}_{e}^\dagger$ (black circles). The wavelength is 10mm, and the radius of the reference circle is 10 mm.   The impedance method (red squares), and the propagation method (blue crosses) described in the text. The vertical dashed line indicates the threshold $2 k_0 R_1$.}
\label{trans_fig2}
\end{figure}

\begin{figure}
%file:///C:/Users/Stafford Withington/Documents/Projects starting in Oxford March 2023/QTNM/PSWF for CRES/CRES_Gaussian/CRES_Gaussian_beam_stratton_test.m
\noindent \centering
\includegraphics[trim = 2cm 1cm 7cm 16cm, width=70mm]{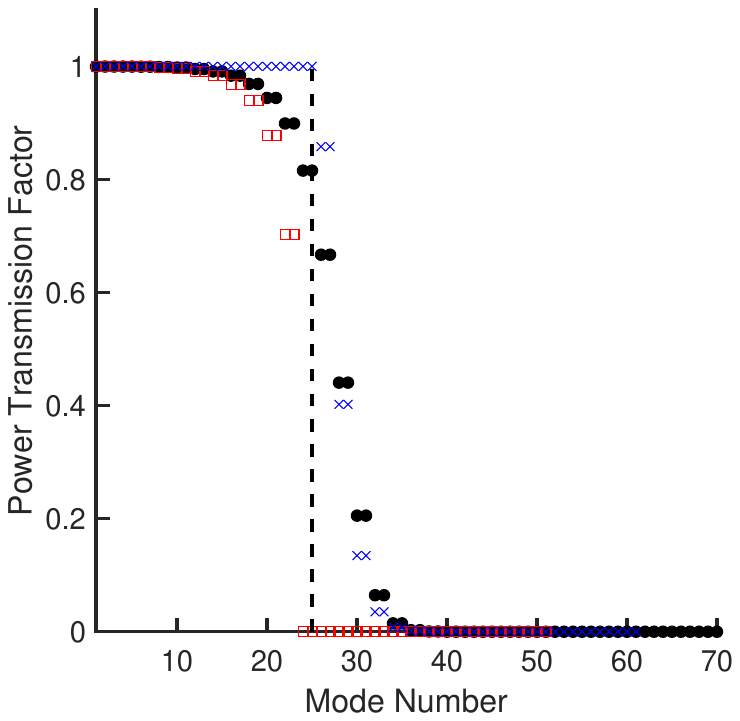}
% left,bottom,right,top
\caption{First few singular values of the embedding scattering matrix loss matrix $\mathcal{I} - \mathcal{S}_{e} \mathcal{S}_{e}^\dagger$ (black circles). The wavelength is 10mm, and the radius of the reference circle is 20 mm.   The impedance method (red squares), and the propagation method (blue crosses) described in the text. The vertical dashed line indicates the threshold $2 k_0 R_1$.}
\label{trans_fig3}
\end{figure}

\begin{figure}
%file:///C:/Users/Stafford Withington/Documents/Projects starting in Oxford March 2023/QTNM/PSWF for CRES/CRES_Gaussian/CRES_Gaussian_beam_stratton_test.m
\noindent \centering
\includegraphics[trim = 2cm 1cm 7cm 16cm, width=70mm]{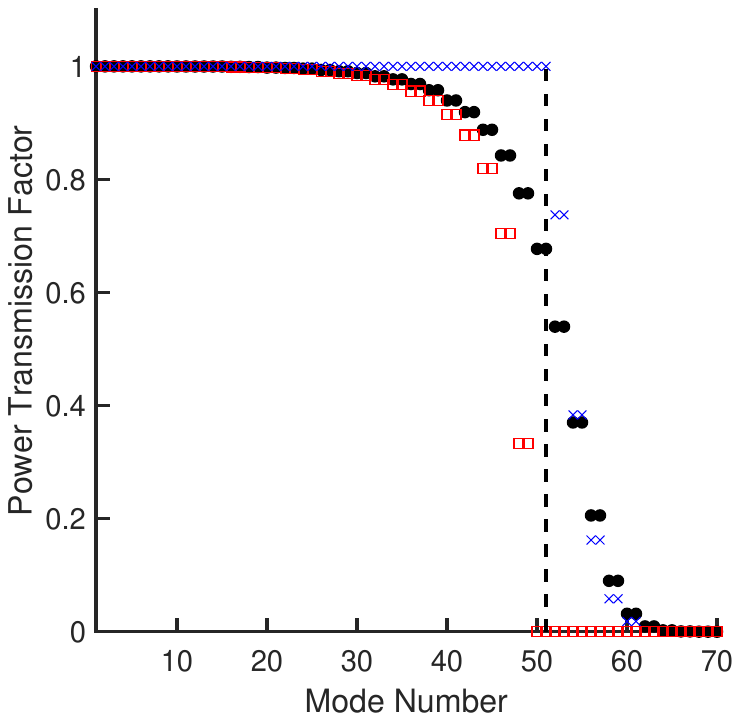}
% left,bottom,right,top
\caption{First few singular values of the embedding scattering matrix loss matrix $\mathcal{I} - \mathcal{S}_{e} \mathcal{S}_{e}^\dagger$ (black circles). The wavelength is 10mm, and the radius of the reference circle is 40 mm.   The impedance method (red squares), and the propagation method (blue crosses) described in the text. The vertical dashed line indicates the threshold $2 k_0 R_1$.
}
\label{trans_fig4}
\end{figure}

Another approach is to plot the singular values of the operator $\mathcal{I} - \mathcal{S}_{e}^\dagger \mathcal{S}_{e}$, which corresponds to scattering an outgoing wave off of the reference surface, performing a time-reversed propagation of the same process, and then counting the number of modes in which power goes missing. The singular values are the same as those of the operator  $\mathcal{I} - \mathcal{S}_{e}\mathcal{S}_{e}^\dagger $, which describes the amount of power that would enter the FoV from a uniform blackbody radiation field. The black circles in Fig.'s \ref{trans_fig1} to  \ref{trans_fig4} show the singular values for $R_1$ = 5 mm,  $R_1$ = 10 mm,  $R_1$ = 20 mm,  $R_1$ = 40 mm respectively. In this case the lowest-order modes correspond to the radiated modes, but otherwise the information is the same.

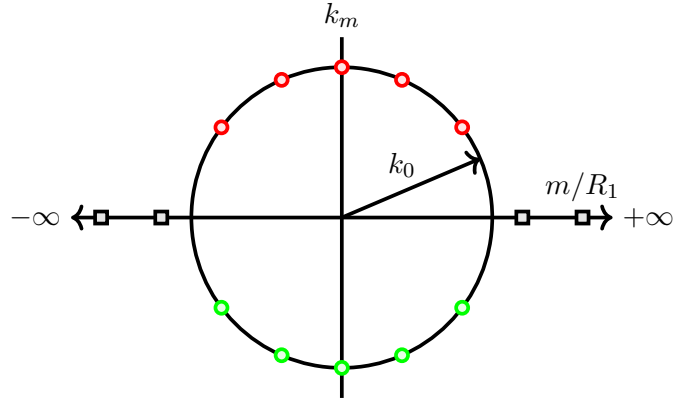
\begin{figure}
\centering
\begin{tikzpicture}[scale=0.8]
%\draw[help lines] (0,0) grid (8,6);
\draw [black,line width=1.4] (4,3) circle (2.5);
\draw [line width=1.4, <->] (-0.5,3) -- (8.5,3);
\draw [line width=1.4] (4,0) -- (4,6);

\draw [line width=1.4,->] (4,3) -- (6.3,3.98);

\draw [red,line width=1.4,fill=red!10] (4,5.5) circle (0.1);
\draw [red,line width=1.4,fill=red!10] (5,5.29) circle (0.1);
\draw [red,line width=1.4,fill=red!10] (6,4.5) circle (0.1);

\draw [red,line width=1.4,fill=red!10] (3,5.29) circle (0.1);
\draw [red,line width=1.4,fill=red!10] (2,4.5) circle (0.1);

\draw [green,line width=1.4,fill=green!10] (4,0.5) circle (0.1);
\draw [green,line width=1.4,fill=green!10] (5,0.71) circle (0.1);
\draw [green,line width=1.4,fill=green!10] (6,1.5) circle (0.1);

\draw [green,line width=1.4,fill=green!10] (3,0.71) circle (0.1);
\draw [green,line width=1.4,fill=green!10] (2,1.5) circle (0.1);

\draw [black,line width=1.4,fill=black!10] (-0.1,2.9) rectangle (0.1,3.1);
\draw [black,line width=1.4,fill=black!10] (0.9,2.9) rectangle (1.1,3.1);
\draw [black,line width=1.4,fill=black!10] (6.9,2.9) rectangle (7.1,3.1);
\draw [black,line width=1.4,fill=black!10] (7.9,2.9) rectangle (8.1,3.1);

\node at (5,3.5) [anchor=south]{$k_0$};
\node at (8,3.1) [anchor=south]{$m /R_1$};
\node at (4,6) [anchor=south]{$k_m$};
\node at (-0.5,3) [anchor=east]{$-\infty$};
\node at (8.5,3) [anchor=west]{$+\infty$};

\end{tikzpicture}
\caption{\label{fig:mode_indexing} Mode indexing. Red circles correspond to modes travelling radially outwards, and the green circles to modes travelling radially inwards. Positive and negative values of $m$ correspond to different azimuthal orders. Black squares indicate evanescant modes of which there is a countably infinite number.}
\end{figure}

In order to know how many antennas are needed to cover a given FoV, it is convenient to have a simple expression for the number of modes that can radiate away from $ {\cal C}_{1}$. It would also be helpful to understand why the spectrum cuts off gradually over some range of mode number. By solving Maxwell's equations in a small differential region in the neighbourhood of $ {\cal C}_{1}$, using azimuthal field solutions in the form of a Fourier series 
\begin{equation}
\label{eqn:fourier}
f(\phi) = \sum_{m = -\infty}^{m = +\infty} f_m e^{i m \phi},
\end{equation}
where the indexing $ -\infty < m < + \infty$ corresponds to counter propagating azimuthal modes on $ {\cal C}_{1}$, it can be shown that that radial wave equation has the localised dispersion relationship
\begin{equation}
k_{m}^{\pm} = \pm k_0 \left[ 1 - \left( \frac{m}{k_{0} R_1} \right)^2 \right]^{1/2},
\end{equation}
which describes a circle for $(m/R_1)^2 < k_0^2$, as illustrated in Fig. \ref{fig:mode_indexing}, where a total of 5 radiative modes are present. It has been assumed that  $k_0 R_1 \gg 1$. The two solutions $(\pm)$, red and green, correspond to outward and inward travelling wave amplitudes $a$ and $b$ respectively. 

The characteristic impedance of each localised mode is given by 
\begin{equation}
Z_m = Z_0  \left[ 1 - \left( \frac{m}{k_{0} R_1} \right)^2 \right]^{-1/2}.
\end{equation}
It is clear that for $m < k_{0} R_1$ the modes have real radial propagation constants and impedances and so propagate, whereas for $m > k_{0} R_1$ the modes have complex propagation constants and impedances and so are cut off. We shall designate the threshold as $M$.

Using the modal impedances it is trivial to calculate reflection coefficients with respect to $Z_0$, and these are shown as  red squares in Figures \ref{trans_fig1} to  \ref{trans_fig4}. This calculation defines an extreme because the differentially small region is effectively only one section of a longer radial transmission structure where the propagation constant and impedance change with distance. In fact, the impedance of every mode that can escape eventually becomes $Z_0$.  This  calculation treats the next differentially small element as a lumped terminating impedance, but nevertheless the calculation is a useful indicator.

Another extreme is given by ignoring the changing impedance and calculating the overall power transmission factor of each mode between  $R_{1}$ and $R_{2} \rightarrow \infty$, taking into account the changing propagation constant:
\begin{equation}
T_m = \lim_{R_2 \rightarrow \infty}   \left| \exp \left[ - i k_0 \int_{R_1}^{R_2} \left[ 1 - \left( \frac{m}{k_{0} \rho} \right)^2 \right]^{1/2} \, d \rho \right] \right|^2.
\end{equation}
This process allows modes that are evanescent over some region to eventually break free once the propagation constant becomes real. Although the integral strictly depends on the outer radius $R_2$, its value can be made large enough so that convergence is reached. These solutions are shown as blue crosses in  Figures \ref{trans_fig1} to  \ref{trans_fig4}, and represent another extreme because wave impedance is not taken into account. Indeed the red squares and blue crosses bound the modal scattering factors calculated through the exact Stratton-Chu-like method. Initially, the mismatch in impedance between differential elements causes the transmission factor to fall, but towards the end of the transition, the mismatch in wavelength dominates.

We conclude that $M = 2 k_{0} R_1 +1$ is a convenient measure of the number of degrees of freedom that can propagate from $ {\cal C}_{1}$ into the far field, and $M = 2 k_{0} R_1$ is shown as a vertical dashed lines in  Fig.'s \ref{trans_fig1} to  \ref{trans_fig4}. Cut off occurs when there is only one modal wavelength around the periphery of $ {\cal C}_{1}$. According to  the vertical dashed lines, $M$ corresponds to an actual power transmission factor of about $0.6$, which tends to 0.5 as $R_1$ increases. It seems that the impedance and propagation methods modify the spectrum above and below the pivot point $0.5$. $M = 2 k_{0} R_1 +1$ will be adopted as a suitable threshold for 2D sampling. Notice that for a FoV having a radius of only one wavelength, 13 degrees of freedom are available, which indicates that a large number of antennas will be needed to cover even a small FoV with high efficiency. Fortunately, the number of antennas scales with the radius of $ {\cal C}_{1}$ rather than its area: consistent with the Stratton-Chu picture, which indicates that the fields in a volume are determined by the values of the fields over the volume's surface.
  
 \subsection{FoV from a current perspective}
\label{sec:fov_curr}
 
The above analysis of the number of degrees of freedom available for observation is based on a field propagation method, but in the 2D model the sources are line currents, and these are our primary target. To understand the number of degrees of freedom at $R_2$ that are driven by source currents, place a densely packed system of line currents around the periphery of  $ {\cal C}_{1}$. In what follows, we used typically 40 line currents per wavelength to ensure convergance. It should be emphasised that these currents are not intended to model the radiation pattern of an electron, but merely to represent, in the spirit of the Equivalence Theorem, the degrees of freedom of any source that may exist within $ {\cal C}_{1}$. 

The field at ${\bf r}_{2}$ on $ {\cal C}_{2}$ as a consequence of the discrete  line currents  $i_{n}$ at ${\bf r}_{1}^{n}$ on $ {\cal C}_{1}$ is 
\begin{eqnarray}
\label{eqn:mapping}
E({\bf r}_{2}) & = \frac{1}{4} \sum_{n=1}^{N} Z_{0} k_{0}   H_{0} (k| {\bf r}_{2} - {\bf r}_{1}^{n} |) i_{n},
\end{eqnarray}
which maps a discrete vector space onto a continuous vector space. Because the source and observation points are distinct,  $|{\bf r}_{1}^{n}| < R_{2}$  $\forall n$,  there is no need to deal with the singularity at ${\bf r}_{2} = {\bf r}_{1}^{n}$. To find the degrees of freedom in the propagating field it is necessary to find the Hilbert-Schmidt Decomposition, which is assured because the currents are square-summable over  $ {\cal C}_{1}$. It is convenient if the singular values are normalised, which is best done with respect to power.  

If $\mathbf{e} = \mathcal{G} \mathbf{i}$ is the discretised form of (\ref{eqn:mapping}), the degrees of freedom available for transferring power can be found through the SVD of $\mathcal{G} = \mathcal{U} {\mathsf \Sigma} \mathcal{V}^{\dagger}$. ${\mathsf V}$ is an  $N \times N$ matrix, the columns of which give a set of orthormal vectors that span all possible currents on  $ {\cal C}_{1}$.  $\mathcal{U}$ is an  $M \times M$ matrix, the columns of which give a set of orthormal vectors that span all possible field distributions on  $ {\cal C}_{2}$. ${\mathsf \Sigma}$ are the singular values that connect current distributions to radiated fields in one-to-one correspondence. If $M > N$, the currents can generate only $N$ orthogonal field distributions in the $M$-dimensional space indicating that  only a limited number of channels exist for transfering energy and information. The actual number of degrees of freedom is less than $N$, due to the filtering effects of propagation.

For the cylindrical geometry shown, each column of  $\mathcal{G}$ is a cyclically shifted copy of its neighbour. $\mathcal{G}$ is therefore a circulant matrix, which by construction imports several key characteristics. Most importantly, square circulant matrices are diagonalised by a discrete Fourier transform, and non-square circulant matrices by right and left Fourier transforms. The columns of  $\mathcal{U}$ and  $\mathcal{V}$ must therefore be discretised Fourier modes. The $j$'th basis vector on $ {\cal C}_{1}$ for $N$ sample points takes the form
\begin{equation}
\label{eqn:sin_cos}
{\mathsf v}_{j} = \frac{1}{\sqrt{N}} \left[1,C^{j}, C^{2j}, \cdots, C^{(N-1)j}   \right]^{T} \hspace{2mm} j \in \left\{0,1,2,\cdots,(N-1)\right\}
\end{equation}
where
\begin{equation}
C = \exp^{2 \pi i / N}.
\end{equation}
are the complex roots of unity. Notice that (\ref{eqn:sin_cos}) is the same as (\ref{eqn:fourier}) but with a different indexing corresponding to sine and cosine in $\phi$ rather than complex exponentials. A similar expression exists for $ {\cal C}_{2}$. This representation also makes intuitive sense because the solutions must be periodic on  $ {\cal C}_{1}$  and $ {\cal C}_{2}$, and so representable as sine and cosine series. This representation describes the way in which multipole moments of the current distribution generate multipole moments of the field distribution. A general current source within $ {\cal C}_{1}$  is being described in terms of mulitpole moments, which transfer power with some efficiency, described by the singular values, to $ {\cal C}_{2}$. The singular values provide a measure of how many field distributions must be captured over $ {\cal C}_{2}$ to collect all of the power that is available for any source distribution within $ {\cal C}_{1}$.

\begin{figure}
\noindent \centering
\includegraphics[trim = 1cm 1cm 8cm 14cm, width=90mm]{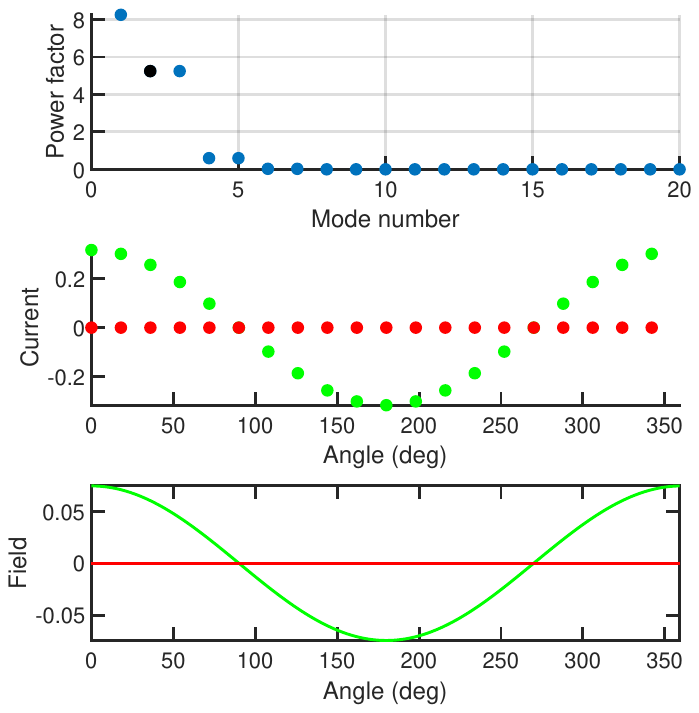}
% left,bottom,right,top
\caption{Typical simulation having a wavelength of 10 mm. 20 line source currents uniformly arranged around an inner circle having a radius of 2 mm. 360 field observation points uniformly placed around the outer circle having a radius of 100 mm. Top shows the power spectrum (singular value squared). Middle shows the real (green) and imaginary (red) parts of the second singular current vector. Bottom shows the real (green) and imaginary (red) parts of the second singular field vector, where an overall propagation phase has been removed. 
}
\label{curr_fig1}
\end{figure}

Figure \ref{curr_fig1} shows a typical simulation with $\lambda$ = 10 mm, $R_{1}$ = 2 mm, $R_{2}$ = 100 mm, M = 360 and N=20.  The top plot shows the singular values squared, $\Sigma_{i}^{2}$, each of which is the total power over  $ {\cal C}_{2}$  generated by an orthonormal current vector over $ {\cal C}_{1}$, compared with the power that would be generated by a unit line current at the origin. Only 5 degrees of freedom show significant coupling to the outer region: the monopole, two spatially orthogonal dipoles, and two spatially orthogonal quadrupoles. The next terms, hexapoles etc., are consistent with the Fourier diagonalisation of the circulant Greens matrix. The middle and bottom plots show the real (green) and imaginary (red) parts of the coupled currents and fields respectively. Care is needed when interpeting plots of this kind as degeneracies mean that the current and field vectors may be rotated within their subspaces to give apparently different basis vectors. This has no physical consequences, and we have found it convenient, when making plots, to introduce a small defect, such as scaling one of the line currents by 0.9, to pin the modes to a specific orientation.  In all of the plots that follow, a constant transmission phase from the inner to outer circles has been removed to make the plots easier to interpret. The current vector associated with the monopole is the in-phase sum of all of the line currents, which effectively gives a single current of $N I$ flowing on a cylinder, producing a cylindrical wave. The second mode, as shown in the middle plot of Figure \ref{curr_fig1}, comprises two sheet currents flowing in opposite directions, giving rise to a longitudinal coil that generates a dipole field in the correct orientation. As $R_{1} \rightarrow 0$, only the  monopole survives, and $\Sigma^{2} \rightarrow N$, corresponding to a total line current of $N$ at the centre. Overall, for this cylindrical case, the source currents are dipole moments, quadrupole moments etc, per unit length. Although not indicated here, as constant current sources are assumed, their inability to radiate power is intimately related to the radiation resistance of the different configurations. Figure \ref{curr_fig1} shows that when the line currents are within a wavelength of each other, only a limited number of degrees of freedom are available for transferring power. It should also be mentioned, that the transmission spectrum changes very little as $R_{2}$ is increased keeping $R_{1}$constant, indicating that the circumference of  $ {\cal C}_{1}$ is the limiting factor.

\begin{figure}
\noindent \centering
\includegraphics[trim = 1cm 1cm 8cm 14cm, width=90mm]{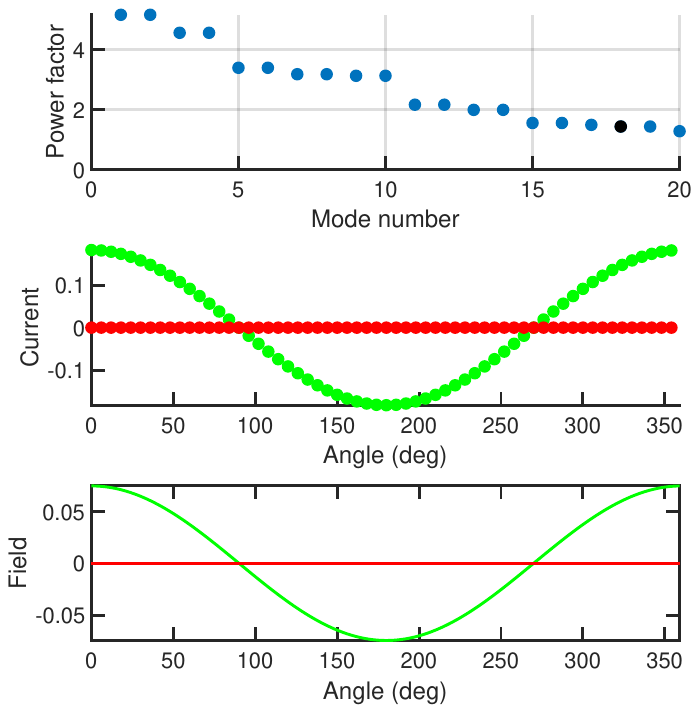}
% left,bottom,right,top
\caption{Typical simulation having a wavelength of 10 mm. 60 line source currents uniformly arranged around an inner circle having a radius of 20 mm. 360 field observation points uniformly placed around the outer circle having a radius of 100 mm. Top shows the power spectrum (singular value squared). Middle shows the real (green) and imaginary (red) parts of the 18th order singular current vector. The bottom plot shows the real (green) and imaginary (red) parts of the 18th order singular field vector. 
}
\label{curr_fig2}
\end{figure}

Figure \ref{curr_fig2} repeats the previous calculation, but now for an inner circle having $R_{1}=$ 20 mm. Notice that the monopole term is actually now mode 17, and the dipole terms modes 18 and 19; suggesting that a cylindrical uniform current is not the same as a line current at the centre. Figure \ref{curr_fig2} should be interpreted in the same way as Fig.~\ref{curr_fig1}, but now many more degrees of freedom are available due to the larger source region. If 4 horn antennas are used on  $ {\cal C}_{2}$, then we would expect to be able to collect the monopole and dipole terms, but wash out the fast Fourier components associated with the higher order multipoles. Another way of thinking about this situation is to imagine a line current at the centre. A small number of antennas can couple well to the source, but if the line current is shifted off centre, it produces a Fourier modulation around the outer circle, which is washed out by the horn's finite aperture size. This is very different to simply collecting power, because here we must use coherent single-mode antennas to achieve high-resolution spectroscopy, and so smoothing out the higher spatial harmonics is intrinsic to the use of coherent receivers. Fundamentally, there is a trade-off between coupling efficiency and FoV unless we add in more antennas to gain access to the high spatial frequencies associated with offset sources. Additionally, if one wants to get the second temporal harmonic of the synchrotron beam, this same reasoning must also be applied at 56 GHz, etc. At this higher frequency, a given FoV, as determined by the fundamental, will couple to many more degrees of freedom and the antenna problem becomes even more severe, actually by a factor of 2. This suggests an additional reason why the harmonic components of the synchrotron field will be difficult to access with high efficiency.

To study the sampling and smoothing effects of the antennas, we can proceed in the same way as \S \ref{sec:horn_antennas}. Suppose that an $M \times P$ matrix $\mathcal{U}$ is formed where column $p$ contains the sampled aperture field reception pattern of antenna $p$.  Each column of  $\mathcal{U}$ should be normalised so that $\mathcal{U}^{\dagger} \mathcal{U} = \mathcal{I}_{P}$, where $\mathcal{I}_{P}$ is the $P$-dimensional identity matrix. If the antennas are identical, $\mathcal{U}$ is block circulant. As described previously, a wide range of antennas can be represented in this way. If the antennas comprise a system of E-plane rectangular horns whose phase fronts follow $ {\cal C}_{2}$, and therefore have radii of curvature $-R_{2}$, the representation is trivial, comprising rows of ones in the blocks where the horns are present and zeros elsewhere. If the E field lies in the $z$ direction, as for the 2D system described here, the aperture field has a cosine form, (\ref{eqn:horn_cosine}).
For more general situations the relevant complex valued aperture field distribution, relative to $\mathcal{C}_{2}$, is contained into the columns of $\mathcal{U}$. 

It is apparent that the new matrix $\mathcal{S} = \mathcal{U}^{\dagger} \mathcal{ G}$ describes a mapping from the $N$ complex-valued source currents to the $P$ complex-valued waves appearing at the terminals of the antennas: such as the amplitudes of waveguide modes. As described, the antenna matrix  $\mathcal{U}$ can be modified to include any beam forming network, or indeed microwave scattering networks. It can also be modified to take into account any ohmic or dielectric loss within the antennas themselves. The SVD of $\mathcal{S}$ can be calculated, which yields a set of current distributions that map in one-to-one correspondence with a set of linearly combined travelling wave amplitudes at the outputs of the horns.

\begin{figure}
\noindent \centering
\includegraphics[trim = 2cm 1cm 12cm 16cm, width=60mm]{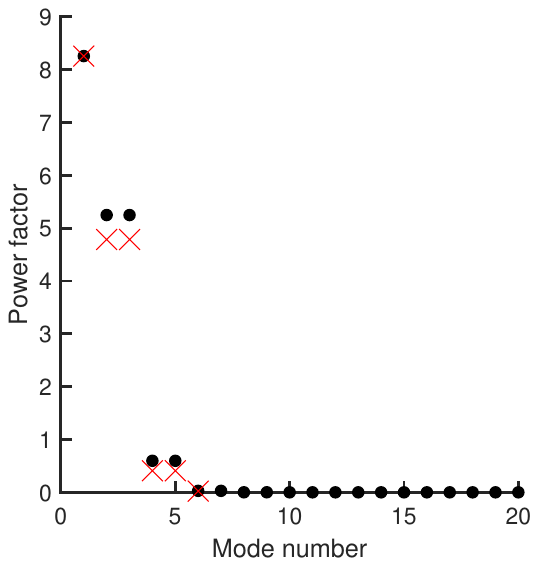}
% left,bottom,right,top
\caption{ 20 line source currents uniformly arranged around an inner circle having a radius of 2 mm. The apertures of 6 E-plane rectangular horns were arranged around a circle having a radius of 100 mm to collect the power available. 93.3 \% of the total power radiated is collected. The remaining power is spread over a large number of high-order Fourier terms, to which the horns are not sensitive.
}
\label{curr_fig3}
\end{figure}

Figure \ref{curr_fig3} shows the result of a simulation where 6 horns,  as shown in Fig.  \ref{figure1}, with uniform aperture fields, are coupled to a CRES region comprising 20 line sources on a 2 mm radius. A uniform aperture field ensures a 100\% filling factor on  $ {\cal C}_{2}$. The black dots show the power transmission factor of each multipole current distribution to $ {\cal C}_{2}$, and the red crosses show the power transmission factors to the horn outputs. The output of a single horn does not correspond to  a particular multipole. Inspection of the left singular vectors of ${\cal S}$ in this case, shows that the ports of the horns are combined in such a way as to create multipole reception patterns; to the order allowed by the number of horns. Fig. \ref{curr_fig3} shows that there is some linear combination of the outputs that couples with high efficiency to the monopole field; likewise for the dipoles and quadrupoles. Fig. \ref{curr_fig3} suggests that 6 horns can be configured very efficiently to get access to the monopole, dipole and quadrupole terms, giving a 93\% efficiency overall. As a check numerical check, for $R_{1} \longrightarrow 0$ an efficiency of 100\% is achieved simply by adding the travelling wave outputs together, which couples to the surviving monopole term. Equally, for $R_{1} \neq 0$, 100\% efficiency can be achieved by increasing the number of horns.

\begin{figure}
\noindent \centering
\includegraphics[trim = 2cm 1cm 12cm 16cm, width=60mm]{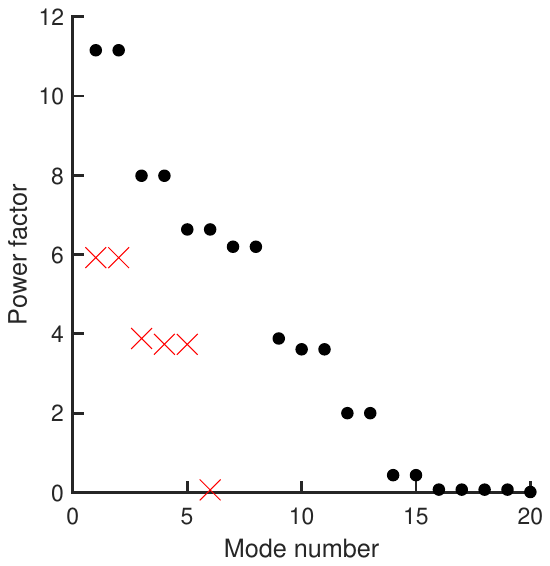}
% left,bottom,right,top
\caption{ 80 line source currents uniformly arranged around an inner circle having a radius of 10 mm. 6 E-plane rectangular horns were arranged around a circle having radius 100 mm to collect the power available. 29 \% of the power is collected.
}
\label{curr_fig4}
\end{figure}

Figure \ref{curr_fig4} shows the same simulation, but with $R_{1}$ increased to 10 mm. Even with this small FoV, the efficiency degrades rapidly. The overall efficiency is now only 29 \%, and only 5 of the chosen 6 degrees of freedom couple to the source region. Interestingly, the monopole term of the power spectrum is no longer the most efficient, and appears as 9th mode in the sequence, as can be seen by looking at the black dots in Fig. \ref{curr_fig3}. It is clear that much of the power is in high-order Fourier terms to which the small number of horns cannot couple. The fact that one of the horn-array degrees of freedom is redundant in this case, and persistently so in simulations, occurs because of the symmetry of the system. This can be corrected by having different sized apertures around the ring, but then we would lose, to some extent, the benefit of having the horns arranged in pairs. The efficiency can be increased to 98 \%, collecting most of the degrees of freedom available, by having 60 horns, which illustrates the problem of achieving a high efficiency and large FoV simultaneously. Overall, the synthesised beam patterns of the horns should be matched to the to the most significant degrees of freedom on  $ {\cal C}_{2}$; the larger the FoV, the more horns are needed. 

\begin{figure}
%fC:/Users/Stafford Withington/Documents/Projects starting in Oxford March 2023/QTNM/PSWF for CRES/CRES_PSWF_CURR_HORN_V1/CRES_EFF_05_04_2023.m
\noindent \centering
\includegraphics[trim = 1cm 1cm 4cm 14cm, width=100mm]{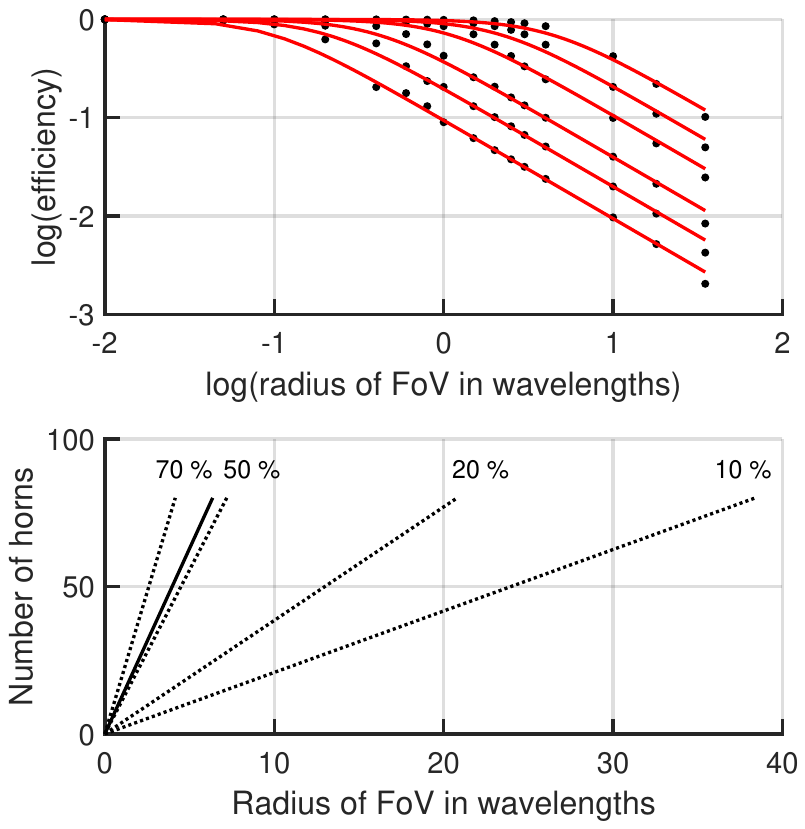}
% left,bottom,right,top
\caption{Top: Log of the total power collection efficiency as a function of the log of the inner source radius in mm. The different curves correspond to 2, 4, 8, 20, 40 and 80 E-plane horns arranged around an outer circle having a radius of  $R_{2}$= 500 mm. Bottom: number of horns needed to sample a given FoV in wavelengths. This data was extracted from the upper plot, and corresponds to total collection efficiencies of 10\%, 20\%, 50\%  and 70\% . The solid line corresponds to the threshold $M = 2 k_{0} R_1 + 1$, and is consistent with \S \ref{sec:fov_field}.}
\label{curr_fig5}
\end{figure}

By repeating simulations of the above kind, it is straightforward to calculate the total power collected as a function of the radius of the source region, $ {\cal C}_{1}$, when different numbers of horns are used: top plot of Fig.~\ref{curr_fig5}.  For  these simulations $\lambda$ = 10 mm, and $R_{2}$ = 500 mm, but the results are insensitive to $R_{2}$. From left to right in the plot, the number of horns used was 2, 4, 8, 20, 40 and 80. The red lines show the functional form
\begin{equation}
\eta = \frac{1}{\sqrt{1 + R_{1}^{2}/\beta}},
\end{equation}
where $\beta$ is a parameter that characterises the radius at which the efficiency starts to fall. There is some small-scale structure in the efficiencies, black dots, particularly for small numbers of horns, because the horns only couple to specific multipoles. The dotted lines in the lower plot in Fig.~\ref{curr_fig5} show the number of horns needed to sample a given FoV in wavelengths. This data was extracted from the upper plot, and corresponds to total collection efficiencies of 10\%, 20\%, 50\%  and 70\%. The solid line corresponds to the threshold $M = 2 k_{0} R_1 + 1$ consistent with the 60 \% value found in \S \ref{sec:fov_field}.

It can be seen that the number of horns needed scales linearly with the radius of the FoV. To achieve a high efficiency over an appreciable FoV, a large number of horns is needed. For example, if 50 antennas and receiver channels can be used, then the radius of the FOV would be 130 mm at $\lambda =10$ mm with 20 \% total collection efficiency. Remember, however, that power not collected will be reflected off of the antennas and may lead to standing wave structure in the response. For this reason efficiencies of order 70 \% are preferable, but with a consequential increase in the number of horns needed. If the allowable efficiency is set at 10 \%, 80 horns will cover a FoV having a radius of 35 wavelengths, which is more in keeping with the needs of a large CRES experiment having a 1 T static field and operating at 27 GHz. Likewise, 8 horns will cover  a FoV having a radius of 35 wavelengths if only 1 \% efficiency is acceptable. Additional antenna sections will, however, be needed in the longitudinal direction, depending on the geometry chosen, \S  \ref{sec:geometry}.

The number of degrees of freedom needed to cover a given FoV has been considered in a number of different ways, and these all give the same self-consistent result for propagating modes. The key message, therefore, is that it is not possible to cover a large FoV without using an appreciable number of antennas, or sacrificing efficiency. It is also clear that, leaving aside other considerations, there is considerable benefit in moving to long wavelengths to monitor a given volume of gas.

\section{Frequency sensitivity}
\label{sec:antenna_sensitivity_factor}

All of the effects of antenna behaviour on frequency sensitivity, contained in (\ref{eqn: ccrb}), are combined into the single factor $\kappa_a$ for given current distribution  $\mathbf{i}$. In CRES, we shall assume, for now, that only a single source is present. The spatial form of the antenna factor, reproduced here for convenience from (\ref{eqn:antenna_factor_map}), is then
\begin{equation}
	\kappa_a (\mathbf{r}_n) = \frac{
		\uvec{i}_n^\dagger \cdot \mathcal{K}^\dagger
		 \cdot \mathcal{L} \cdot
		\mathcal{H} 
		\cdot \mathcal{L}^\dagger \cdot \mathcal{K}
		\cdot \uvec{i}_n^{\vphantom{\dagger}}
	}{
		\text{Tr}[\mathcal{H}]
	}.
\end{equation}

Likewise, (\ref{eqn:h_factor}) can be written
\begin{equation}
	\mathcal{H} = \Bigl[ \mathcal{I} + \tfrac{1}{1 + r} \mathcal{E}^2 \Bigr]^{-1}
\end{equation}
where for brevity we have assumed that the backward noise radiated by the amplifiers is much smaller than the physical temperature of the environment $T_b \ll T_p$, and therefore $r = T_a / T_p$. A sequence of plots were calculated on this basis.

\begin{figure}
     \centering
     \begin{subfigure}[b]{0.48\textwidth}
         \centering
         \includegraphics[width=\textwidth]{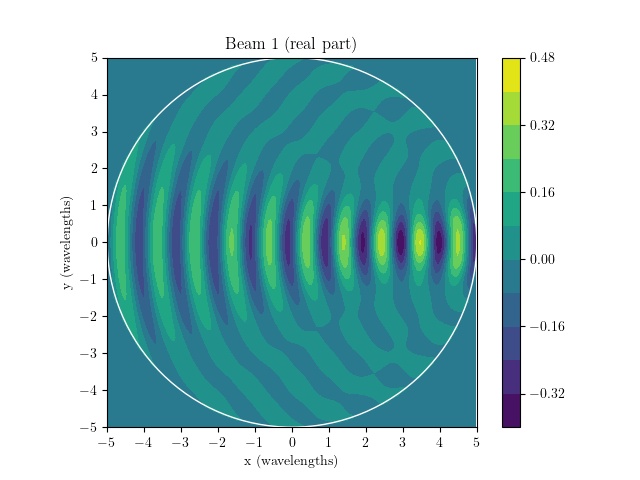}
         \caption{Beam of first antenna (real part).}
         \label{fig:y equals x}
     \end{subfigure}
     \begin{subfigure}[b]{0.48\textwidth}
         \centering
         \includegraphics[width=\textwidth]{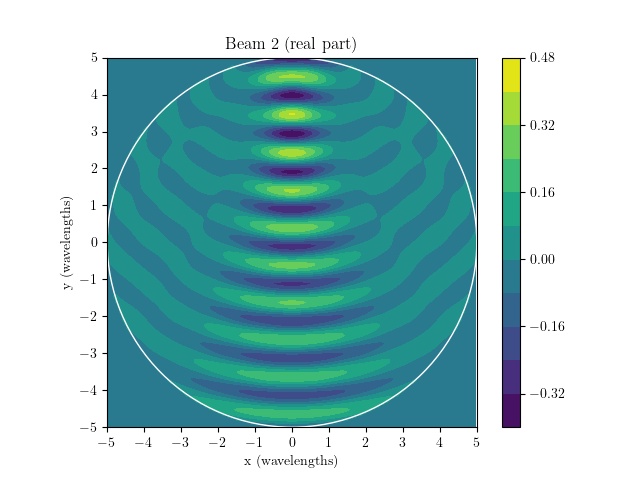}
         \caption{Beam of second antenna (real part)}
     \end{subfigure}
     \\
     \begin{subfigure}[b]{0.48\textwidth}
         \centering
         \includegraphics[width=\textwidth]{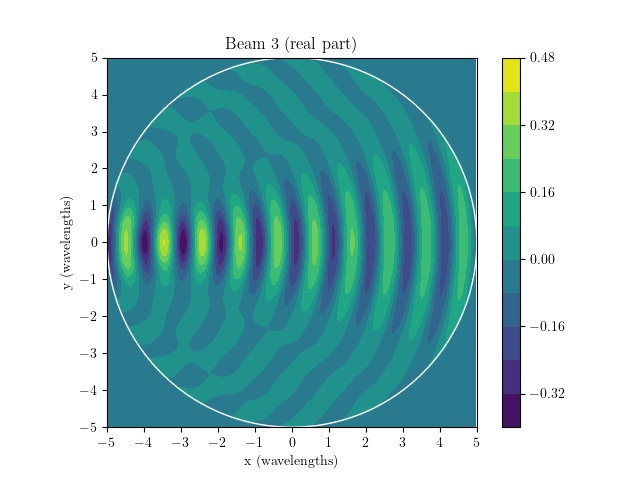}
         \caption{Beam of third antenna (real part)}
     \end{subfigure}
     \begin{subfigure}[b]{0.48\textwidth}
         \centering
         \includegraphics[width=\textwidth]{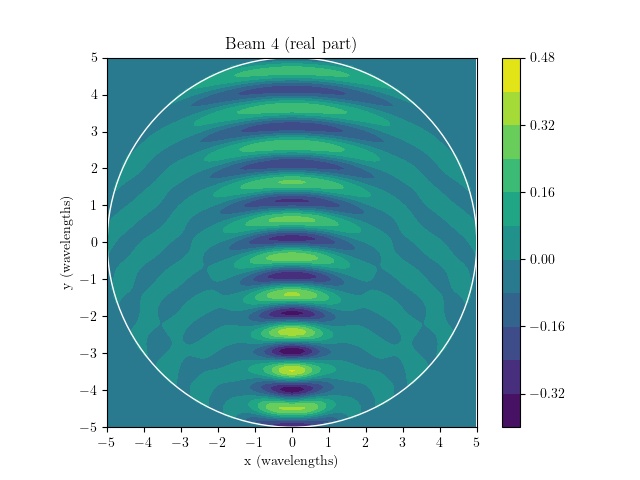}
         \caption{Beam of fourth antenna (real part)}
     \end{subfigure}
     \\
     \begin{subfigure}[b]{0.48\textwidth}
         \centering
         \includegraphics[width=\textwidth]{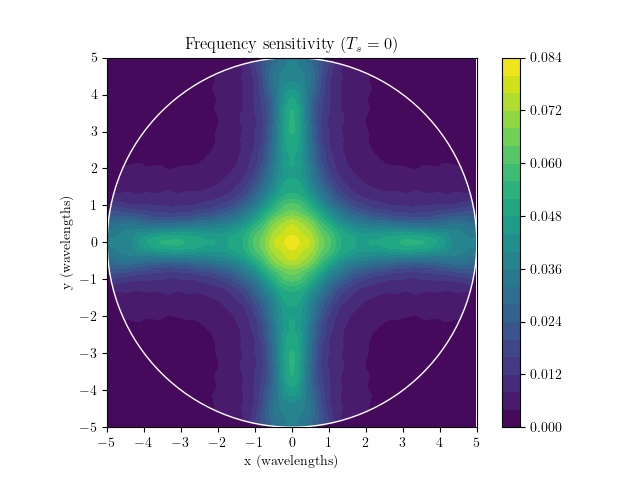}
         \caption{Antenna factor when $T_p = 0$.}
     \end{subfigure}
     \begin{subfigure}[b]{0.48\textwidth}
         \centering
         \includegraphics[width=\textwidth]{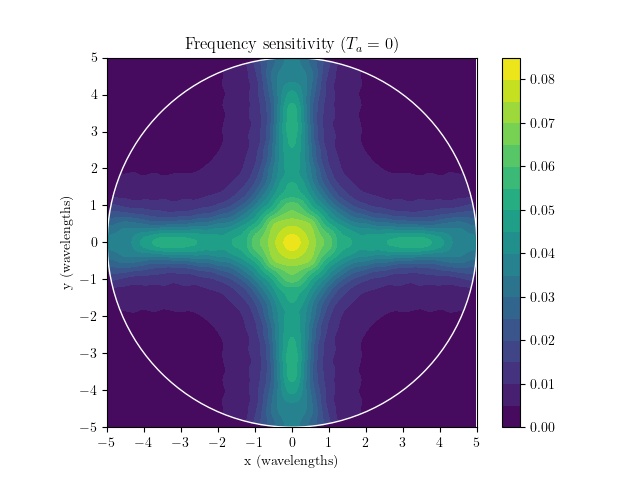}
         \caption{Antenna factor when $T_a = 0$.}
     \end{subfigure}
\caption{\label{fig:narrow_horns} Beam patterns and antenna factors of a set of 4 `narrow' horns.}
\end{figure}

\begin{figure}
     \centering
     \begin{subfigure}[b]{0.48\textwidth}
         \centering
         \includegraphics[width=\textwidth]{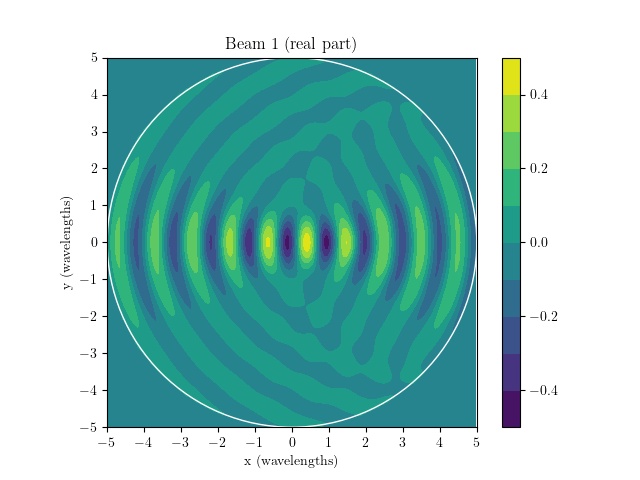}
         \caption{Beam of first antenna (real part).}
         \label{fig:y equals x}
     \end{subfigure}
     \begin{subfigure}[b]{0.48\textwidth}
         \centering
         \includegraphics[width=\textwidth]{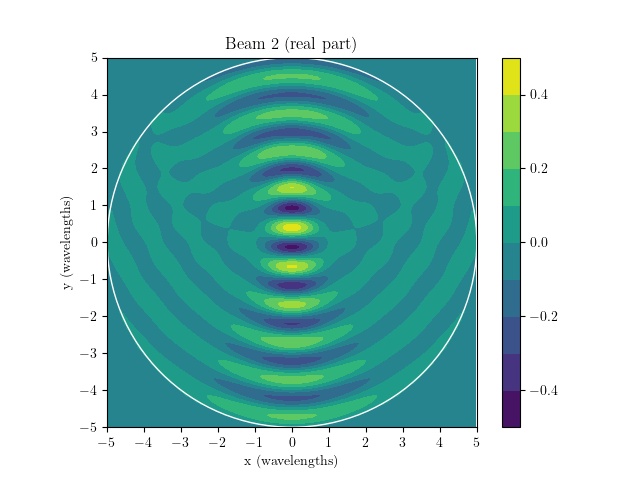}
         \caption{Beam of second antenna (real part)}
     \end{subfigure}
     \\
     \begin{subfigure}[b]{0.48\textwidth}
         \centering
         \includegraphics[width=\textwidth]{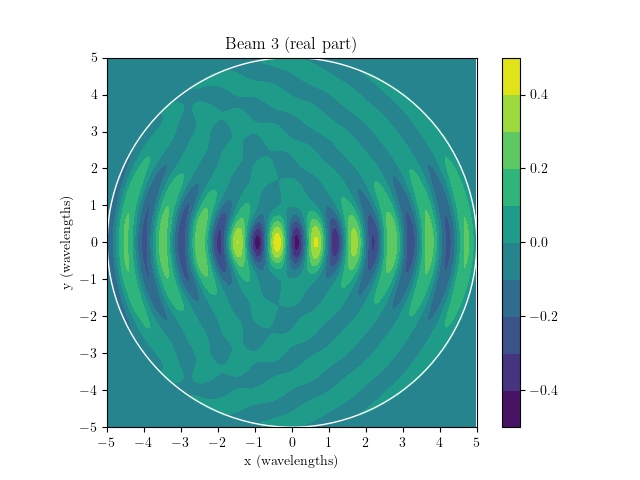}
         \caption{Beam of third antenna (real part)}
     \end{subfigure}
     \begin{subfigure}[b]{0.48\textwidth}
         \centering
         \includegraphics[width=\textwidth]{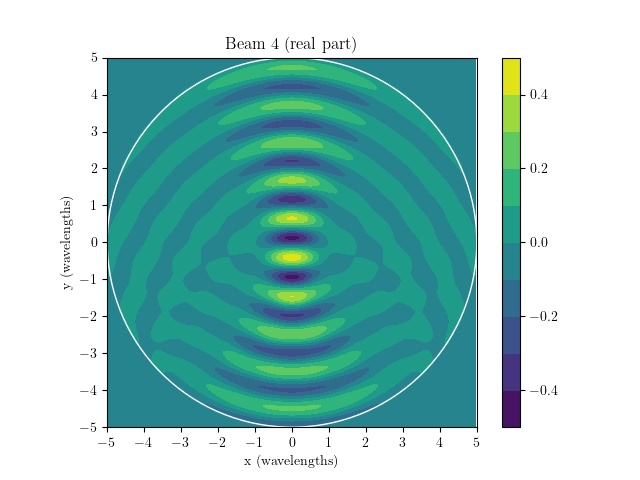}
         \caption{Beam of fourth antenna (real part)}
     \end{subfigure}
     \\
     \begin{subfigure}[b]{0.48\textwidth}
         \centering
         \includegraphics[width=\textwidth]{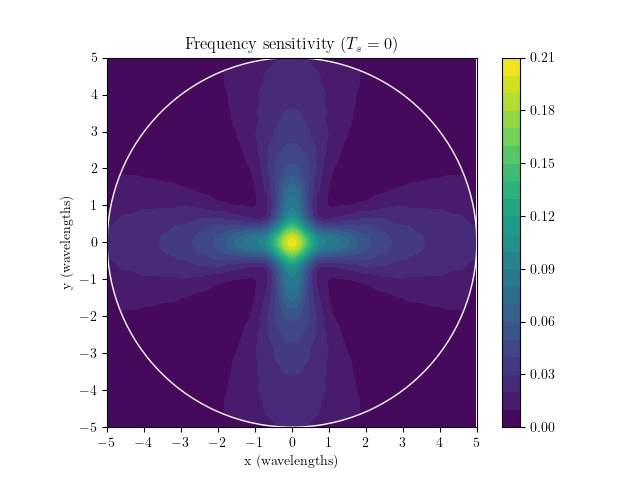}
         \caption{Antenna factor when $T_p = 0$.}
     \end{subfigure}
     \begin{subfigure}[b]{0.48\textwidth}
         \centering
         \includegraphics[width=\textwidth]{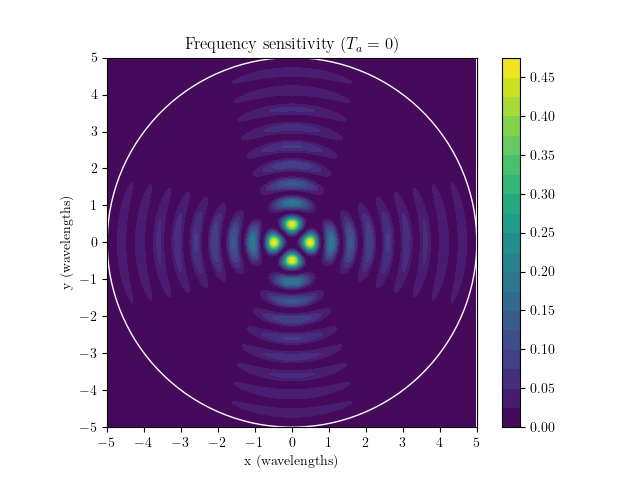}
         \caption{Antenna factor when $T_a = 0$.}
     \end{subfigure}
     \caption{\label{fig:wide_horns} Beam patterns and antenna factors of a set of 4 `wide' horns.}
\end{figure}

\begin{figure}
     \centering
     \begin{subfigure}[b]{0.48\textwidth}
         \centering
         \includegraphics[width=\textwidth]{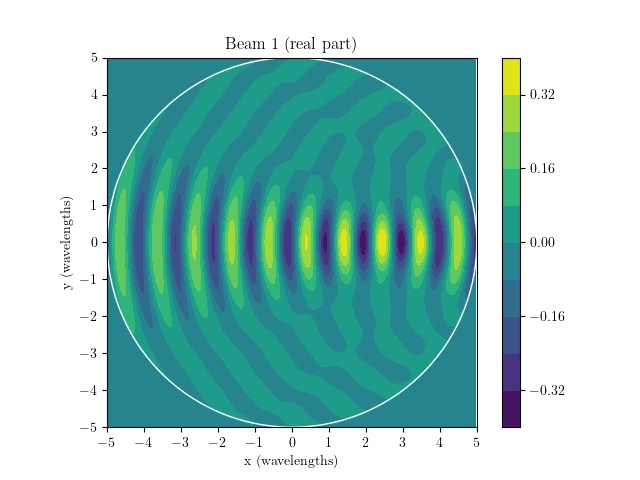}
         \caption{Beam of first antenna (real part).}
         \label{fig:y equals x}
     \end{subfigure}
     \begin{subfigure}[b]{0.48\textwidth}
         \centering
         \includegraphics[width=\textwidth]{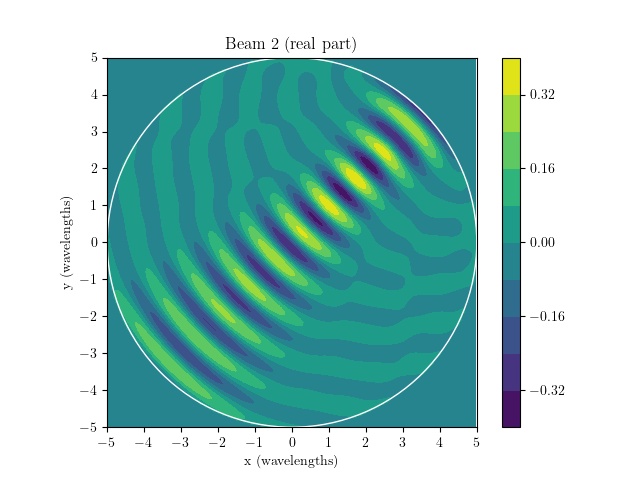}
         \caption{Beam of second antenna (real part)}
     \end{subfigure}
     \\
     \begin{subfigure}[b]{0.48\textwidth}
         \centering
         \includegraphics[width=\textwidth]{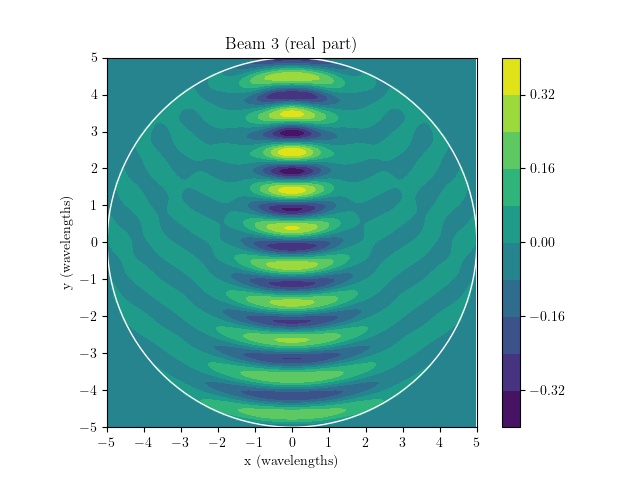}
         \caption{Beam of third antenna (real part)}
     \end{subfigure}
     \begin{subfigure}[b]{0.48\textwidth}
         \centering
         \includegraphics[width=\textwidth]{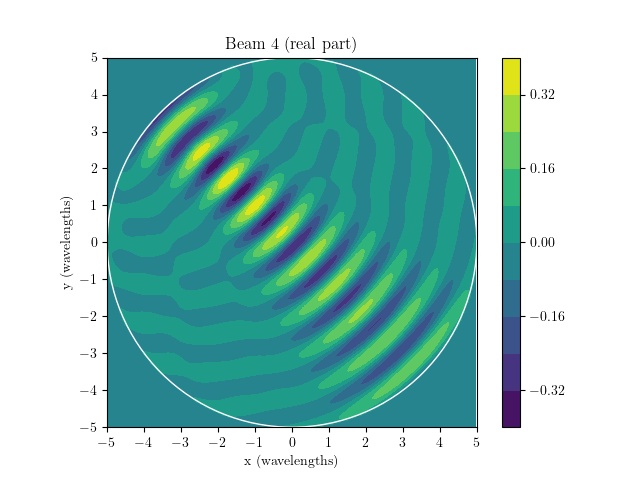}
         \caption{Beam of fourth antenna (real part)}
     \end{subfigure}
     \\
     \begin{subfigure}[b]{0.48\textwidth}
         \centering
         \includegraphics[width=\textwidth]{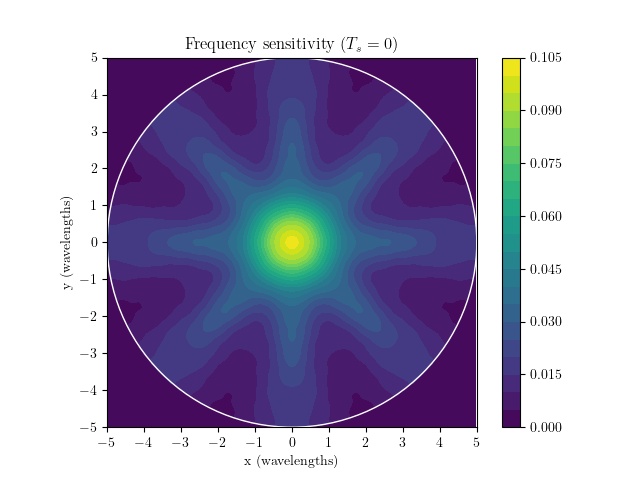}
         \caption{Antenna factor when $T_p = 0$.}
     \end{subfigure}
     \begin{subfigure}[b]{0.48\textwidth}
         \centering
         \includegraphics[width=\textwidth]{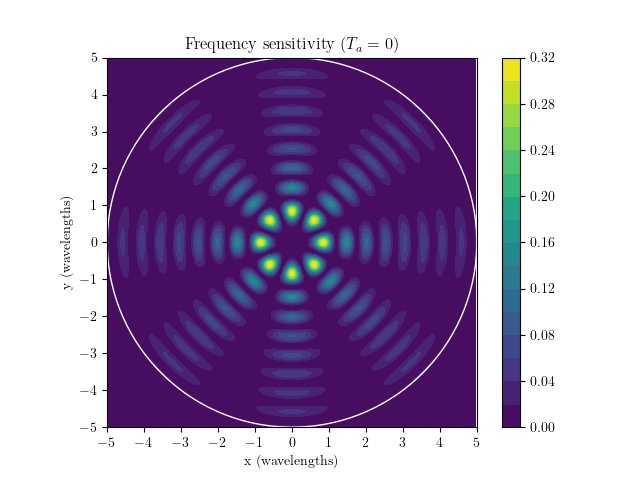}
         \caption{Antenna factor when $T_a = 0$.}
     \end{subfigure}
     \caption{\label{fig:eight_horns} Beam patterns and antenna factors of a set of 8 horns. Only the first 4 beam patterns are shown.}
\end{figure}

The following figures show the results of a number of simulations where a set of cosine-aperture-field antennas were spaced evenly around a circular boundary having $R_2=5 \lambda$. Figure \ref{fig:narrow_horns} shows results for a set of 4 `narrow' horns, where the aperture of each subtends an angle of $\pi / 5$ radians at the centre of the circle. Figure \ref{fig:wide_horns} shows equivalent results for a set of 4 `wide' horns, where each antenna subtends an angle of $\pi / 2$. Figure \ref{fig:eight_horns} shows equivalent results for a set of 8 close-packed horns, where each antenna subtends an angle of $\pi / 4$. In all cases, (a)--(d) show the calculated real part of response patterns, or `beams', of each antenna;  in the last case, only the first 4 beams are shown. These results indicate expected behaviour, including the fact that the focus is not at the centre of ${\cal C}_2$ as a consequence of the phase front of the horn being $R_2$: refer to the discussion in \S \ref{sec:can_arr}. Subplots (e) and (f) show the forms of the spatial antenna factor  $\kappa_a (\mathbf{r}_n)$ in two extreme cases where either the amplifier noise or environmental noise dominates.

The antenna factor reveals interesting behaviour. First, the wider horns give a higher overall antenna factor, which is expected because they collect more power, but unfortunately the response is concentrated in a smaller area, which is again expected because the beam is narrower. Second, there is more structure in the energy resolution when environmental  thermal noise dominates. This effect is particularly pronounced in Fig.s \ref{fig:wide_horns}(f) and \ref{fig:eight_horns}(f), and seems to occur because evanescent modes can achieve extremely high signal to noise ratios and so can dominate the sensitivity pattern. The `smoother' behaviour, when amplifier noise dominates, results from an averaging over all mode patterns. This subtle effect is important and illustrates why electromagnetic response, noise and data analysis methods must all be considered together in order to understand behaviour. 

\section{Determining position}
\label{sec:det_pos}

Fisher information characterises what is possible using the system; it places a bound on the best energy resolution possible; it does not indicate what phasing network should be used to achieve this resolution. In fact $\mathcal{P}$ does not appear in (\ref{eqn: ccrb}). Once the data from all of the antennas is stored, multiple off-line processing techniques become possible. In this section we consider the problem of determining an electron's position. Once this is known, a software-defined beamforming algorithm can be used to put as much of the signal as possible into a single output, which can be used to determine the frequency of the radiator. In this section we consider the application of the MUSIC algorithm \cite{Schmidt_1986} to CRES, which has been used successfully in other direction-finding and frequency-identification applications. 
 
For brevity, the output of the receiver chain (\ref{eqn:signal_chain}) can be written
\begin{equation}
\label{eqn:signal_chain_generic}
\mathbf{o} = \mathcal{T} \mathbf{i} + \mathbf{n},
\end{equation}
$\mathbf{o}$ is a vector containing the complex travelling wave amplitudes of the outputs of the $N$ receivers, and likewise $\mathbf{n}$ is a vector containing the noise travelling wave amplitudes.  $\mathbf{i}$ is an $M$-dimensional vector of possible source line currents; although generally, only one, possibly two, of the elements will be active at any one time in accordance with the lack of knowledge about where an electron will appear. If two or more electrons appear, the sources will be incoherent. $\mathcal{T}$ is an $N \times M$ matrix that maps line currents onto outputs: the details of which were derived in previous sections. We shall assume, although not strictly necessary, that the frequency of the radiators is approximately known, and the problem is how to find their positions, allowing for the unlikely event that two electrons having similar frequencies may appear simultaneously. Each of the terms in (\ref{eqn:signal_chain_generic}) is a complex analytic signal having some central frequency $\omega$, although this is not indicated explicitly.

Ultimately, the only quantities that can be measured easily are the cross correlations between the receiver outputs:
\begin{align}
\label{eqn:signal_chain_corr}
\langle \mathbf{o}  \mathbf{o}^{\dagger} \rangle & =   \mathcal{T} \langle \mathbf{i} \mathbf{i}^{\dagger} \rangle \mathcal{T}^{\dagger}  + \langle \mathbf{n} \mathbf{n}^{\dagger} \rangle \\ \nonumber
\mathcal{O} & = \mathcal{T}  \mathcal{J} \mathcal{T}^{\dagger} + \mathcal{N},
\end{align}
where $\mathcal{O}$, $\mathcal{J}$ and  $\mathcal{N}$ are covarience matrices of zero-mean processes.

It is important to appreciate that $\mathcal{O}$ now corresponds to actual data, which in reality is an estimator of the true correlations.  $\mathcal{T}  \mathcal{J} \mathcal{T}^{\dagger}$ is Hermitian and so can be written in diagonal form. However, the number of radiators present over the observation period is much smaller than the number of receivers, and so   $\mathcal{T}  \mathcal{J} \mathcal{T}^{\dagger}$ is highly singular. $\mathcal{J}$ is singular not least because the number of actual sources is much smaller than the number of possible sources. Assume that the noise sources are uncorrelated and of equal strength, $\mathcal{N} =\sigma^2  \mathcal{I}$, which occurs for an ideal receiver-noise limited system, or for a system where all sources of environmental and ohmic noise are at similar temperatures. For a background-noise limited system, with large temperature differentials, which is usually not the case in practice, this assumption can be relaxed, at the cost of increased complexity. Those very few eigenvectors of  $\mathcal{T}  \mathcal{J} \mathcal{T}^{\dagger}$ that have appreciable eigenvalues correspond to the signal subspace: they definitely contain signal. Those eigenvectors corresponding to the null space, definitely do not contain signal, and we shall refer to them as the noise subspace. Resolving the the identity matrix  $\mathcal{I}$ into the eigenvectors of  $\mathcal{T}  \mathcal{J} \mathcal{T}^{\dagger}$ it can be appreciated that the eigenvectors of $\mathcal{O}$ are the same as the eigenvectors of $\mathcal{T}  \mathcal{J} \mathcal{T}^{\dagger}$. The eigenvalues of the noise subspace are all $\sigma^2$, and so the noise subspace is degenerate. Overall, it follows that the observed covariance matrix $\mathcal{O}$ can be diagonalised to identify the signal and noise subspaces of any particular meaurement, under the apriori assumption that very few localised sources are present, and that the SNR is reasonable. Due to the Hermitian form of  $\mathcal{O}$, the signal and noise subspaces are orthogonal.

Now form a new matrix $\mathcal{E}$ whose columns contain the eigenvectors of the noise subspace. If $D$ sources are present, usually only one,  $\mathcal{E}$ is a $N \times (M-D)$ matrix. Each column of $\mathcal{T}$ corresponds to the outcome of a possible sighting, and therefore we can search for the column that has the smallest projection onto the noise subspace. A suitable measure of orthogonality is the length of the projection vector, giving
\begin{equation}
d_{n}^2 = \mathbf{t}_{n}^{\dagger} \mathcal{E} \mathcal{E}^{\dagger}  \mathbf{t}_{n} \hspace{5mm} \forall \, n,
\end{equation}
where  $\mathbf{t}_{n}$ is the $n$'th column vector of  $\mathcal{T}$. If the noise subspace has been correctly identified, then any signal vector that lies wholly in the signal subspace must be orthogonal to the noise subspace.  If the source resides at one of the locations $n'$, then $d_n'=0$ when the location has been found. A suitable position estimator is
\begin{equation}
\label{eqn:MUSIC_est}
P_n = \frac{1}{ \mathbf{t}_{n}^{\dagger} \mathcal{E} \mathcal{E}^{\dagger}  \mathbf{t}_{n}} \hspace{5mm} \forall \, n,
\end{equation}
which is sharply peaked at $n=n'$. Because the data is a sample variance, subject to change, the $P_n$ is not singular.  By plotting $P_n$ as a function of $n$ the most likely position of the electron can be found. The method uses the whole of the noise subspace to identify the most likely position, rather than projecting the measured eigenvectors of the sources onto the $\mathbf{t}_{n}$. The MUSIC algorithm determines the system operator from the data, provides an unbiased estimate of position, and approaches the CRB on position.

Once the most likely  $\mathbf{t}_{\rm act} =\mathbf{t}_{n'}  $ is known, or are known if there are multiple sources, the actual system operator $\mathcal{T}_{\rm act}$ can be reconstructed, omitting the noise subspace. The source coherence matrix can then be determined though 
\begin{align}
\label{eqn:rec}
\mathcal{J}_{\rm act} & = \mathcal{T}^{-1}_{\rm act} \left[  \mathcal{O}  - \mathcal{N} \right] \left( \mathcal{T}^{\dagger}_{\rm act} \right)^{-1},  
\end{align}   
where $\mathcal{N}$ is known through knowledge of $\sigma^2$, taken from the eigenvalues of $\mathcal{O}$ in the noise subspace. Or if the system is stable, $\mathcal{N}$ can be measured when it is certain that no radiating electrons are present. $\mathcal{T}^{-1}_{\rm act}$ denotes the Moore-Penrose inverse of the actual system matrix, reconstructed from the MUSIC estimator. Ordinarily, for a long enough sampling time, any multiple electron events will not be correlated and $\mathcal{J}_{\rm act}$ will be diagonal.

It could be argued that if $\mathcal{T}$ is known from modelling or calibration, (\ref{eqn:rec}) could be calculated anyway using $\mathcal{T}^{-1}$ rather than $\mathcal{T}^{-1}_{\rm act}$, and the Moore-Penrose inverse would isolate that part of the space containing signal, and ensure best projection.  However, MUSIC can be applied when the test vector $\mathbf{t}_{n}$ is a continuous function of some parameter, say position $\mathbf{t}({\bf r})$, and so can be used when there is an analytical description of the system's response function in terms of the continuous position vector ${\bf r}$. This method provides an efficient way of finding an electron's position within the CRES region.  The algorithm provides software-based weighting vectors that can be used to put as much of the signal as possible, from each of the electrons, into individual channels, where their frequencies can be measured. The possibility of using MUSIC to determine position and frequency simultaneously is another line of consideration \cite{Hongtao_2009}.

\section{Beam synthesis}
\label{sec:synth}
 
In this final section, we consider CRES antenna systems from the perspective of beam-sythesis. We continue to take the view that it is best to use a software-based phasing or weighting network, rather than trying to tackle the daunting task of engineering a cryogenically cooled low-loss phasing network ahead of the amplifiers. We also assume that the best arrangement is a phasing network that maps as best as possible each source position onto a single output, such that a single frequency measurement yields a result with the needed accuracy. We will use the parlance of beam synthesis and frame theory throughout.

The first question is what representation should be used for the source. It is straightforward to show that the radius of a cyclotron orbit is $r_e / \lambda = \beta /2 \pi$ , which for an 18.6 keV  electron evaluates to $r_e / \lambda = $0.043, showing that the orbit is much smaller than the wavelength. It follows that the orbit cannot ordinarily be resolved. Whilst using a delta function as the current density distribution is reasonable, there are an infinite number of delta functions over any FoV and so strictly the problem of finding the location can never be well conditioned. Using a localised displaced set of basis functions, such as wavelength-sized Gaussians (Gabor set) \cite{Berry_2004_1}, is an attractive alternative, but many other possibilities such as Shapletts \cite{Berry_2004_2} are also of interest. All of the basis functions have the property that they are complete or over-complete with respect to some vector space \cite{Withington_2004}, but they are not orthogonal, and therefore it is beneficial to describe beam-synthesis in terms of frame theory \cite{Withington_2006, Withington_2008}.

For a suitable set of basis functions with respect to some FoV, there can be a one-to-one mapping because enough antennas are available, by definition. If the source lies outside of the FoV insufficient antennas are available but frequency information may still be obtainable in certain locations, and so what is the best phasing network in this case? If the source is within the FoV, the mapping between the position subspace associated with the FoV and the antenna outputs is unitary. The unitary subspace of the Morse-Penrose inverse $\mathcal{K}^{-1}$ achieves the best possible positional mapping, and can be realised using a unitary phasing network: $\mathcal{P} = \left[ \mathcal{K}^{-1}\right]_{\rm sub}$. If the source lies outside of the FoV, the  pseudo inverse $\mathcal{K}^{-1}$ is still the best choice because it achieves the best least-squares fit to position with the number of degrees of freedom available. For example, 5 horns could recover the monopole, dipole and quadrupole terms, and so could uniquely couple to 5 localised spatial regions. In this case, the phase network merely reverses the linear combinations that give rise to the multipoles. This choice is equivalent to the Butler phasing network used in linear arrays, which has the property that it maps directions onto ports.

For a noiseless system, (\ref{eqn:signal_chain_generic}) becomes
\begin{equation}
\label{eqn:signal_chain_generic_nonse}
\mathbf{o} = \mathcal{T} \mathbf{i}.
\end{equation}
We shall now call $\mathcal{T}$ the {\em synthesised beam operator} because each row is a synthesised `reception pattern' that maps a current distribution onto a specific receiver output. 

$\mathcal{T}$ maps current in a position basis onto the basis of transmission line outputs. $\mathcal{T}$ does not necessarily have a left inverse, because there are generally fewer antenna channels than spatial points to be observed, but the pseudo-inverse can be used to recover the information available:
\begin{eqnarray}
\mathbf{i}'  & = \mathcal{T}^{-1} \cdot \mathbf{o} \\ \nonumber
 & = \mathcal{T}^{-1} \, \mathcal{T} \cdot \mathbf{i}
\end{eqnarray}
$\mathbf{i}' =  \mathbf{i}$ only to the extent to which the set of synthesised beams forms a complete set. The derived current distribution is therefore the best fit to the actual current distribution given the information available. The SVD and the pseudo-inverse are given by 
\begin{equation}
\mathcal{T} \approx \sum_{n} \mathbf{u}_{n} \sigma_{n} \mathbf{v}^{\dagger}_n
\end{equation} 
then 
\begin{equation}
\mathcal{T}^{-1} \approx \sum_{n} \mathbf{v}_{n} \sigma_{n}^{-1} \mathbf{u}^{\dagger}_n.
\end{equation} 
The summation has to be limited to the subspace for which $|\sigma_{n}| > 0$: the observable currents. The synthesised reception patterns become 
\begin{align}
\mathbf{t}_{p}^{\dagger}  & =  \mathbf{o}_{p}^T \cdot \mathcal{T} \\ \nonumber
                        & =   \sum_{n} \mathbf{o}_{p}^T \cdot  \mathbf{u}_{n} \sigma_{n} \mathbf{v}^{\dagger}_n \\ \nonumber
\mathbf{t}_{p}   & =  \sum_{n} \mathbf{v}_n    \sigma_{n}   \mathbf{u}_{n}^{\dagger}  \cdot \mathbf{o}_{p},
\end{align}
where $\mathbf{o}_{p}$ are the unit basis vectors associated with the individual ports. This operation extracts the rows of $\mathcal{T}$, which are the sythesised reception patterns $\mathbf{t}_{p}^{\dagger}$. Equivalently, $\mathbf{t}_{p}$ are the current vectors onto which the current distribution must be projected to get the individual outputs, $o_p = \mathbf{t}_{p}^{\dagger} \cdot \mathbf{i}$. Likewise
\begin{align}
\widetilde{\mathbf{t}}_{p}  & =  \mathcal{T}^{-1} \cdot  \mathbf{o}_{p}  \\ \nonumber
                        & =   \sum_{n}  \mathbf{v}_{n}  \sigma_{n}^{-1}   \mathbf{u}_{n}^{\dagger} \cdot  \mathbf{o}_{p} 
\end{align}
This operation extracts the columns of $\mathcal{T}^{-1}$, which are the current vectors (beam patterns) that must be used to reconstruct the total current distribution, as best as possible, once the individual ouputs are known: $\mathbf{i}'  =  \sum_{p} o_p  \widetilde{\mathbf{t}}_{p} $. The vectors $\widetilde{\mathbf{t}}_{p}$ are formally the duals of the vectors $\mathbf{t}_{p}$ even when the current basis is non-orthogonal. Thus we refer to $\mathbf{t}_{p}$ and $\widetilde{\mathbf{t}}_{p}$ as the beams and dual beams respectively. As such they resolve the identity over any current distribution within the FoV: $ \mathcal{I}_{p} = \sum_{p} \widetilde{\mathbf{t}}_{p}  \mathbf{t}_{p}^{\dagger}$. Using this resolution, the source-recovery process, written in terms of frame theory becomes
\begin{eqnarray}
\mathcal{J}'  & = \sum_{pq} \widetilde{\mathbf{t}}_{p} \left\{ \mathbf{t}_{p}^{\dagger}   \mathcal{J}     \mathbf{t}_{q} \right\}  \widetilde{\mathbf{t}}_{q}^{\dagger} \\ \nonumber
& = \sum_{pq}  J_{pq}  \widetilde{\mathbf{t}}_{p}  \widetilde{\mathbf{t}}_{q}^{\dagger}
\end{eqnarray} 
where the quantity in braces $J_{pq}$ is the covariance between the outputs of two channels in the noiseless case. These coefficients  are used to reconstruct  the covariance matrix of the current distribution through the use of the dual beams. When noise is included
\begin{eqnarray}
& = \sum_{pq}  \left[ O_{pq}  - N_{pq} \right] \widetilde{\mathbf{t}}_{p}  \widetilde{\mathbf{t}}_{q}^{\dagger}.
\end{eqnarray} 

In general terms, there is a close relationship between the MUSIC algorithm and the frame-theory of phased arrays. Whereas MUSIC recovers the system response operator from the data of the actual measurement being analysed, frame theory inverts the whole the system operator including those locations where sources may appear, but for any single observation are mostly unoccupied.

\section{Discussion}
\label{sec:disc}

A range of topics relating to the design, analysis and optimisation of single-electron CRES experiments comprising arrays of inward-looking microwave antennas has been considered. Whilst there is a great deal of literature relating to the design of outward looking phased arrays for applications such as radar and telecommunications, there is very little coverage of the new issues that come into play when designing inward looking phased arrays for volumetric imaging. Our paper presents a single framework that allows signal, noise and signal processing to be considered. The framework allows many different effects to be included, and gives conceptual insights into those factors that drive system-level performance. A few design considerations are as follows:

\begin{enumerate}

\item When observed at its fundamental frequency, a CRES electron in a cylotron orbit is best described by a rotating electric dipole, or equivalently two orthogonal out-of-phase electric dipoles. A magnetic dipole is not appropriate. The radius of the orbit is much smaller than a wavelength, and so the orbit cannot be resolved by far-field electromagnetic observations.

\item Single-electron CRES events are short lived, and so it is essential to achieve high SNR's with short observation times. Cryogenically cooled amplifiers are necessary, and a low-temperature radiometric environment is highly desirable. A quantum noise-temperature-limited system is a realistic goal, giving 1.3 K at 27 GHz for a 1T field.

\item For narrow-band observations, an inefficiency occurs with respect to the total radiated power, because power is lost to harmonics associated with the rotating synchrotron beam. The fundamental frequency contains 87 \% of the total radiated power. The inclusion of a magnetic trap causes additional power to be lost into nearby sidebands, unless these can also be measured with high signal-to-noise. 

\item The geometrical arrangement of the antennas, to some extent independent of their individual forms, imposes a further inefficiency. If the antennas are placed on a cyclindrical surface, only some fraction of the radiated power flows normally across the surface; any power travelling through the end caps is lost.  For example, for a cylinder whose length is equal to its diameter, 62 \% of the total power radiated by a central electron flows out radially through the surface. For a length to diameter ratio of 0.1, typical of a single ring of antennas, this efficiency falls to 10 \%. An additional inefficiency occurs for directive antennas, which further degrades the amount of power collected. Two flat, parallel, arrays of on-axis antennas, with holes to allow beam access, is an interesting arrangement that can be more efficient for certain FoV's. A spherical surface is a natural solution, but does not lead to a significant reduction in the total number of antennas needed, and the polarisations of the antennas reception patterns would need to vary over the array, or dual-polarisation receivers used, leading to channel redundancy.

\item To achieve a high coupling efficiency to a single CRES event, regardless of where it appears within the FoV, a large number of antennas and receivers is needed. There is a fundamental trade off between the point-source coupling efficiency, the size of the FoV, and the number of receivers used. The number of antennas needed in a single ring to cover a FoV having radius $R_1$ is approximately $2 k_{0} R_1 + 1$; which corresponds to the circumference of the FoV in wavelengths. A large number of antennas is needed to cover even a small FoV, and so it is essential to be able to manufacture and operate large arrays of near quantum-noise-limited receivers. Equivalently, if the number of antennas is fixed, the FoV can be increased by working at a longer wavelength, but then other considerations come into play. 

\item For antennas that face each other, the input impedance of any single antenna is generally not the same as the impedance seen when the antenna is radiating into free space. This complicates the design process considerably, and makes it very difficult to interface the outputs of an inward looking array to low-noise electronics. 

\item Weakly coupled resonant trapped modes appear as a consequence of the fact that a coherent receiver can only absorb energy through one electromagnetic mode, and therefore an antenna tends to reflect any signal that is in any of the orthogonal modes.  This problem is intimately related to the notion of a {\em minimum-scattering antenna} \cite{Andersen}, which seeks to make an antenna invisible when illuminated by any general incoming field. Realising an antenna that absorbs all of the energy in all of the spatial modes that are orthogonal to the reception pattern is a topic of considerable interest. Different antenna types have different scattering cross sections, and it would be interesting to apply the principles of minimum scattering antennas to the design of CRES systems. 

\item High Q-factor trapped modes that are isolated from the antennas reception patterns will necessarily be generated, and these can lead to back action on the electron, even if they are not apparent in the signals appearing at the antenna ports. 

\item It should be mentioned that photometric arrays operate in a very different way, and have some optical properties that would be ideal for CRES systems; however, they are not capable of achieving the high frequency-resolution required. An intermediate sensor, such as a homodyne detection scheme, does not solve the problem because the pump induces coherence on the reception pattern, and we are back to the antenna problem.

\item For many types of antenna, the single-electron source will be in the near field of the antenna, and the antennas may be in the near fields of each other, and this complicates behaviour considerably because of the increased number of spatial modes involved. When evanescent modes are available for coupling the source to the antennas, there seems to be a marked difference in the uniformity of response of background-noise and amplifier-noise limited systems, which needs to be investigated further. The precise behaviour depends on how the evanescent modes couple to the environment.

\item It is possible, with constraints on the configuration, to engineer the individual antennas so that the array only couples weakly to the thermal noise of the background.  This makes it possible to achieve receiver-limited sensitivity even with high thermal backgrounds. There are benefits of using aperture-phase corrected horns so that the beam of each antenna couples well to the other antennas. In this way the coupling to the thermal background can be minimised. For small fields of view involving just a small number of antennas ($<$10) this scheme works well. With care, background noise rejection can also be achieved when a large number of antennas is used. A -20 dB coupling loss would allow a quantum-noise limited system to work in a 70 K environment. This level of isolation could open the door to the use of squeezing, and noise temperatures below the quantum limit. Ensuring that a single antenna is well coupled tp the beams of some subset of the others also means that the input impedance of the antennas can be easily matched to 50 $\Omega$.

\item All amplifiers radiate noise from their inputs, which may or may not be comparable with the noise temperature. In CRES, this noise will travel back through the antenna system, and become a source of noise for the other receivers. Also, if any of this noise scatters back into the amplifier from which it came, it will lead to changes in noise temperature which can vary rapidly with frequency depending on the length of the path involved. Usually, a circulator is placed ahead of a low-noise amplifier. The circulator ensures that the noise travelling back into the CRES antenna system is uncorrelated with the noise sources in the amplifier, removing the potential for noise detuning. Importantly, the load of the circular can be cooled, to say 100 mK, ensuring that the backward travelling noise has a radiometric  temperature equal to the physical temperature of the load. A subtle point is that if the circulator is at the same temperature as the background environment, and the background has a uniform temperature, the noise waves travelling away from the ports of the antennas are uncorrelated, and this has benefits when analysing data, because only the signal is coherent across the outputs: MUSIC makes this assumption. If the system is background noise limited, and the ports are not all perfectly matched, or the physical temperature is appreciably not everywhere uniform, the noise generated by the background at the antenna outputs is correlated, which makes data analysis more awkward.

\item There is a lower bound, regardless of the data analysis algorithms used, on the error of any frequency measurement, and this error depends on the SNR and therefore where the electron is located. The aim of any data analysis method is to achieve the CRB. The priority is on measuring frequency, but position measurement is also valuable. If the position of the radiating electron is known, weightings can be found that synthesise a beam that has a high coupling efficiency to this single event, or indeed two simultaneous events.

\item If an array of receiver channels having similar characteristics can be used, software-based beam forming does not lead to any degradation over an ideal hardware-based beam forming network placed between the amplifiers and first-stage amplifiers. 

\item Because it is not practical, however, to place a large unitary phasing network, with multiple inputs and the same number of outputs, ahead of the first-stage amplifiers, or indeed even after, the most effective approach  to data analysis is to digitise the outputs of all of the receivers, to form the cross-correlation products, and then to use position-search and frequency-determination software-based algorithms. For example, if two electrons are present in the system at the same time, perhaps dephasing each other, it should be possible to separate out their individual behaviours. If a physical unitary phasing network is used, then a Butler matrix would be close to ideal.

\item The exact form of the individual antennas requires careful  consideration. It seems beneficial to use aperture phase corrected antennas, which may take the form of phase-corrected  horns, profiled horns, or line-fed cylindrical antennas. These allow power to be collected over a large surface area, even when there is a small number of receivers. Patch antennas tend to be of the order of $\lambda /2$ in size, and so do not lend themselves individually, to making phase corrected antennas, although integrated arrays may be possible. Single-dipole and even few-element Yagi antennas have many advantages, and it may be possible to exploit the use of minimum-scattering design techniques. Because a significant fraction of the power is radiated axially, and is circularly polarised, it may be possible to use arrays of on-axis helical antennas, having different hands of polarisation at the two ends of the FoV. 

\item It might be argued that the number of receivers needed can be minimised by using a waveguide detection scheme, where cyclotron electrons are held in a metallic waveguide, and a waveguide probe used to couple the radiation pattern of the electron to an output transmission line. This does not however reduce the number of antennas needed for a given FoV. All that happens is that the free-space Green's function used in \S \ref{sec:full_elec_model} must be replaced by the waveguide Green's function. A single probe will only couple to a single mode made up of a weighted linear combination of waveguide modes, and this will determine the FoV and coupling efficiency. If one wishes to achieve high-efficiency coupling regardless of the position of the electron, multiple waveguide probes are needed, the outputs of which must span the same degrees of freedom present in the waveguide modes themselves, whilst retaining a good input match on each. Those modes that are not read out, including evanescent fields, become trapped as resonant modes whose quality factors are determined by losses: such as radiation from the ends of the waveguide. Many of the waveguide modes are cut off, meaning that they do not radiate if the waveguide is long enough, leading to complex behaviour and backaction. Fundamentally, antenna-based schemes behave in the same way as waveguide-based schemes, the only difference being the shape and nature of the boundary. In fact, the analysis procedure described in this paper could be used to create an elegant model of a waveguide detection system.  The sampling problem exists even if one considers more adventurous schemes such as free-space loop-antenna coupling. A single probe will only achieve a small FoV, depending on what coupling efficiency can be tolerated. The number of receivers needed is a fundamental issue regardless of the configuration used.

\item We have described a 2D model, which captures the essential features of CRES antenna systems. For detailed design purposes, the 2D model can be extended to 3D using exactly the same concepts. The 3D model does not however bring anything fundamentally new to the basic principles already discussed, but introduces polarisation into the analysis. The antennas may be arranged over the surface of a cylinder, on parallel plates or indeed over the surface of a sphere. In the case of a cylinder, antennas must also be packed in the longitudinal direction, and this adds a third direction to the modes already discussed. Now, there is some cylinder that represents the surface of the FoV, and some cylinder that represents the surface over which the antennas are packed. These do not have to be the same length. Indeed, if the outer is in the far-field of the inner, then by analogy with far-infrared optics, it can be expected that longitudinal forms of the modes will be prolate spheroidal wavefunctions. Some compromises will need to be made; for example,  in the case of a spherical array, polarisation becomes a challenging issue, and dual polarisation antennas will be needed. Long cylindrical arrays of linearly polarised antennas or parallel plates of circularly polarised antennas seem to be good choices.

\item Energy resolutions of order meV are in principle achievable over a wide range of parameters, but microwave loss is a significant factor in determining the CRB. The CRB  used in this report should be regarded as a measure of the information available from the apparatus. Many other artefacts will degrade this intrinsic capability: magnetic field nonuniformity, sampling and timing errors, etc. In fact, the matter of how best to determine the {\em initial} energy of a decaying relativistic electron has not to our knowledge been considered in detail. Even for an otherwise perfect system, the matter of waveform sampling is of crucial importance because it relates to whether the CRB is achieved in practice. It is well known that the error on a measurement of frequency, assuming a harmonic tone in noise, achieves the CRB as long at the SNR is greater than some critical value, SNR$_{\rm crb}$, below this critical value the error increases rapidly until some other well defined threshold is reached SNR$_{\rm det}$, where no clear evidence exists about whether the signal is unique or exists at all \cite{Serbes_2022}. SNR$_{\rm det}$ can be regarded as the signal detection limit. Above the threshold SNR$_{\rm crb}$, maximum likelihood estimators achieve the CRB relatively easily, and then the CRB falls as the SNR is increased: consistent with the 5th term in (\ref{eqn: ccrb}). The critical thresholds depend on sampling rate, and for even a modest number of samples, the CRB can be achieved even when the signal is well buried in noise.

It should be appreciated that the matter of achieving a high enough SNR to enable the visualisation of an event, through plotting a temporal sequence of spectra in a waterfall plot,  is not the same as the SNR needed to achieve a precision measurement of energy.

The CRB also arises in the context of determining a source's location, as mentioned in \S~\ref{sec:synth}, but now the sampling relates to the local field sampling imposed by the antenna array. Although the position finding algorithm described is well suited to achieving the CRB on position, it will not be achievable if the number of antennas falls below some critical value. Achieving sufficient spatial field sampling for the SNR available may turn out to impose a bigger demand than the problem of achieving sufficient temporal sampling. 

\item We have not considered the possibility of measuring position and frequency simultaneously through a single data-analysis algorithm: as is done in applications such as radar. This way of proceeding takes us beyond the purpose of this overview, but it should be noted that hierarchical decomposition methods may be valuable in this respect. The data-analysis methods need optimising to take into account the way in which radiative decay causes the frequency to shift, and more care is needed to include the dynamics of turn on. The key point, however, is that it should be possible to extract the energy of an electron to within meV levels using optimised experiments and data-analysis methods.

\end{enumerate}

Finally, we emphasise that although the above concepts have been described in terms of a generic 2D system, they apply equally well to 3D systems. Creating a wholistic model of a 3D system using dyadic Green's functions is straightforward, and would be based on the same method and considerations, and would give rise to the same general behaviour. The application of the method to modelling the overall behaviour of waveguide systems will be decribed in an upcoming paper.

\section{Acknowledgments}

This study was carried out in the context of the UK project Quantum Technology for the measurement of Neutrino Mass: UKRI Grant ST/T006439/1. The authors are grateful to Prof. Saakyen at UCL and the national QTNM team for numerous discussions relating to neutrino mass measurement.  

%\bibliographystyle{ieee} 
%\bibliography{stafford}

\end{document}